\documentclass[%
 reprint,
 amsmath,amssymb,
 aps,
prstab,
]{revtex4-1}

\usepackage{subfigure}
\usepackage{graphicx}
\usepackage{epstopdf}
\usepackage{dcolumn}
\usepackage{bm}

\begin{document}

\preprint{APS/123-QED}

\title{ Thermal loading on crystals in an X-ray free-electron laser oscillator}

\author{Nanshun Huang$^{1,2}$}


\author{Haixiao Deng$^{3,1}$}%
\email{denghaixiao@zjlab.org.cn}

\affiliation{%
$^1$Shanghai Institute of Applied Physics, Chinese Academy of Sciences, Shanghai 201800, China}%
\affiliation{%
$^2$University of Chinese Academy of Sciences, Beijing 100049, China}%
\affiliation{%
$^3$Shanghai Advanced Research Institute, Chinese Academy of Sciences, Shanghai 201210, China}%


\date{\today}

\begin{abstract}
     X-ray free electron laser oscillators (XFELO) is future light source to produce fully coherent hard X-ray pulses. The X-rays circulate in an optical cavity built from multiple Bragg reflecting mirrors, which has a high reflectance in a bandwidth of ten meV level. The X-ray crystal mirrors exposed to intense X-ray beams in the cavity are subjects to thermal deformations that shift and distort the Bragg reflection. Therefore, the stability of the XFELO operation relies on the abilities of mirrors to preserve the Bragg reflection under such heat load. A new approach was used to analyze the heat load of mirrors and the XFELO operation. The essential light-matter interaction is simulated by the GEANT4 with a dedicated Bragg-reflection physical process to obtain the precise absorption information of the XFELO pulse in the crystals. The transient thermal conduction is analyzed by the finite-element analysis software upon the energy absorption information extract from GEANT4 simulation. A simplified heat-load model is then developed to integrate the heat load in the XFELO. With the help of the heat-load model, the analysis of XFELO operating with several cryogenically cooled diamond mirrors is conducted. The results indicate that the heat load would induce an oscillation when XFELO operates without enough cooling.


\begin{description}

\item[PACS numbers] 41.60.Cr

\end{description}
\end{abstract}

\maketitle

\section{\label{sec:introduction} Introduction}

In the hard X-ray regime, the operating free electron lasers (FELs) are based on the SASE (self-amplified spontaneous emission) mode \cite{Ackermann.2007,Emma.2010,Ishikawa.2012,Kang.2017,milne2017swissfel}, which generates X-ray pulses with unique characteristics, such as high spatial coherence, ultra-high peak power, and ultrashort pulse duration. However, the stochastic nature of SASE leads to low temporal coherence and poor pulse-to-pulse stability in the intensity of produced X-ray pulses. In order to overcome these drawbacks, an X-ray free electron laser oscillator (XFELO) has been proposed \cite{kim2008proposal,Lindberg.2011, Dai.2012}, where the X-ray pulses circulate in a low-loss optical cavity formed by multiple Bragg-reflecting crystals. The crystal is used because crystals could reflect hard X-rays with close to 100\% reflectivity and a \~10~meV bandwidth~\cite{AUTHIER.2003b, Shvydko.2011}, whereas conventional mirrors are unavailable in the short wavelength region.

For the cavity having a reasonable length, the repetition rate of the electron bunch has to be about 1 MHz or higher. With intra-cavity pulse energy of about 800 $\mu$J and a beam radius of about 50 $\mu$m at a repetition rate of MHz level, the intense x-ray pulses would impose a high heat load to the mirrors, which results in the lattice distortion. The lattice distortion would decrease the peak of Bragg reflectivity, shift the wavelength range, and increase sidebands~\cite{Rousse.2001, Bargheer.2006}, as the Bragg reflection originates from the X-rays scattering of a periodic atomic lattice. Therefore, the realization of the optical cavity with Bragg reflecting mirrors relies highly on the stability of the crystal lattice. Thus, the heat load of the crystal mirror under such intense X-ray pulses is an essential issue for building an XFELO.

In order to investigate the influence of heat load in the crystal , two types of experiment methods are used to test performance limitations of the crystal thermal deformation~\cite{Kojima.1994, Perrin.2006, vanaerenbergh.2010, Stoupin.2014,  Kozak.2015,Kolodziej2018, Samoylova.2019D}. The first one is to utilize a high power conventional laser in the long-wavelength regime for modelling laser-mirror interactions. The advantage of this method is the possibility of precisely controlling the pulse duration and pulse energy, which is crucial for light-material interactions. However, this approach produces a penetration depth at the nanometer level~\cite{Perrin.2006}, which is far shorter than those produced by X-rays. Besides, the repetition rate of the high power conventional laser is challenging to reach MHz or higher. Another method is to apply the synchrotron radiation light sources for the heat load experiments. The synchrotron radiation could reach the hard X-ray regime with enough time-averaged power. However, the synchrotron radiation is polychromatic, and the monochromator is required to obtain high spectral purity. With the monochromator, the pulse energy of the synchrotron radiation light source decreases significantly to the level that is far lower than XFELO~\cite{vanaerenbergh.2010}. Besides, if the X-ray beam is micro-focused to the FEL beam level~\cite{Kolodziej2018}, its angular spreads would be much larger than Bragg reflection acceptance. Thus, in general, the experimental conditions do not meet the requirements of XFELO operation, and it is still necessary to numerically model the heat load of mirrors and its feedback on XFELO operation.

In previous studies~\cite{Song.2016, Yang.2018, Liu.2019}, the numerical evaluation of the X-ray absorption in the crystal uses an exponential attenuation model that describes the intensity of the X-rays decreases to 1/e of initial value through one attenuation length~\cite{Seltzer.1995}. The exponential attenuation model is practicable in most of the X-ray interactions but not satisfied in the X-ray Bragg reflection~\cite{Henke.1993}. Instead, a simulation model describing the physical process of Bragg reflection in GEANT4, a powerful particle tracker, is developed in this paper. Combining the other implemented light-matter physical process in GEANT4, a comprehensive set of interactions is provided into the currently simulating processes. As the heat source in the crystal is obtained, a start-to-end like transient thermal analysis can be conducted with the finite-element analysis (FEA) software~\cite{ANSYS17}. FEA simulations can determine the temperature map of the crystal, and therefore the strain field in a heat-distorted crystal. Furthermore, by coupling the thermal loading into the XFELO simulation, it can gain the ability to investigate the continuous influence of crystal thermal loading on XFELO operation. 

This paper is organized as follows. The Sec.~II presents the characteristics of Bragg reflections relating to the thermal load. In Sec.~III, the light-matter interaction is simulated within GEANT4, and the following thermal behavior is analyzed. In Sec.~IV, a simplified model to couple the thermal loading in XFELO simulation is presented. Then the simulation of thermal loading coupled XFELO is shown. A summary is presented in Sec.~V.

\section{\label{sec:BRAGG_example} Characteristics of Bragg Reflections}

An X-ray free electron laser oscillator is a low-gain device, in which the X-ray pulse is amplified by a gain $G$ for each round trip when it overlaps with an electron bunch in the undulator. In addition, the X-rays circulate in a low-loss optical cavity formed by multiple Bragg mirrors, corresponding to a total reflectance $R$. In general, when $(1+G)R > 1$, the X-rays could evaluate from the initial noise to intense coherent radiation. And the oscillator reaches steady state when $(1+G)R = 1$. 

Overall, the Bragg reflectivity should be sufficiently high so that the initial exponential gain of the intra-cavity pulse energy can be sustained for a reasonable set of the electron beam and undulator parameters. As the single-pass gain is fixed by the electron beam qualities and undulator configuration, the ability to achieve the stable XFELO operation relies on the reflectivity $R$ that is sensitive to the lattice structure of the crystals. Adverse effects, including the thermal expansion of the crystal and mounting vibration, may emerge to affect the lattice and to disrupt the Bragg reflection. Thus, the relationship between the distortion of the crystal lattice and the Bragg reflection is discussed because of the unique characteristics of the Bragg reflection. 


\begin{figure}[!htb]
     \centering
     \subfigure{\includegraphics*[width=240pt]{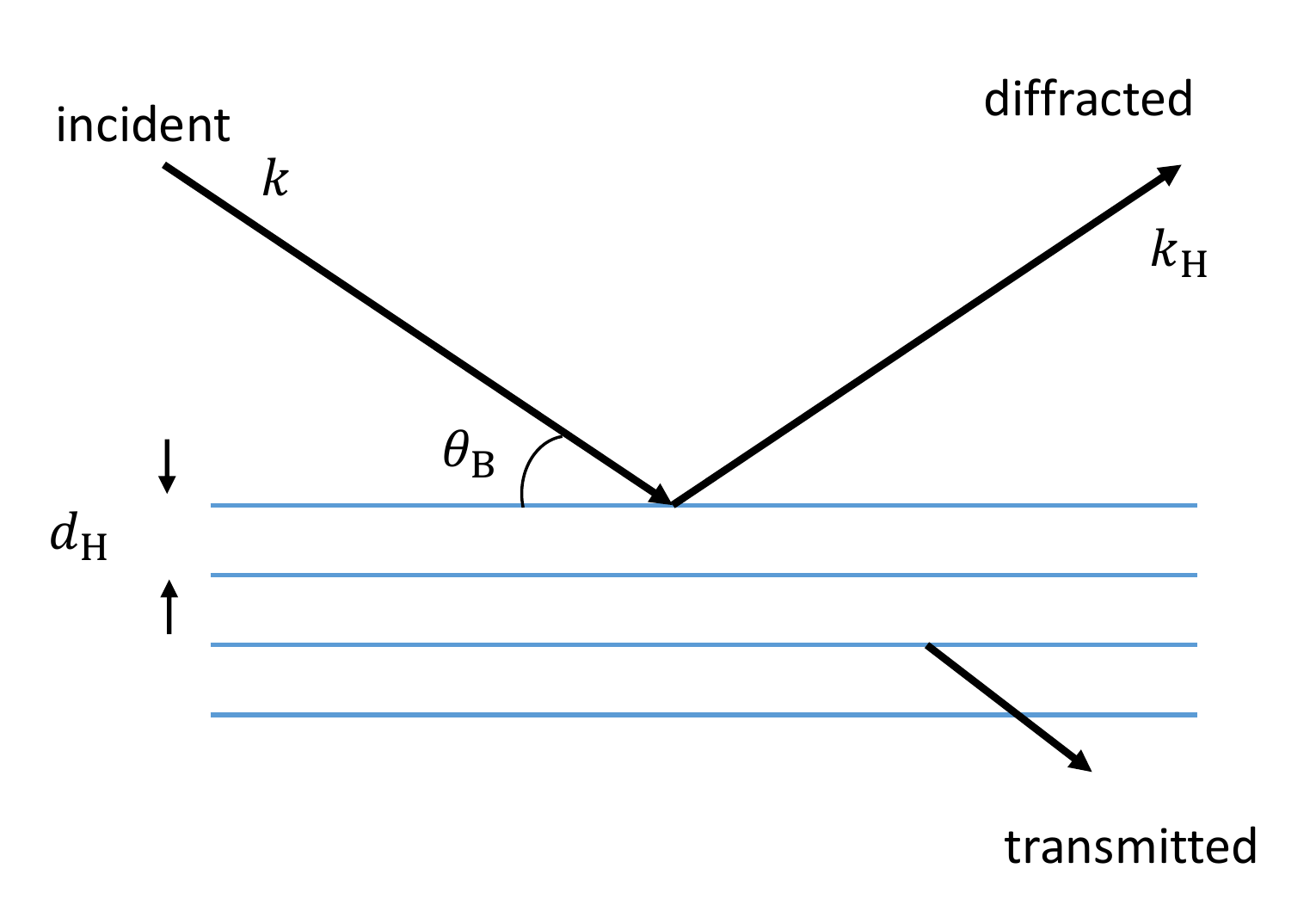}}
     \caption{A symmetric X-ray diffraction. The blue lines are the virtual atomic planes. $k$ is the wave number of incident X-ray, and $k_{\mathrm{H}}$ is reflected one. $\theta_{\mathrm{B}}$ is the Bragg angle. }
     \label{fig:bragg_scheme}
\end{figure}

The Bragg reflection or diffraction originates from the X-rays scattering of atoms in a periodic structure \cite{Hubbell.1979}, which can be imagined as a reflection of imaginary "mirrors" formed by atomic planes in the crystal lattice, seen in Fig.~\ref{fig:bragg_scheme}. The Bragg condition can be expressed as
\begin{equation}
     n \lambda_{\mathrm{B}}=2 d_{\mathrm{H}}(T) \sin \theta_{\mathrm{B}} ,
 \end{equation}
 where $n$ is the diffraction order, $\theta_B$ is the angle of incidence, $\lambda_B$ is the radiation wavelength at which the reflection occurs, $d_H$ is the distance between the atomic planes and $T$ the temperature. The lattice planes is described by the Miller indices, $h,k,l$, three integers. The spacing between the lattice planes is given by
 \begin{equation}
     d_{\mathrm{H}}(T)=\frac{a(T)}{\sqrt{h^{2}+k^{2}+l^{2}}},
 \end{equation}
 where $a$ is the lattice constant. For a diamond crystal, a cubic lattice, the lattice constant $a$  is 3.567~\cite{Stoupin.2011}. Diamond is the preferred material employed by a high power XFELO to form the X-ray cavity. Because it is a unique combination of outstanding thermal and optical properties, including a high thermal diffusivity, a low thermal expansion, and a high Bragg reflectivity of x-rays. In the following discussions, the crystal mirror is assumed to be diamond for the design and the optimization of the XFELO configuration.

\begin{figure}[!htb]
     \centering
     \subfigure{\includegraphics*[width=240pt]{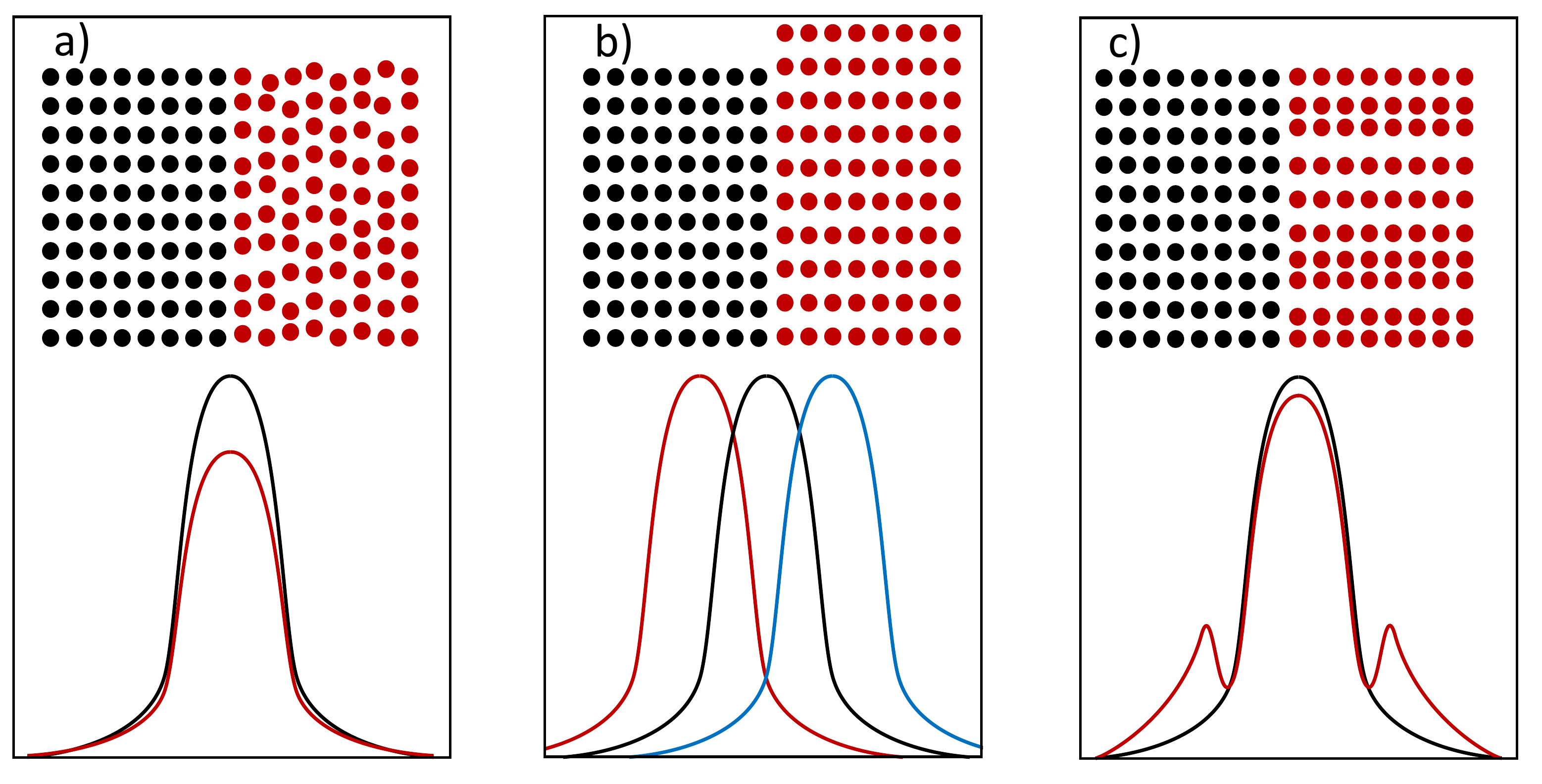}}
     \caption{ Three schematic presentations of lattice distortion. In each case, the strained crystal (red dot) is compared with undistorted lattice (black dot). The lower panel shows the impact of the distortion on the reflectivity curve of Bragg reflection.  a) Random disorder decreases the Bragg peaks. b) Homogeneous deformations lead to spectral shifts depending on the relative change of the lattice parameter. c) Longitudinal non-linear strains create sidebands and decrease the Bragg peaks.}
     \label{fig:Bragg_disorder}
\end{figure}

As expected, the high reflectivity in Bragg reflection is only possible if the lattice structure is perfect for a large amount of diffracting plane. In general, the change of crystal lattice acts on both the position and the amplitude of the Bragg reflectivity~\cite{Rousse.2001, Bargheer.2006}. Fig.~\ref{fig:Bragg_disorder} shows three kinds of lattice distortions (up panel) and corresponding effect (bottom panel). In each condition, the undistorted lattice (black dot) is compared with the strained lattice (red dot). The lower plots present the influence of these distortions on the reflectivity curve over the spectrum. a) Random disorder decreases the peak reflectivity of the Bragg reflection. b) Homogeneous linear strains shift the reflection. c) Longitudinal non-linear strains create sidebands and decrease the peak value of the Bragg reflectivity.

When a high-brilliance X-ray pulse is reflected by a Bragg reflecting crystal, a certain part of the energy is deposited. The energy from light is first absorbed by the electrons and then transferred to the crystal lattice. As the position and the amount of deposited energy are different, various thermal distortions of the lattice would be presented in the crystal. The type a) distortion occurs for a very high energy density that imposes radiation damage upon crystal to lose the long-range order, e.g., ionization. Far below the damage threshold, the absorbed energy due to the light-matter interactions translate into a thermal strain that leads to the other disorders, type b) and c), which are the major issues in the XFELO heat load. The deformed reflectivity curve of the large thermal deformation can be obtained by solving the Takagi--Taupin equations with the strain profile owing to the X-ray heating \cite{Honkanen.2018}. 

In principle, the relative change of the lattice constant can be described by the thermal expansion coefficient, $\beta$. For a cubic crystal like the diamond, the relative change owing to the thermal expansion can be expressed as:
\begin{equation}
     \delta_a = \exp\left( \int_{T_0}^{T} \beta(T^{\prime}) T^{\prime} d \right) - 1 
     \label{eq:thermal_exp}
\end{equation}

While the temperature change is small, $\beta$ can be assumed constant and relative change of lattice constant can be approximated to $ \frac{a(T)-a(T_0)}{a(T_0)} \approx \beta \times (T-T_0)$ with the first order Taylor series. The thermal expansion coefficient $\beta(T)$ can be obtained by an empirical formula from Ref.~\cite{Stoupin.2011}, which can be written as $\beta = 4.25\times 10^{-14} T^3$. 

As equation (\ref{eq:thermal_exp}) predicted, a small expansion coefficient is expected to reduce the change of the crystal and to increase the tolerance of the heat load on the crystal. It is therefore preferable to operate the mirrors at a low temperature, below 100~K. Because in the cryogenic temperature, the thermal expansion coefficient drops by a few orders of magnitude compared with the room temperature value of $\approx 1\times10^{-6}$. Besides, the thermal conductivity of diamond would reach the peak value in the temperature range from 60~K to 100~K~\cite{Inyushkin.2018}.

\section{Light-matter interactions and heat conductions}

In order to investigate the lattice distortion of mirrors under heat load, it is fundamental to obtain the accurate and reliable information of the absorbed part of the FEL X-ray pulse. This means that the interaction between the XFELO radiation and the crystal mirror is an essential limitation in the realization of thermal loading calculation. In this section, with the help of a dedicated Bragg reflection model, the light-matter interactions have been numerically simulated by using the GEANT4 toolkit~\cite{Allison.2016}, which is developed and actively used for particle experiments and detector designs. 

GEANT4 is a Monte Carlo based toolkit for simulating the particle interacting with matter. It includes many functionalities like tracking, geometry, physics models, and hits. For convenience, it already implements many widely used physics models to describe the fundamental light-matter interactions. Additionally, many other physics processes and low-energy models are included, covering a wide range of interactions, e.g., electromagnetic, hadronic, and a large set of long-lived particles. 

Based on the Low Energy Livermore Model in the GEANT4~\cite{Cirrone.2010}, an additional Bragg reflection process has been implemented to extend the available range of particle interactions. The implemented Bragg reflection model could change the direction of the photon in the crystal, which is satisfied with the Bragg condition. Thus, this model naturally enjoys the future of the multiple reflecting, which is the most important consideration for Bragg reflection and is absent from the previous exponential attenuation model. The crystal is built in geometry construction with its orientations, lattice information. As the Bragg reflection occurs in a range of ten meV level, the table that stores the cross section of each interaction in the GEANT4 kernel must be extended to reach the meV level. The cross section of the Bragg reflection, the most important factor in the GEANT4 simulation, is calculated with the help of the dynamical theory of X-ray diffraction. The cross section is given by
\begin{equation}
     \begin{aligned}
          \sigma_{\mathrm{B}}(\mathrm{b} / \mathrm{atom}) =   \mu / \rho\left(\mathrm{cm}^{2} \mathrm{g}^{-1}\right)  \left\{m_{\mathrm{u}}(\mathrm{g}) A\right\} \times 10^{24},
     \end{aligned}
\end{equation}
where $1/\mu$ is the excitation length calculated by the dynamical theory \cite{Shvydko.2012,AUTHIER.2003b}, $m_{\mathrm{u}}=1.66053886 \times 10^{-24} \mathrm{g}$ is the atomic mass unit, and $A$ is the relative atomic mass of the target element. Beyond the Darwin width, the Bragg reflection filter out the spectral components. Therefore, it is assumed for simplicity that the cross section beyond the Bragg width is zero in the GEANT4 simulations. As an example, the Bragg width is about 11~meV while the photon energy is 14.33~keV at the diamond (3 3 7) reflection. Besides, a parameter defined as the probability of the reflected X-ray to be reflected again can be used to control the multiple Bragg reflection and to correct the reflectivity in a very thick crystal so as to satisfy the predictions of the dynamical theory. 

\begin{figure}[!htb]
     \centering
     \subfigure{\includegraphics*[width=200pt]{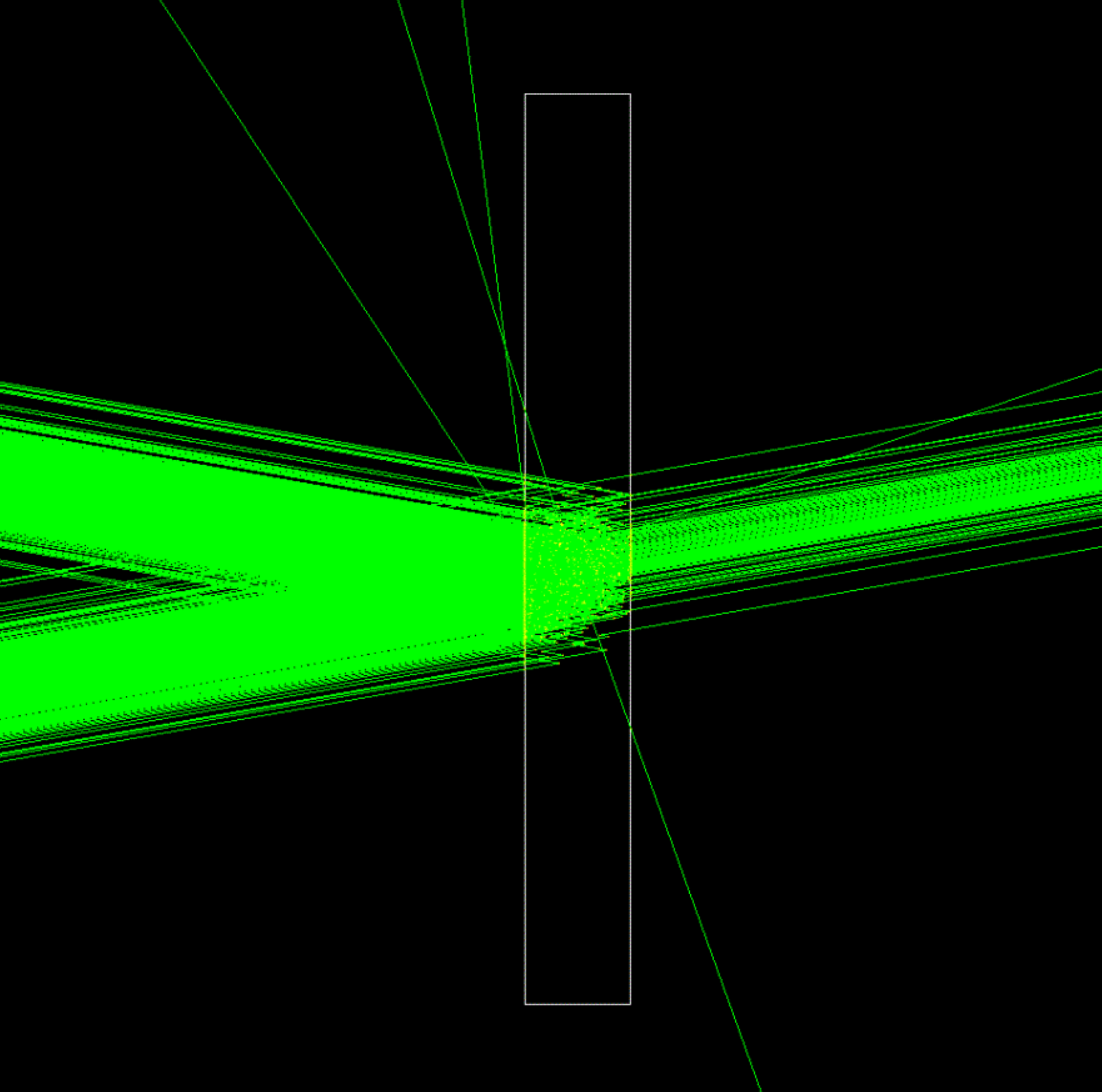}}
     \caption{A snapshot of the light-matter interactions in the GEANT4 simulation. The green line is the X-rays. The white box is the frame of the crystal. A large part of incident X-rays is reflected, with crystal thickness of 70 $\mu$m, photon energy of 14.33~keV and Bragg angle of about 84$^{\circ}$. }
     \label{fig:bragg_G4}
\end{figure}

According to the earlier simulation results of XFELO mode for SHINE (Shanghai HIgh repetition rate XFEL aNd Extreme light facility), the first hard X-ray free electron laser in China~\cite{Zhao:2018lcl, Yan.2019M}, a set of baseline parameters are utilized for the generation of primary photons in the GEANT4 simulation. The cavity is formed by four diamond mirrors with (3 3 7) diffracting planes while the cavity configuration is optimized to amplify the X-ray pulse with a photon energy of 14.33~keV. The intra-cavity pulse energy is assumed to be 600~$\mu$J with a 1~MHz repetition rate \cite{Li.2018}. Due to the excellent longitudinal coherence, constant phonon energy of 14.33~keV is adopted. The temporal distribution of X-ray photons is assumed to a Gaussian distribution with an RMS length of 12~$\mu$m. The total length of the X-ray pulses, defined as the distance from the first photon to the last one, is fixed to be 100~$\mu$m in the photon generator. Meanwhile, the pulse has a Gaussian transverse profile with an RMS size of 25~$\mu$m, while the divergence angle is assumed to be 0.7~$\mu$rad, which is much smaller than Bragg reflection widths. In addition, the crystal size is 1000$\mu$m$\times$1000$\mu$m$\times$70$\mu$m in the GEANT4 simulation. 

Overall, in the view of implemented and defined components, all aspects of the Bragg reflection simulation have been involved: 1) the geometry of the diamond crystal, 2) the primary particle (photons), 3) the generation of particles, 4) the tracking of particles through materials, 5) the physics processes modeling particle interactions (photoelectric effect, Compton scattering, pair production, bremsstrahlung, and Bragg reflection.).

Fig.~{\ref{fig:bragg_G4}} shows a snapshot of the particle tracking in the crystals. As it is shown, the Bragg reflections govern the interaction of the X-rays with the crystal. The absorptions of the X-rays mainly result from the photoelectric effect and Compton scattering. The Bragg reflectivity could be optimized by adjusting the thickness of the crystal. According to the statistical data from the GEANT4 simulation at 14.33~keV, a diamond mirror with a thickness of 70~$\mu$m allows reflecting $87\%$ of the incident X-ray while the transported part remains $11.7\%$. In other words, only $1.3\%$ of the incident X-ray has been absorbed. Since the dynamical theory of X-ray diffraction predicts the reflectance to be 88\% and transmittance to be 11.7\%, the GEANT4 simulation results are reasonable and acceptable.


\begin{figure}[!htb]
     \centering
     \subfigure{\includegraphics*[width=240pt]{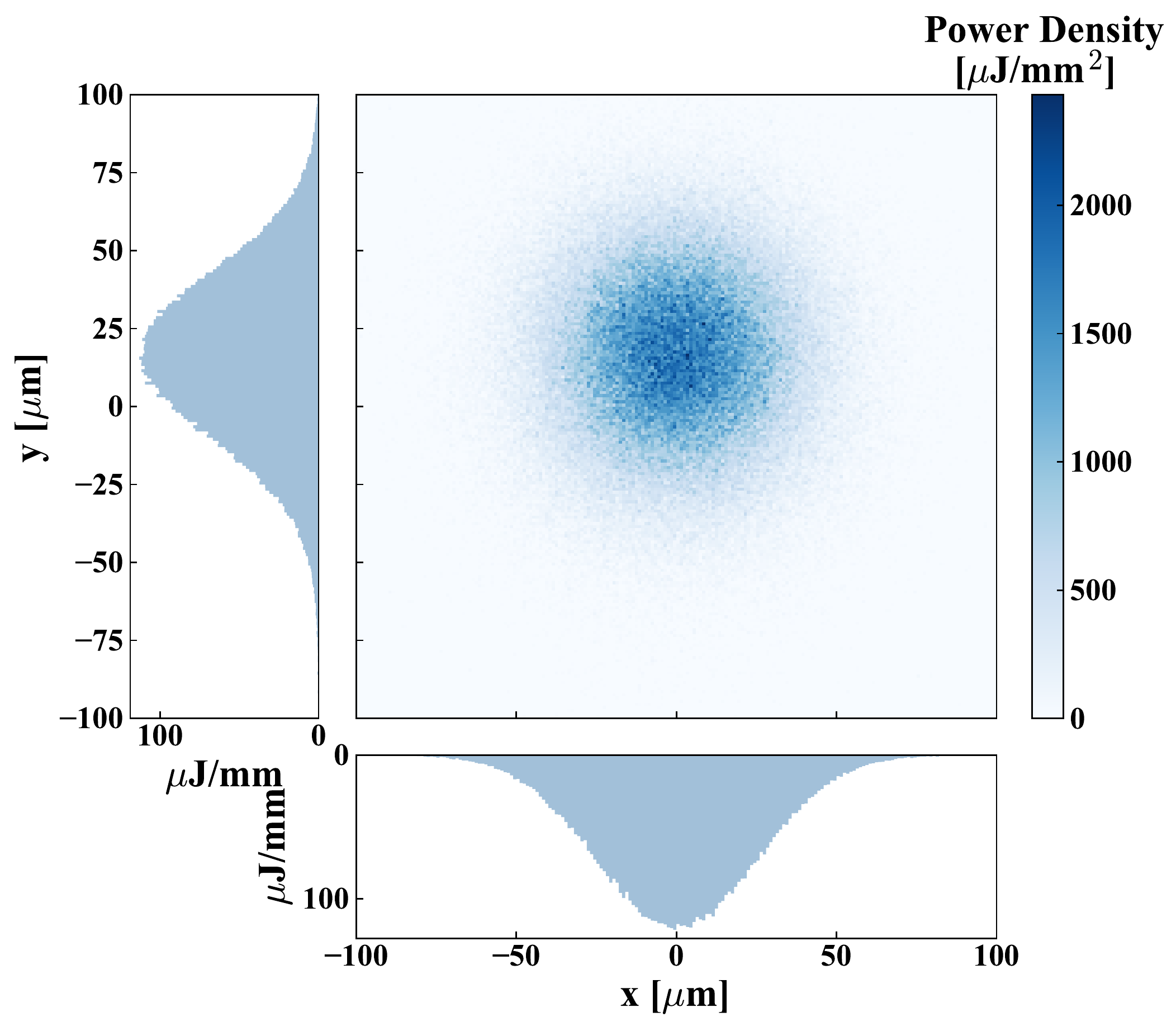}}
     \caption{The density of absorbed energy over the transverse plane. The corresponding marginal distribution is also shown. A very slight deformation of marginal distribution in y-direction occurs, for the crystal is rotated $6^{\circ}$ along the x-direction to fulfill the Bragg condition.}
     \label{fig:G4_transverse}
\end{figure}

Consider an intra-cavity X-ray FEL pulse with a pulse energy of 600~$\mu$J at 14.33~keV reflecting by a diamond, the total absorbed energy is about 7.3~$\mu$J in the diamond crystal during the reflection. For the diamond (3 3 7) reflection at 14.33~keV, the Bragg angle is $83^\circ$ nearing the exact backscattering. In this condition, the transverse distribution of the absorbed energy is mainly fixed by the transverse profile of the incident X-ray pulse, seen in Fig.~\ref{fig:G4_transverse}. With an RMS spot size of about 25~$\mu$m, the maximum energy density reaches 2400~$\mu$J/mm$^{2}$ corresponding to time-averaged power of 2.4~kW/mm$^{2}$ at 1~MHz. Since the crystal rotates about $7^{\circ}$ along the x-direction to fulfill the Bragg condition, the irradiation spot size in Fig.~\ref{fig:G4_transverse} on y-direction is a little larger than that of x-direction. Because the reflection of a non-normal incidence would cause a long tail with relating to transverse shift distance as radiation penetrated in crystal. For a small Bragg angle, this effect would significantly affect the transverse distribution of the absorbed energy. Moreover, it is insufficient to evaluate the multiple reflections in a theoretic formula but is simple in this GEANT4 model.

\begin{figure}[!htb]
     \centering
     \subfigure{\includegraphics*[width=240pt]{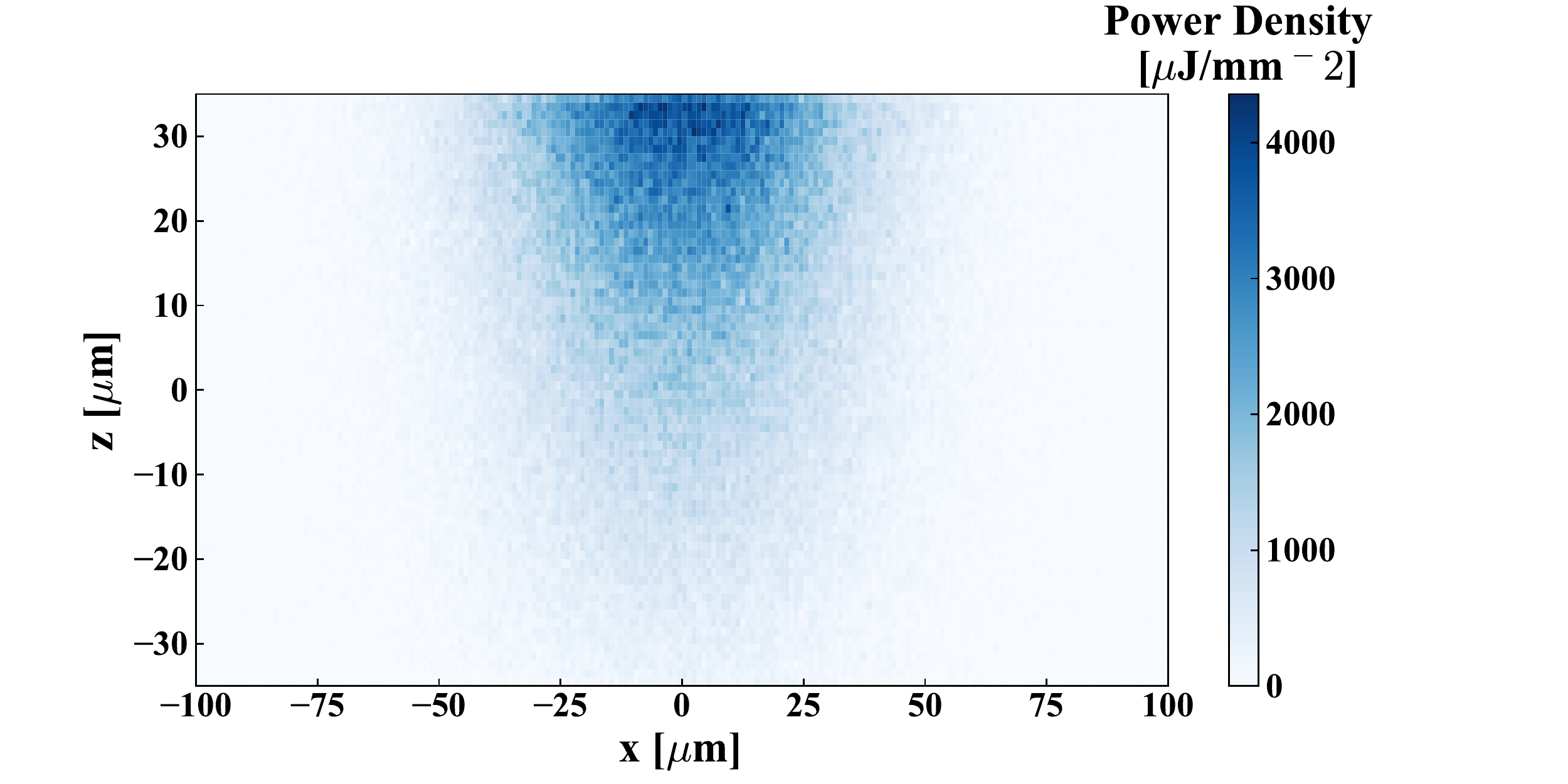}}
     \subfigure{\includegraphics*[width=240pt]{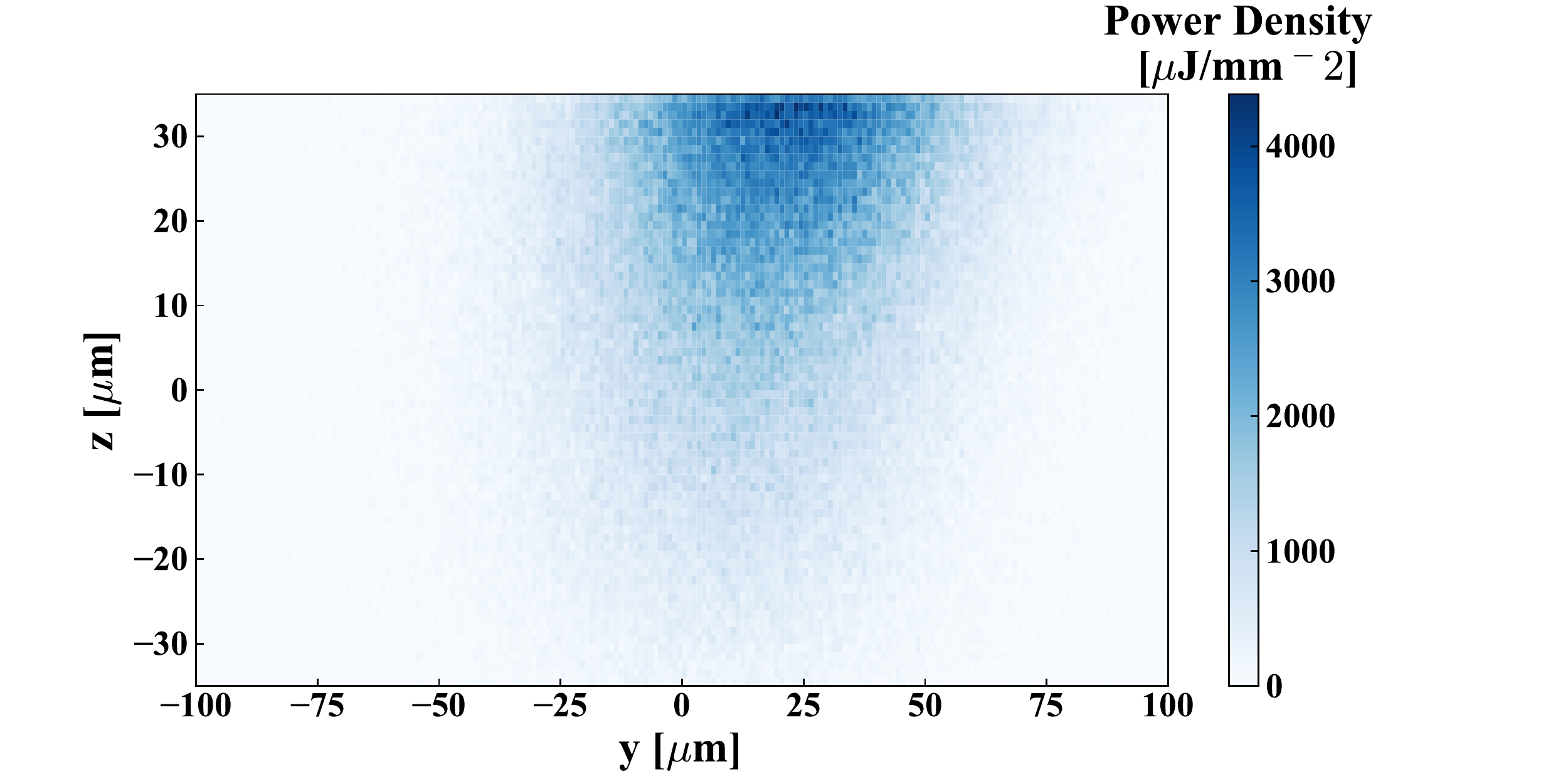}}
     \caption{The distribution of absorbed energy in x-z plane (top) and y-z plane (bottom).}
     \label{fig:G4_x_y_z_plane}
\end{figure}

One advantage of this model is the ability to obtain the three-dimensional absorption information of the X-ray pulse while Bragg reflection is considered. The distributions of the absorbed energy on x-z plane (top) and y-z plane (bottom) are shown in Fig.~\ref{fig:G4_x_y_z_plane}. The incident direction of the X-rays points to the negative value of the z-axis. The tilted penetration satisfying the Bragg condition is demonstrated in the bottom plot of Fig.~\ref{fig:G4_x_y_z_plane}. Due to the short excitation length related to a high reflectance, the incident X-ray pulse is attenuated significantly. Since it can give more details about the absorbed energy, this model could be a powerful tool to investigate the thermal loading of XFELO.

\begin{figure}[!htb]
     \centering
     \subfigure{\includegraphics*[width=200pt]{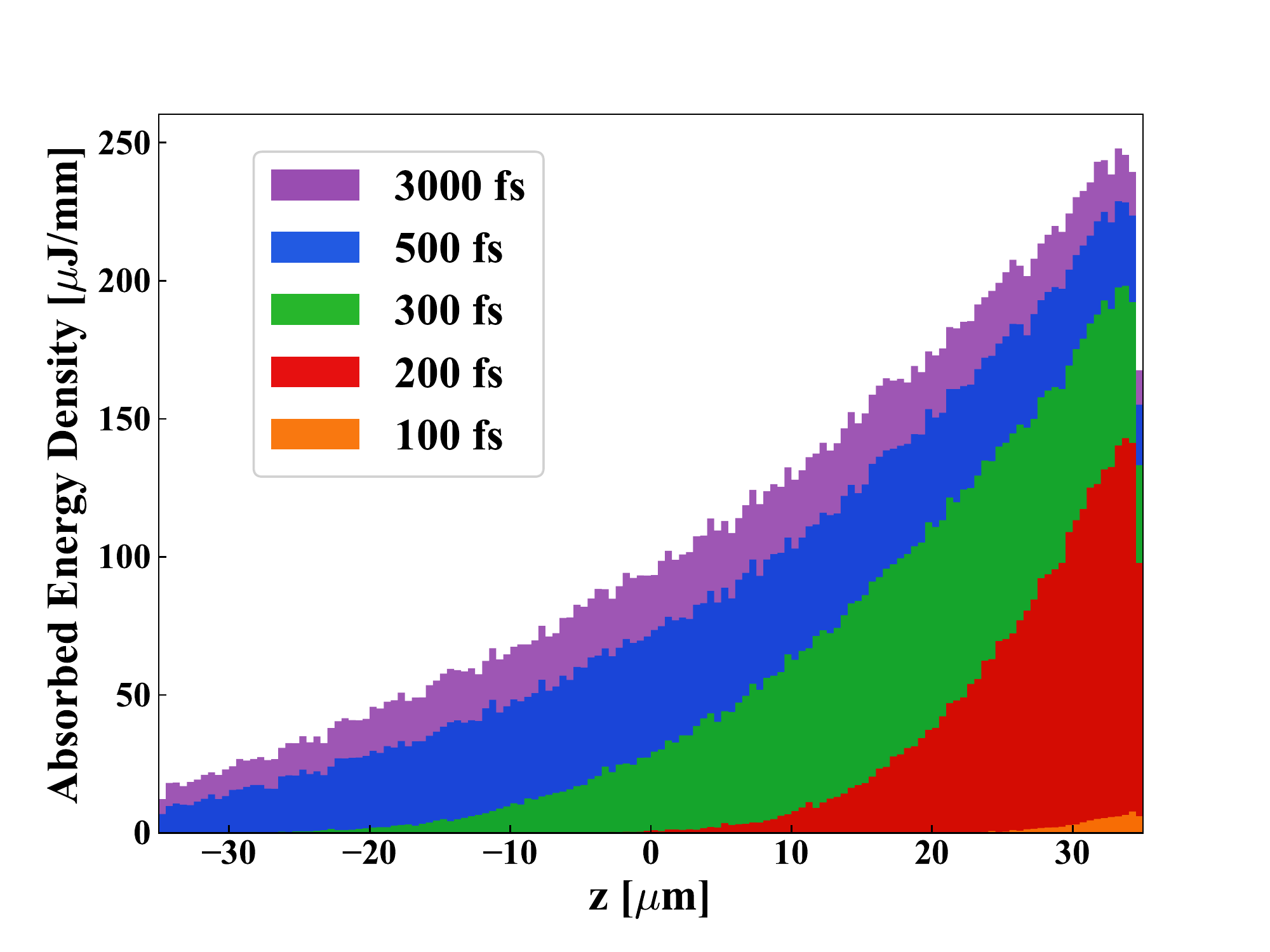}}
     \caption{The absorbed energy density along thickness at different time. }
     \label{fig:G4_z_time}
\end{figure}

Another advantage is that this solution could study dynamical absorption, the absorbed energy at different times, seen in Fig.~\ref{fig:G4_z_time}. In this case, the time of zero is the time when the first photon enters crystal. At the time of 500~fs, most of the interactions of X-rays with crystal are finished. The time as light-matter interaction finished is determined by the X-ray pulse duration and the crystal thickness. The peak value of the absorbed energies is inwards crystal, as the absorption always occurs after penetration. Unlike the exponential attenuation model, which predicts a near-uniform trend for the thickness of only 70~$\mu$m and a large attenuation length of the diamond, this model indicates that the absorptions concentrate on the part near the incident surface of the crystal owing the high reflectance. For a thinner crystal, the X-ray would penetrate the crystal with few multiple reflections, and the absorptions are quite similar to the prediction of the exponential attenuation model. Further, with the help of molecular dynamics simulation software, the dynamical absorption can be extended to investigate the possible dynamical change of the crystal over the time of the pulse duration.

Since the XFELO pulses may be too short on the time scale to influence their own Bragg reflection~\cite{Medvedev.2018, SokolowskiTinten.2004}, one can assume that the crystal temperature change acts only on the subsequent pulses at present research. Therefore, the crystal temperature seen by the next pulse is much more crucial in order to maintain the stability of the cavity. Additionally, the time between sequential pulses is far longer compared to the response time of crystal thermalization. Thus, it can be assumed that the X-ray pulses induce instantaneous heating in the crystal. In other words, the energy depositions resulting from light-matter interactions translate into the transient temperature increase.

In order to study the thermal behavior of the diamond crystal after the intense X-rays irradiation, the commercial finite element analysis software ANSYS is used~\cite{ANSYS17}. The initial temperature is obtained from the absorption information of the GEANT4 simulation. In ANSYS transient thermal analysis, the crystal size is 800$\mu$m$\times$800$\mu$m$\times$70$\mu$m, and the environmental temperature is 70~K. The work in Ref.~\cite{Inyushkin.2018,Wei.1993,Reeber.1996} is adopted to obtain the thermal conductivity and specific heat capability, and peak value of thermal conductivity can be found between 60~K and 100~K. As it is various as a function of the temperature, a fitting linear polynomial formula is utilized for simplicity. The mesh size is optimized, and the time step is fixed to be 0.5~ns.

\begin{figure*}[!htb]
     \centering
     \subfigure{\includegraphics*[width=160pt]{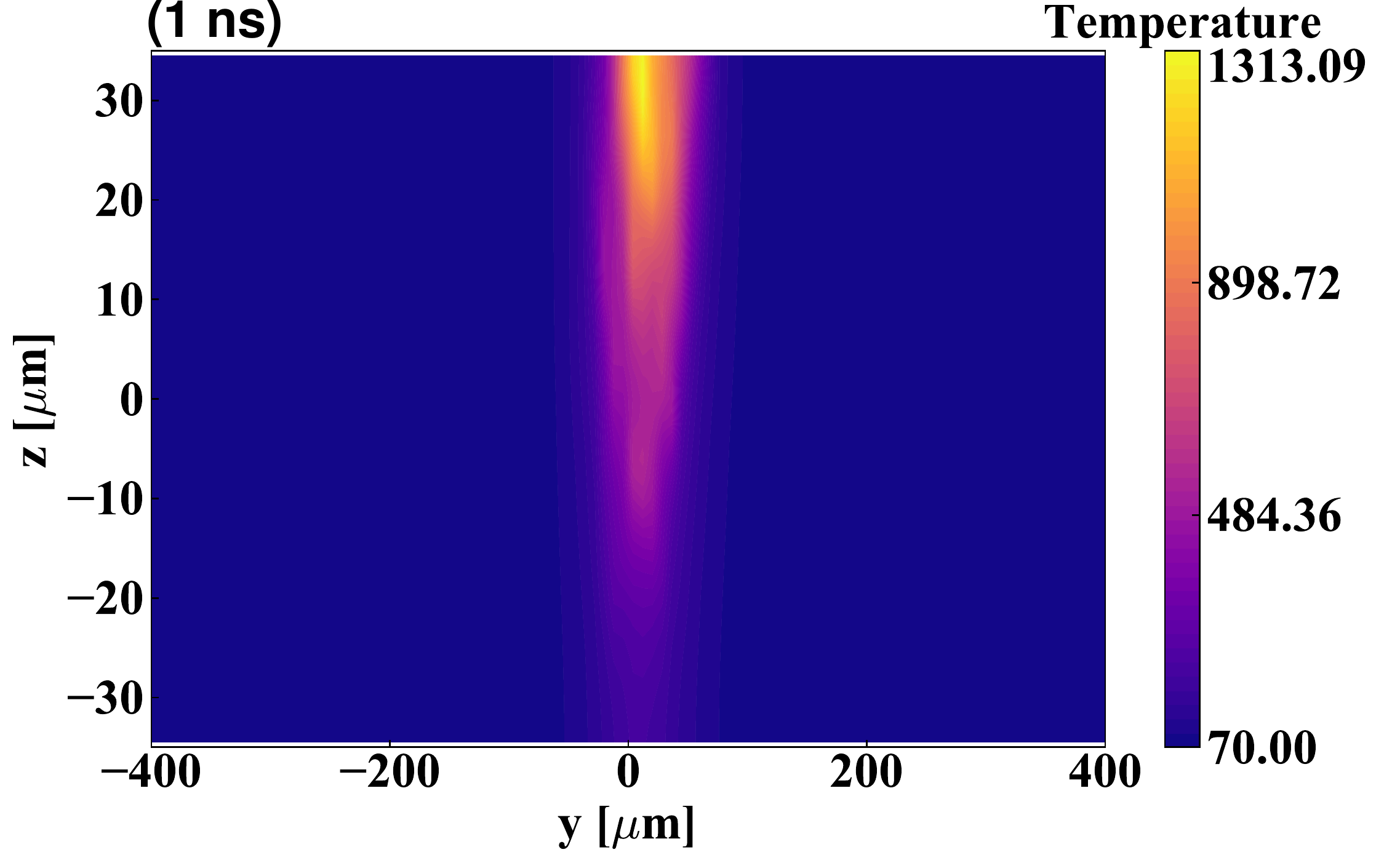}}
     \subfigure{\includegraphics*[width=160pt]{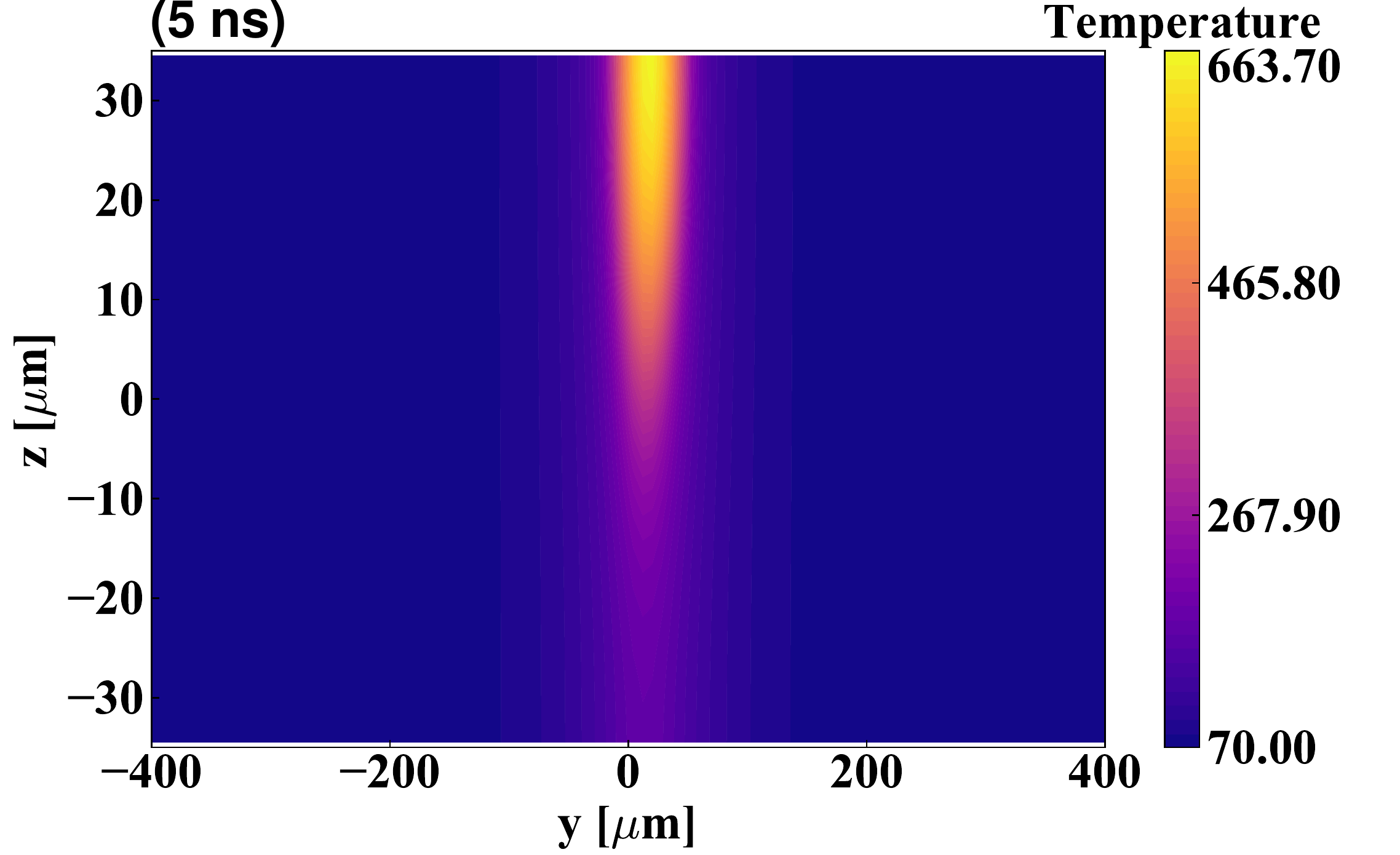}}
     \subfigure{\includegraphics*[width=160pt]{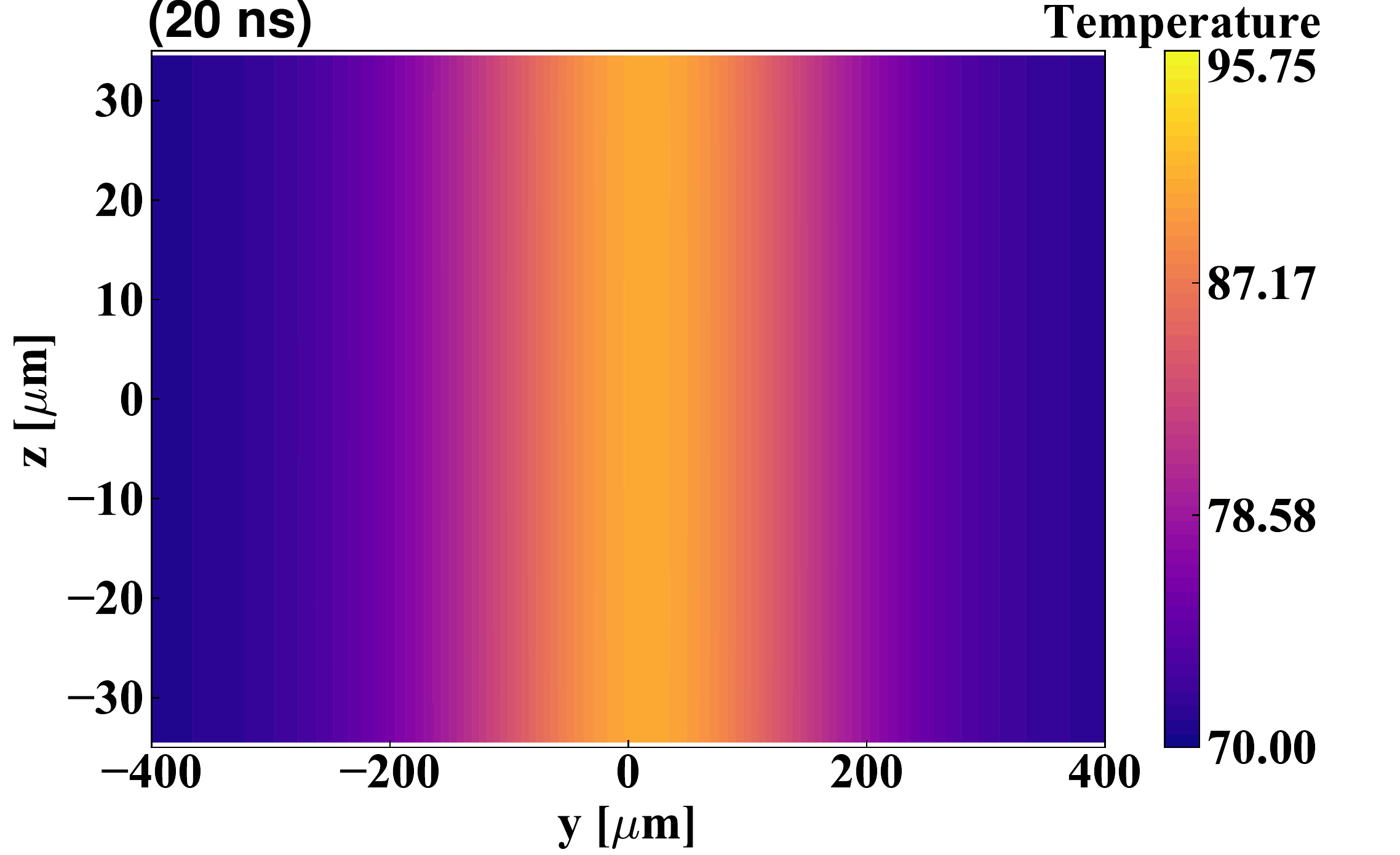}}
     \subfigure{\includegraphics*[width=160pt]{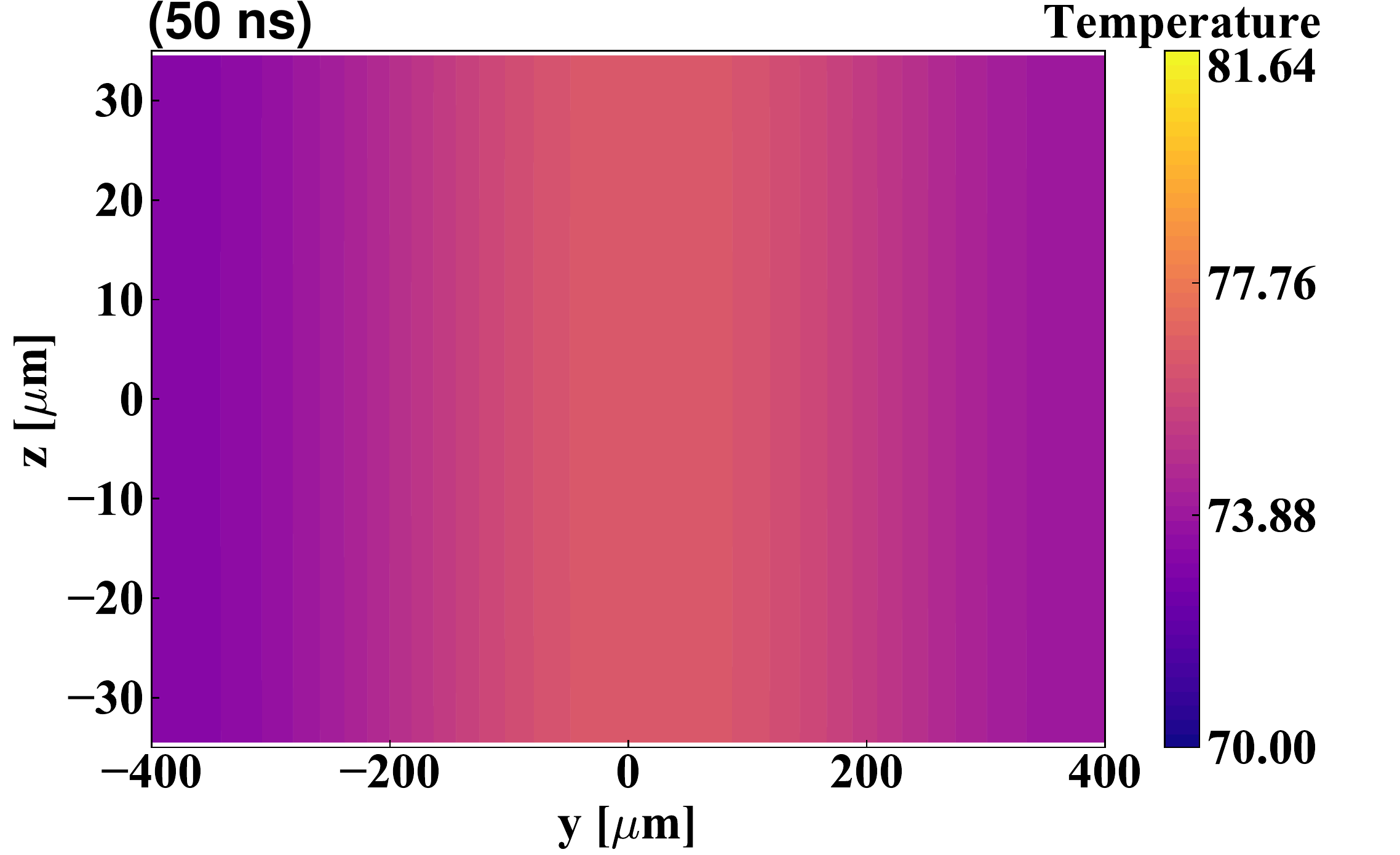}}
     \subfigure{\includegraphics*[width=160pt]{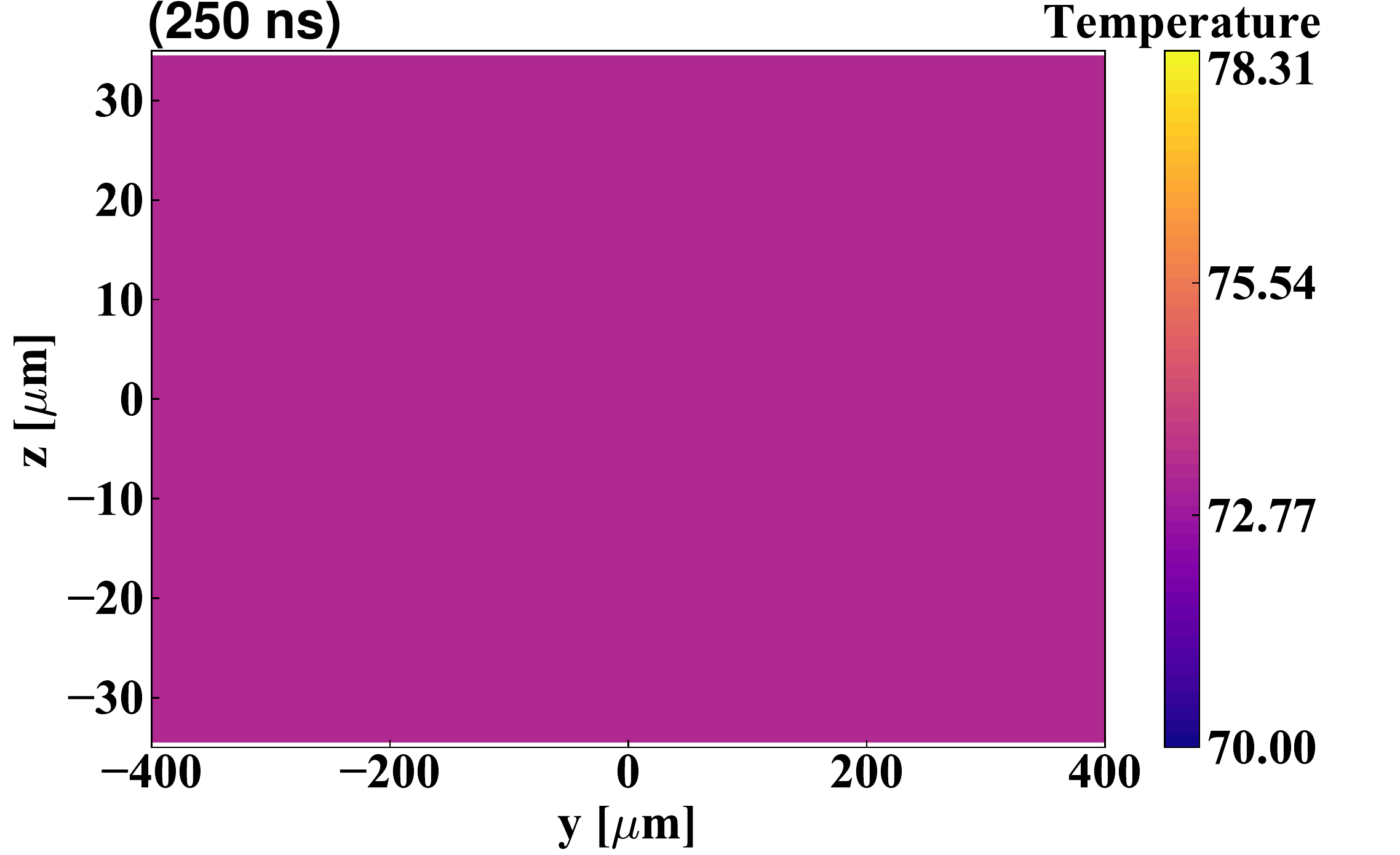}}
     \subfigure{\includegraphics*[width=160pt]{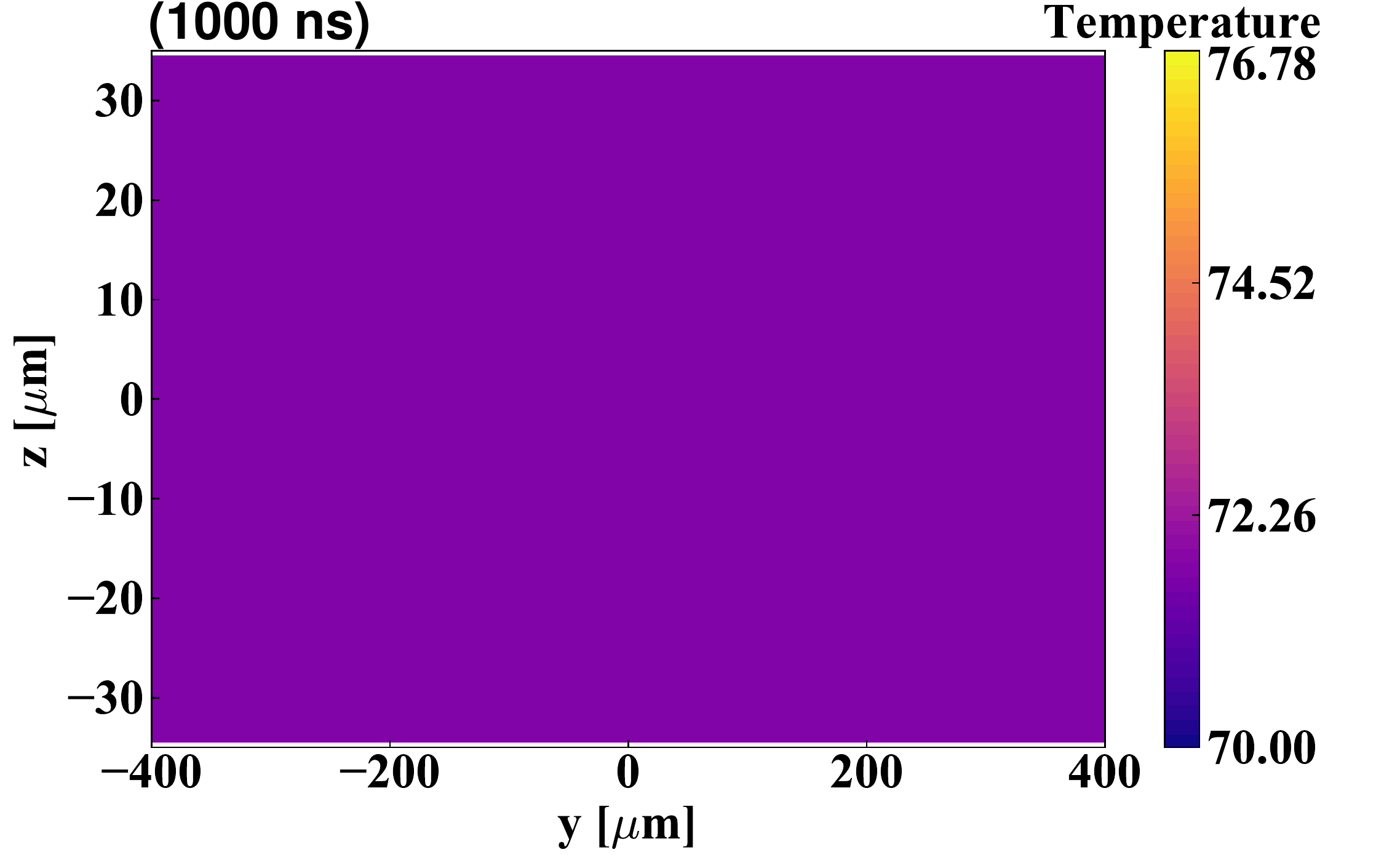}}
     \caption{Snapshots of temperature field calculated with a single-pulse input. It can be seen that the early temperature is much higher on the incident surface, and the highest temperature drops rapidly to 95~K within 20~ns. At 20 ns, it can be found that the temperature along the thickness becomes homogeneous. The temperature is nearly uniform throughout the entire crystal at 250~ns and drops to about 71.8~K at 1000~ns.  }
     \label{fig:single_pulse_thermal_y_z}
\end{figure*}

\begin{figure}[!htb]
     \centering
     \subfigure{\includegraphics*[width=200pt]{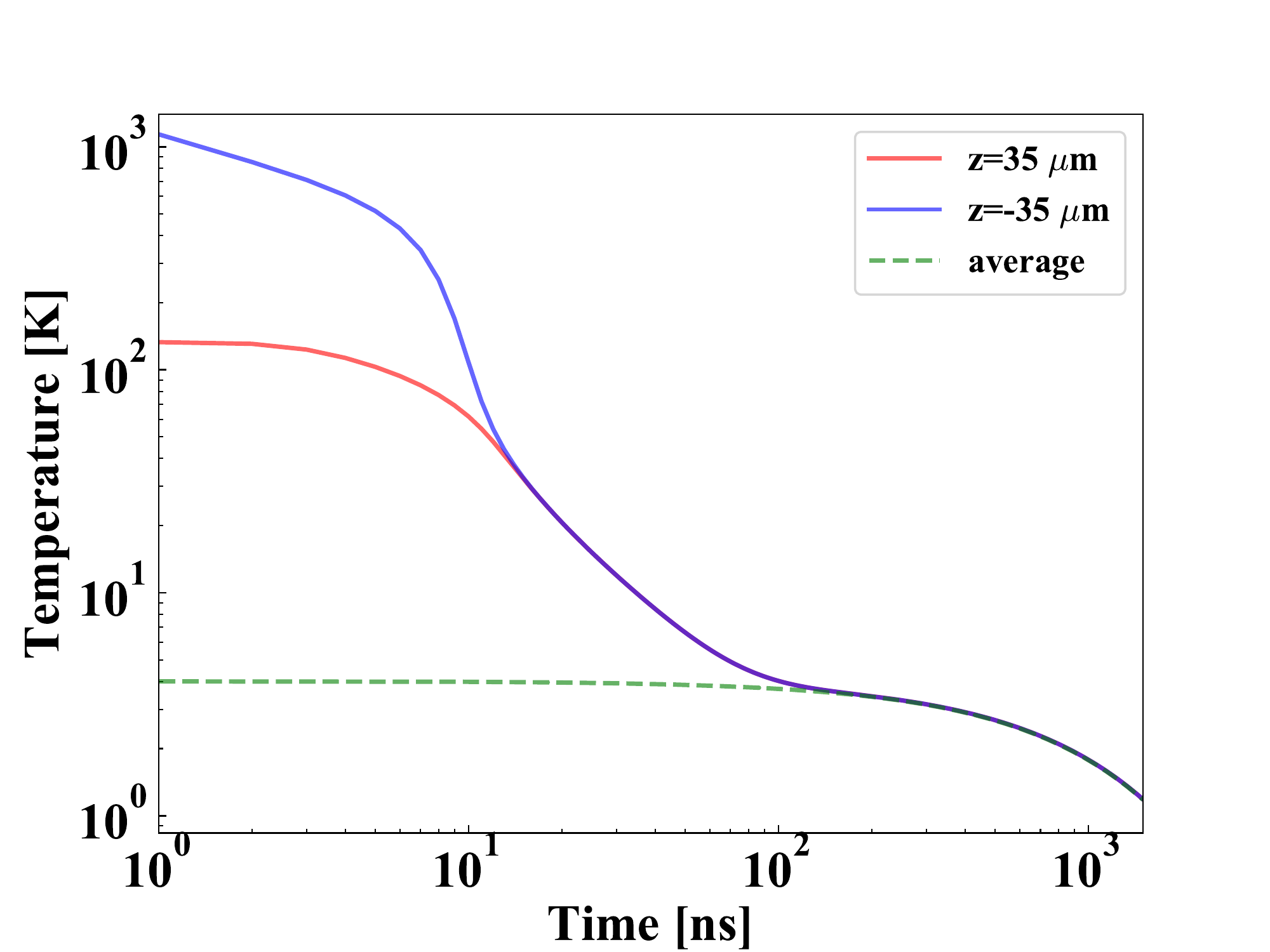}}
     \caption{The temperature (minus 70~K) at the center of incident surface (z=35~$\mu$m) and back surface (z=-35~$\mu$m) is shown. The temperature averaged over the whole crystal is presented by a green dash line.}
     \label{fig:single_pulse_thermal_A}
\end{figure}

Calculated results of the heat conduction are presented in Fig.~\ref{fig:single_pulse_thermal_y_z} and Fig.~\ref{fig:single_pulse_thermal_A}. Fig.~\ref{fig:single_pulse_thermal_y_z} shows snapshots of the temperature filed at various time of 1, 3, 5, 10, 250 and 1000~ns after irradiation. Fig.~\ref{fig:single_pulse_thermal_A} shows the evolution of the temperature (minus 70~K) at the center on the incident ($z=35$~$\mu$m) and back ($z=-35$~$\mu$m) surface, and the temperature (minus 70~K) averaged over entire crystal volume with time up to 2000~ns. 

As shown in Fig.~\ref{fig:single_pulse_thermal_y_z}, the initial temperature is higher on the incident surface than on the back surface due to the Bragg reflection. The highest temperature is about 1300~K near the center of the incident surface. It drops rapidly in the first 50~ns, and then gradually slow in following 100~ns. Because the thermal conductivity increases rapidly when the temperature decreases in the first 50~ns, and then heat diffusion is slowing when the temperature gradient decreases gradually. The thermal relaxation time across the whole thickness is considered to be about 20~ns, as suggested by the nearly identical temperature on the incident and back surface in Fig.~\ref{fig:single_pulse_thermal_y_z} and Fig.~\ref{fig:single_pulse_thermal_A}. Meanwhile, the thermal relaxation time in the radial direction is over 150~ns, as the temperature on the incident and back surface reaches the volume-averaged temperature. Indeed, the temperature averaged over the whole crystal is nearly homogeneous after 200~ns. The maximum temperature difference over the entire crystal is about 0.1~K at 250~ns, and it reduces to 0.05~K at 500~ns. In fact, the averaged temperature increase to about 74~K, with a total absorbed energy of 7.3~$\mu$J. In the following 1000~ns, the averaged temperature decrease to about 71.8~K, corresponding to a relative change of lattice parameter of $1 \times 10^{-8}$ that meets the requirement of cavity stability. At 2000~ns, the averaged temperature decreases to 0.6~K. It still have a downward trend but would continue to decrease over a longer period.

\begin{figure}[!htb]
     \centering
     \subfigure{\includegraphics*[width=200pt]{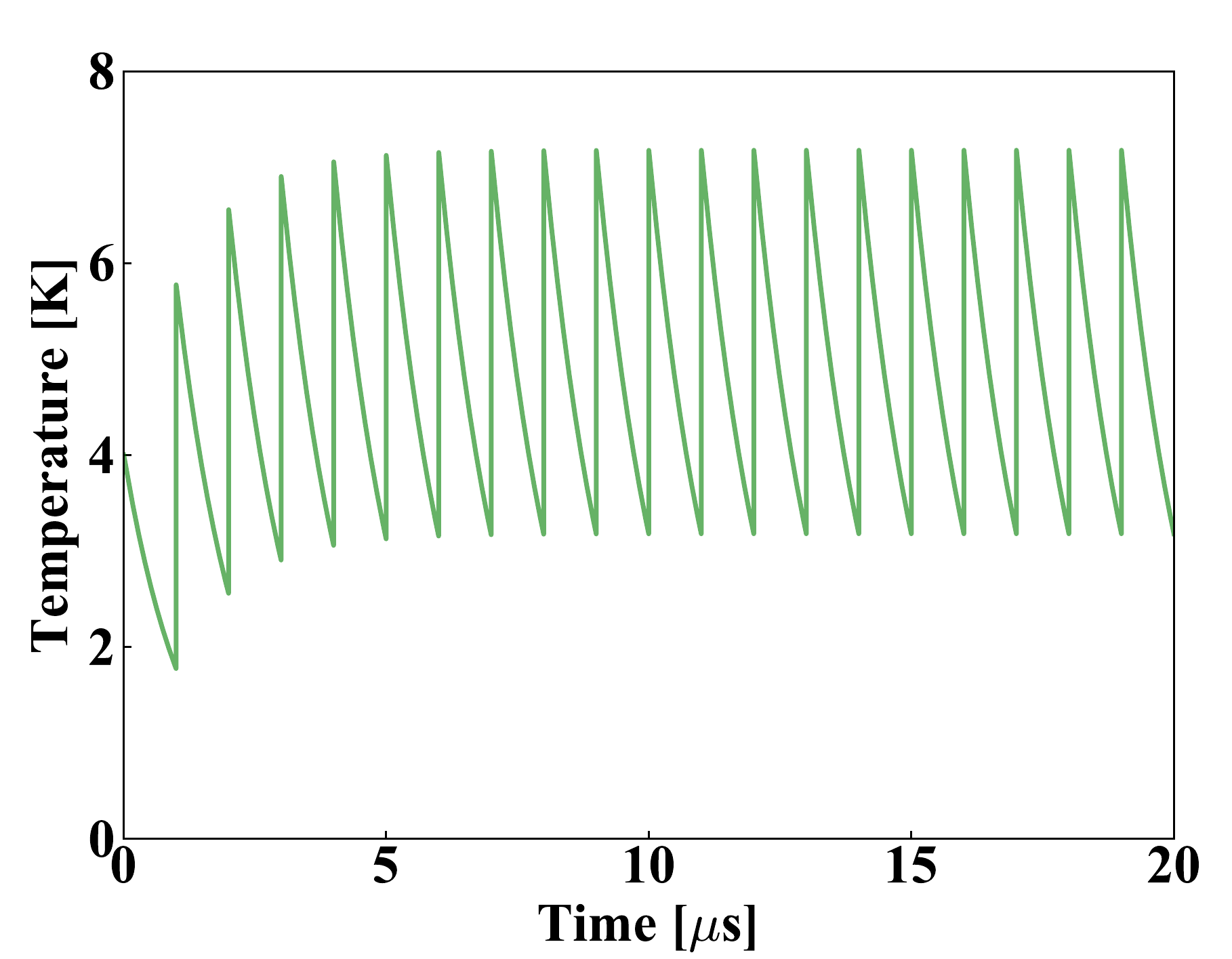}}
     \caption{Time histories of the averaged temperature (minus 70~K). The repetition rate is 1~MHz, and the periodic input laser pulse is the same as the single-pulse case. }
     \label{fig:multiple_pulse_thermal_A}
\end{figure}

The multiple-pulse result is present in Fig.~\ref{fig:multiple_pulse_thermal_A}. All the parameters except time step are the same as those in the above case, and the energy of individual laser pulses is added periodically. The time step is fixed to be 1~ns. The simulations are undertaken over 20~$\mu$s, 20 pulses at 1~MHz. The averaged temperature rises instantaneously upon each pulse. Its evolution over the interpulse period is similar to that in the above single-shot case. However, the averaged temperature keeps on rising, for some residual heat is left behind in the crystal and the finite thermal conductivity. Thus, the temperature increment consists of two parts: 1) the nearly instantaneous heating due to the X-rays irradiation; 2) long-term accumulation of the remaining heat in crystal. Since heat transmission depends on the temperature difference, the rising temperature would improve the capability of transferring heat to the cooling system. Thus, the remnant temperature drops against temperature increase, and the system would eventually evolve to a steady-state with a stable accumulated temperature increment. In our case, the temperature increment averaged in the whole crystal is about 4~K, and the accumulated one at the steady state is about 3.5~K. The heat accumulation raises a practical requirement for the cooling system that ensures the accumulated temperature is trivial. The solution should be established in a specific case where an XFELO operates, and it is not the purpose here. The specific cooling system design can take part experiences from the monochromator design of the synchrotron radiation light sources.

\section{thermal loading coupled XFELO simulation}

Since it is complicated to run the time-dependent XFELO simulation considering the thermal loading, a simplified model to describe the thermal loading is needed to be developed. The thermal loading mainly affects the Bragg reflection through the distortion of the lattice. As mentioned in Sec.~II, the pulse energy determines that the type 1) distortion is negligible in XFELO. Furthermore, based on the transient thermal simulation above, the type 3) distortion that needs a temperature gradient across the thickness could also be negligible. The thermal relaxation time across the thickness at a cryogenic temperature is very short compared to the interpulse time. Thus, in the long-term accumulation the temperature gradient across the thickness is negligible, or even the next pulse would see the temperature distribution is nearly uniform upon the entire crystal. Thus, it can be assumed that the most crucial issue of the thermal loading is the shift of the reflectivity curve due to the homogeneous expansion through the crystal thickness.

Therefore, the simplified model could focus on the temperature change averaged over the whole crystal or over the effective volume of the irradiation. It is reasonable because the variation of the temperature inside the diamond is negligible when the thermal relaxation time is much shorter than the cooling time. This assumption is based on the fact that the thermal conductivity of diamond at cryogenic temperature is very high. Thus, a simplified model of the averaged temperature can be expressed as~\cite{Shorr.2015}:
\begin{equation}
     \frac{d\Delta T}{dt} = -\frac{h_{eff}S}{mC_p} \times \Delta T,
\end{equation}
where $\Delta T = T -T_0$ is the relative change of the averaged temperature, $h_{eff}$ is the heat transfer coefficient related to the design of the cooling system, $S$ is the surface area of heat transfer, $m$ is the mass of diamond and $C_p$ is the specific heat~\cite{Reeber.1996}. When the temperature does not change significantly, the value of $\frac{h_{eff}S}{mC_p}$ can be assumed to be a constant. Then the solution can be written as
\begin{equation}
      \Delta T = \Delta T_0 e^{- \alpha t},
\end{equation}
where $\alpha = \frac{h_{eff}S}{mC_p}$. With a stable repetition rate, the time for cooling is a constant. Thus, we can define $\eta = e^{-\alpha \Delta t}$ as the cooling efficiency at the repetition rate of $1/\Delta t$. It describes how much temperature is left behind after a time of $\Delta t$. Its approximation value can be obtained from the transient thermal analysis of single-pulse input. 

Using the parameter $\eta$, the averaged temperature change of multiple-pulse input can be written as:
\begin{equation}
     \Delta T_{n+1} = (\Delta T_n + \Delta T_p) \eta \quad (\eta < 1),
\end{equation}
where $n$ indicates nth pulse. At steady state, the averaged temperature change become stable and we can get a solution, $\Delta T_{n+1} = \Delta T_{n} = \eta \Delta T_p / (1 - \eta)$. In the single-pulse thermal analysis, $\eta$ is about $1.8/4 = 0.45$. And the steady-state temperature change is 3.27~K, which is a little lower than that produced by multiple-pulse thermal analysis but reasonable for the simplicity. Then, the shift of reflectivity can be written as $\Delta E \approx E_H \beta \Delta T_n$, where $E_H$ is the Bragg energy.




With the help of the simplified thermal loading model, an XFELO operation of SHINE coupled with heat load is conducted. The parameters of SHINE is presented in Tab.~\ref{tab:SHINE_para}. The main accelerator of SHINE uses the superconducting technology to produce electron bunches with the energy of 8~GeV, a 100~pC charge compressed to the peak current of 700~A, and 1~MHz repetition rate. 

\begin{table}[!htb]
     \centering
   \caption{\label{tab:SHINE_para}%
   The main parameters of XFELO operation for SHINE.
   }
   \begin{tabular}{lcr}
   \hline
   \textrm{Parameter}&
   \textrm{Value}\\
   \hline
   Beam Energy         & 8~GeV    \\ 
   Relative Energy Spread     & 0.01\%     \\ 
   Normalized Emittance        & 0.4~mm$\cdot$mrad     \\
   Peak Current      & 700~A     \\
   Charge            & 100~pC \\
   Undulator Period Length    & 16~mm     \\
   Undulator Segment Length & 4~m \\
   Photon Energy     & 14.33~keV     \\
   Mirror Material   & Diamond     \\
   Peak Reflectivity & 87\% \\
   Darwin Width                           & 11~meV \\
   Mirror Thickness                          & 70~$\mu$m \\
   \hline
   \end{tabular}
   \end{table}

   \begin{figure}[!htb]
     \centering
     \subfigure{\includegraphics*[width=240pt]{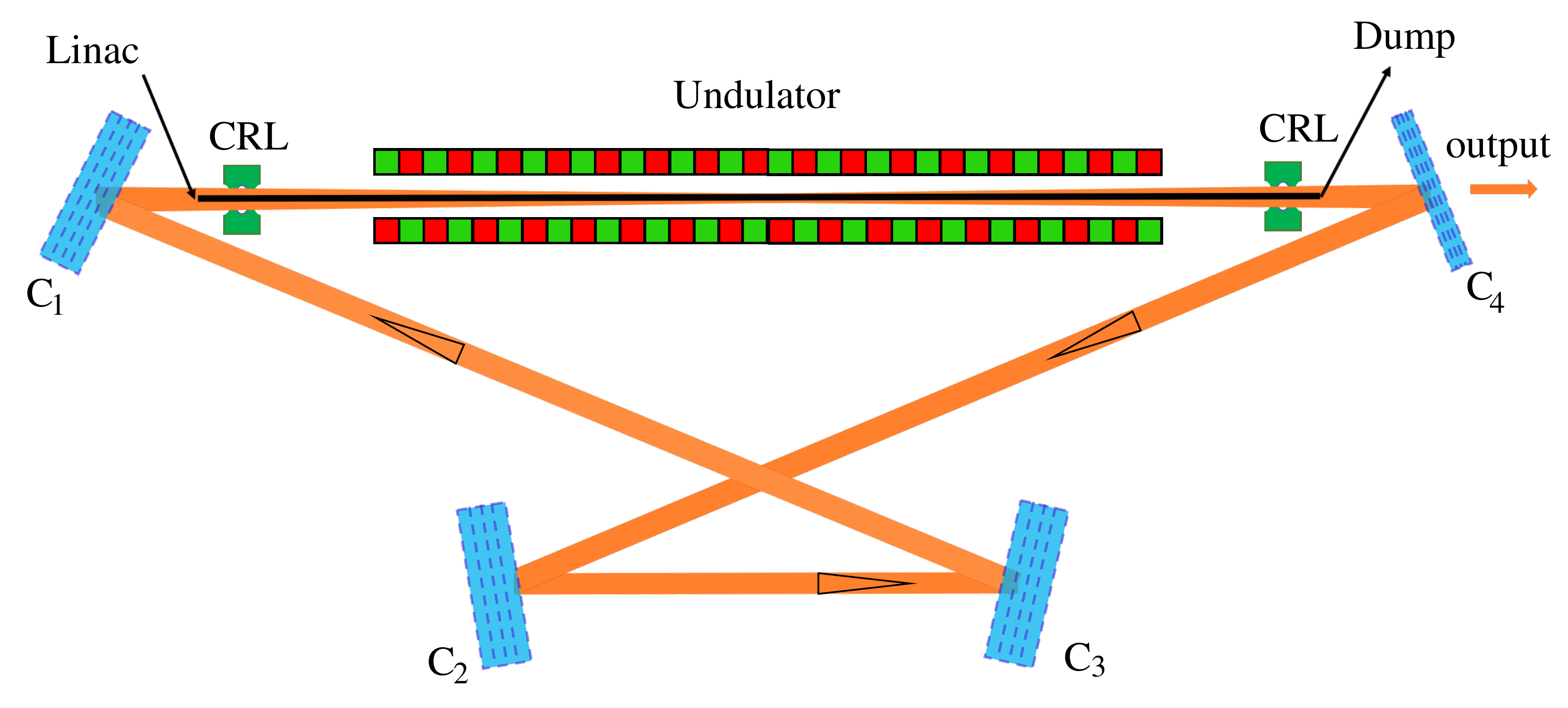}}
     \caption{
      A schematic XFELO configuration for the SHINE. Four diamond mirrors are used. The phonon energy is 14.33~keV. $C_1$, $C_2$, and $C_3$ are thick crystal to reach a total reflection. The X-ray coupled out from the downstream mirror $C_4$ which peak reflectivity reach 87\%. }
     \label{fig:XFELO_cavity}
\end{figure}

The optical cavity is built from four diamond (3, 3, 7) crystal mirrors, seen in Fig.~\ref{fig:XFELO_cavity}. With a thickness of 70~$\mu$m, the downstream diamond mirror $C_4$ at the Bragg energy of 14.33~keV reaches 87\% (coupling output is 12\%). The other three mirrors is expected to reach the total reflection with thick crystals. To simplify the calculation, the temperature of the three crystals is appropriate to be fixed by 70~K. The heat load is mainly considered in the downstream mirror $C_4$ due to its thinner thickness. 

The simulations are conducted by using the combination of a time-dependent FEL code GENESIS~\cite{Reiche.1999}, a field propagation simulation code OPC~\cite{Karssenberg.2006}, and a Bragg reflection simulation code BRIGHT~\cite{Huang.2019}. The model of simplified thermal loading is integrated into the BRIGHT as a part of Bragg reflection. In the following analysis, the X-ray pulse is indicated to be the intra-cavity one to present how much energy is induced to the mirrors. The output value needs to be multiplied by an output-coupling factor, the transmittance.

\begin{figure}[!htb]
     \centering
     \subfigure{\includegraphics*[width=200pt]{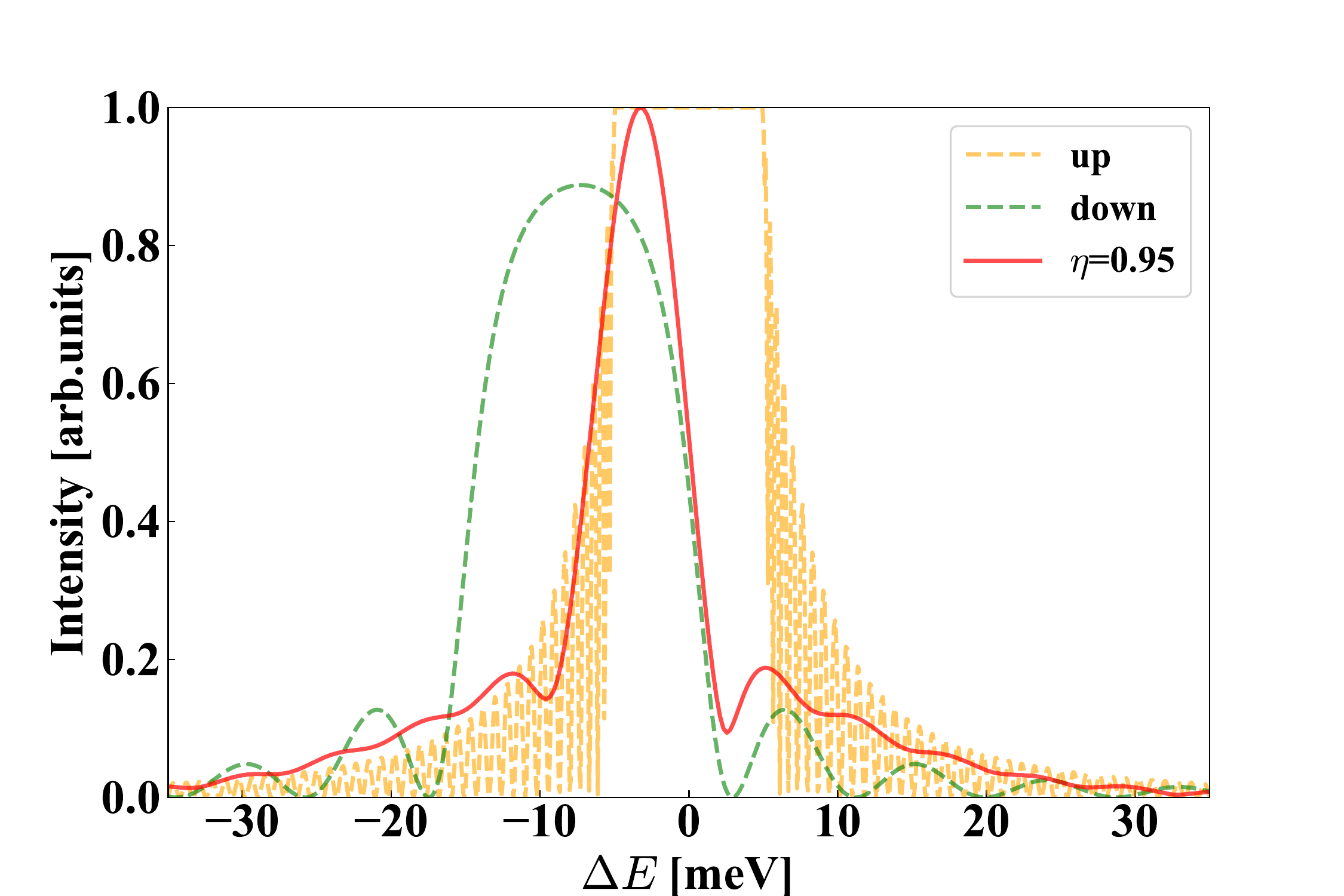}}
     \subfigure{\includegraphics*[width=200pt]{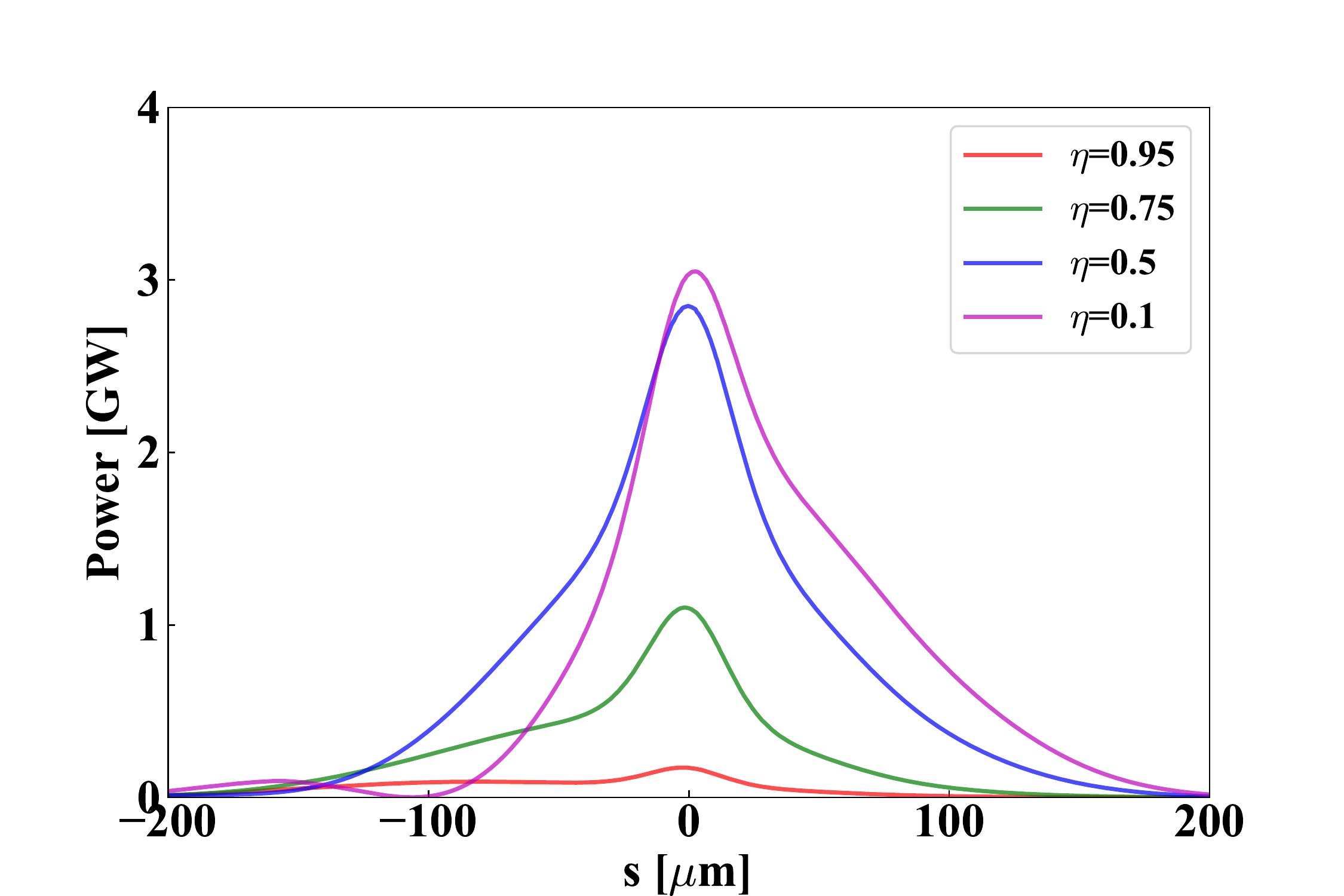}}
     \caption{The spectrum of XFELO pulse is shown in left plot while the eta is $0.95$. The Bragg reflectivity as a function of photon energy for downstream mirror (green) and upstream mirror (orange) is shown by a dash line. The shift of the spectrum and the downstream reflectivity can be found. The power profile in each case $\eta$ is shown in right plot. The peak power exceeds 3~GW. }
     \label{fig:ref_400}
\end{figure}

\begin{figure*}[!htb]
     \centering
     \subfigure{\includegraphics*[width=200pt]{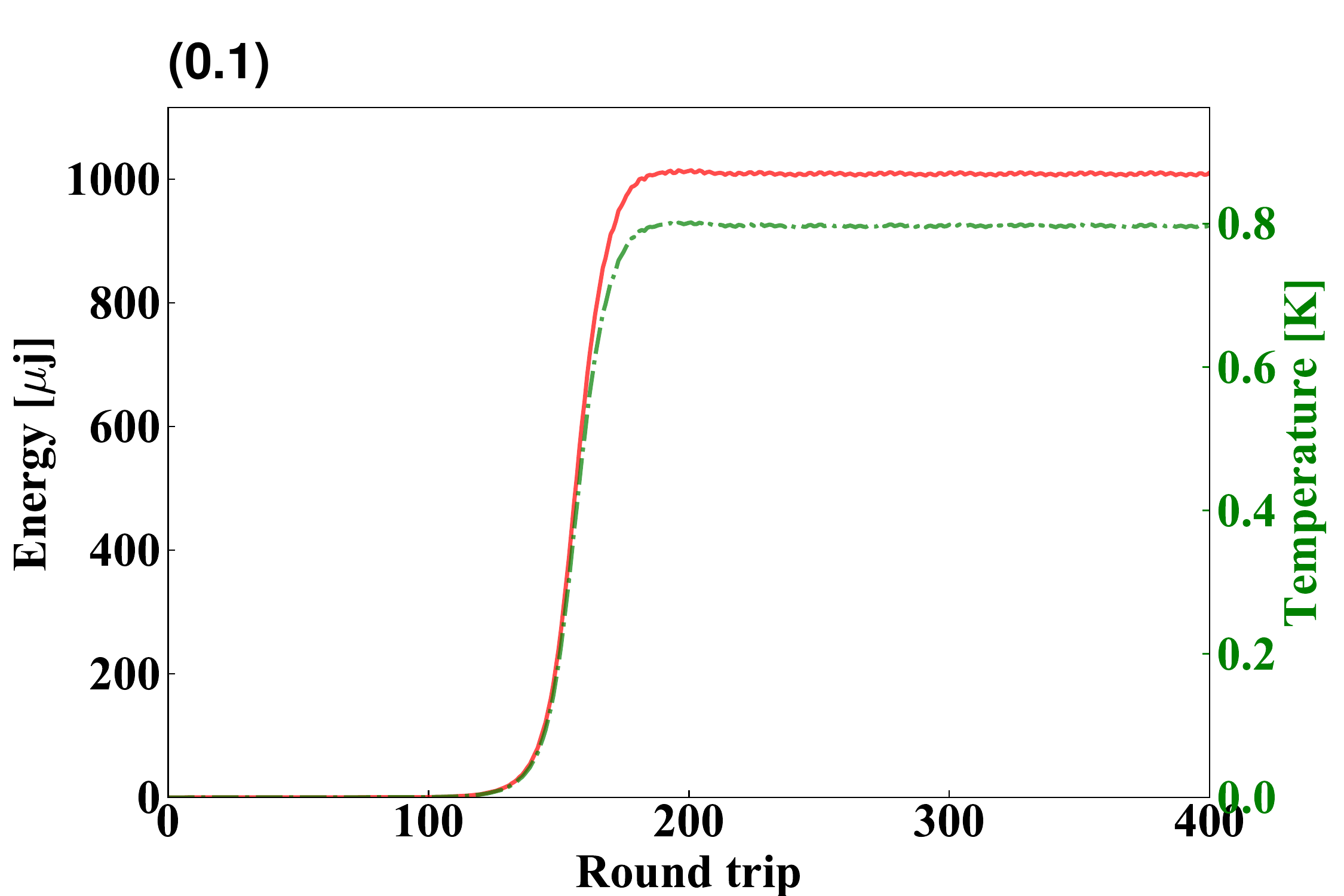}}
     \subfigure{\includegraphics*[width=220pt]{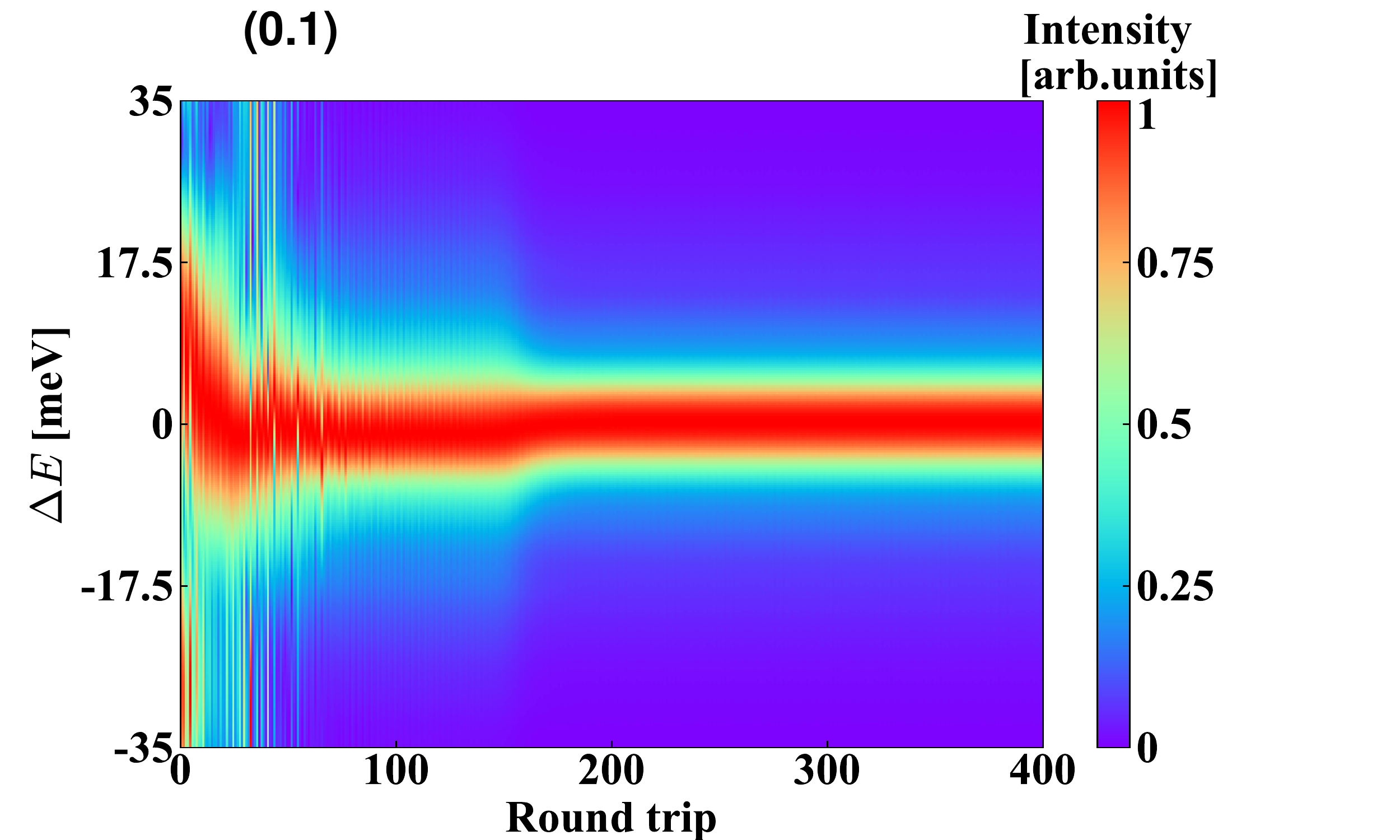}}
     \subfigure{\includegraphics*[width=200pt]{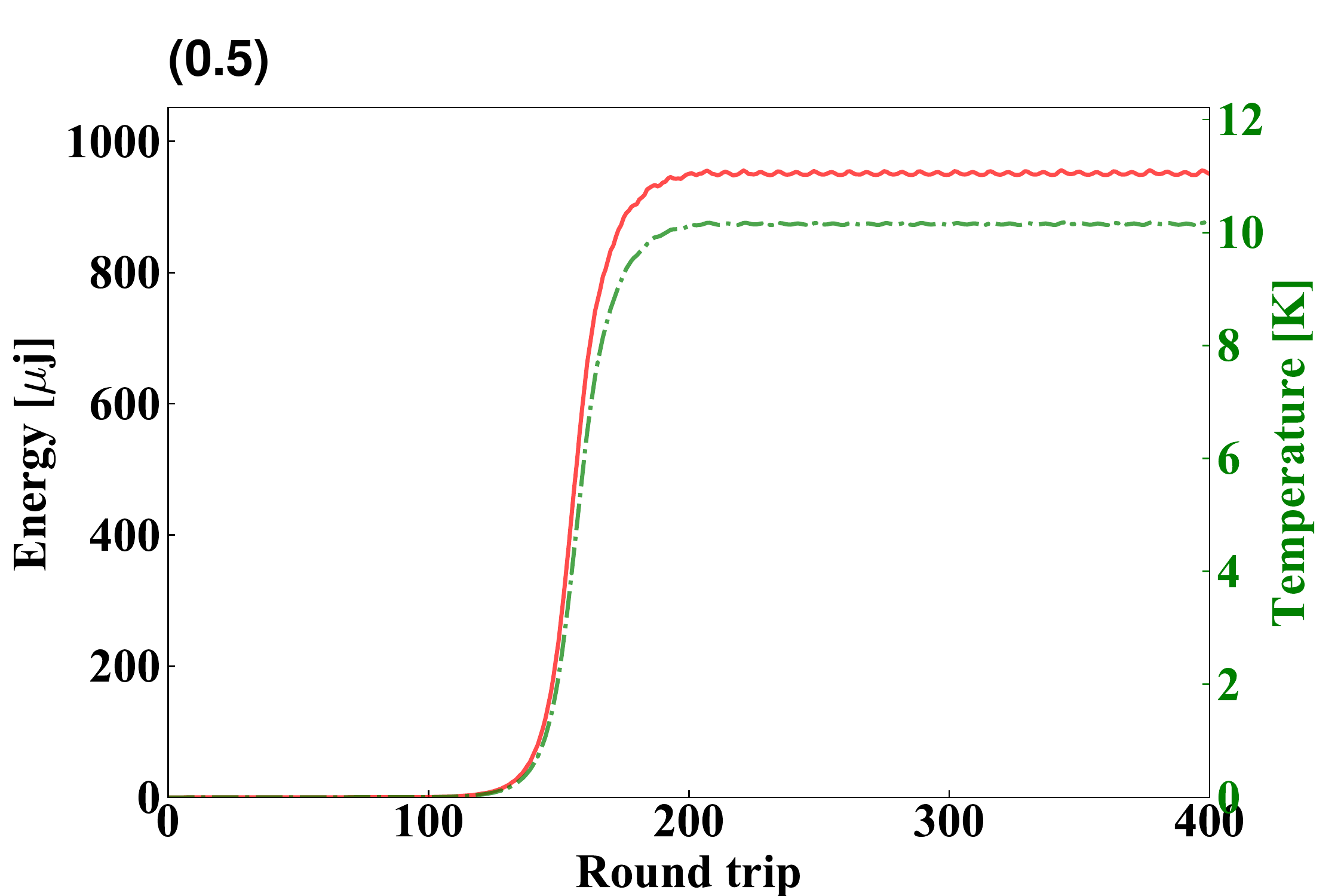}}
     \subfigure{\includegraphics*[width=220pt]{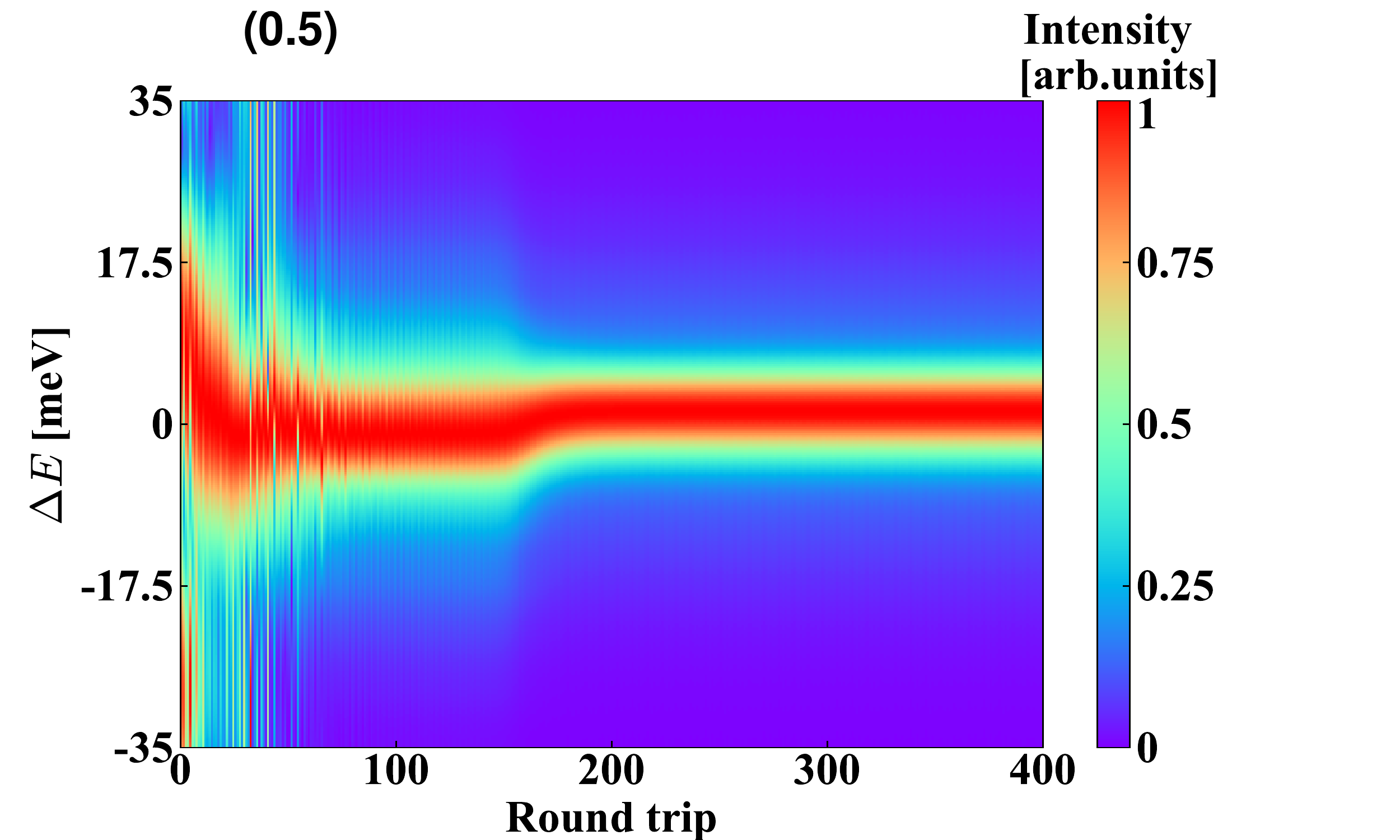}}
     \subfigure{\includegraphics*[width=200pt]{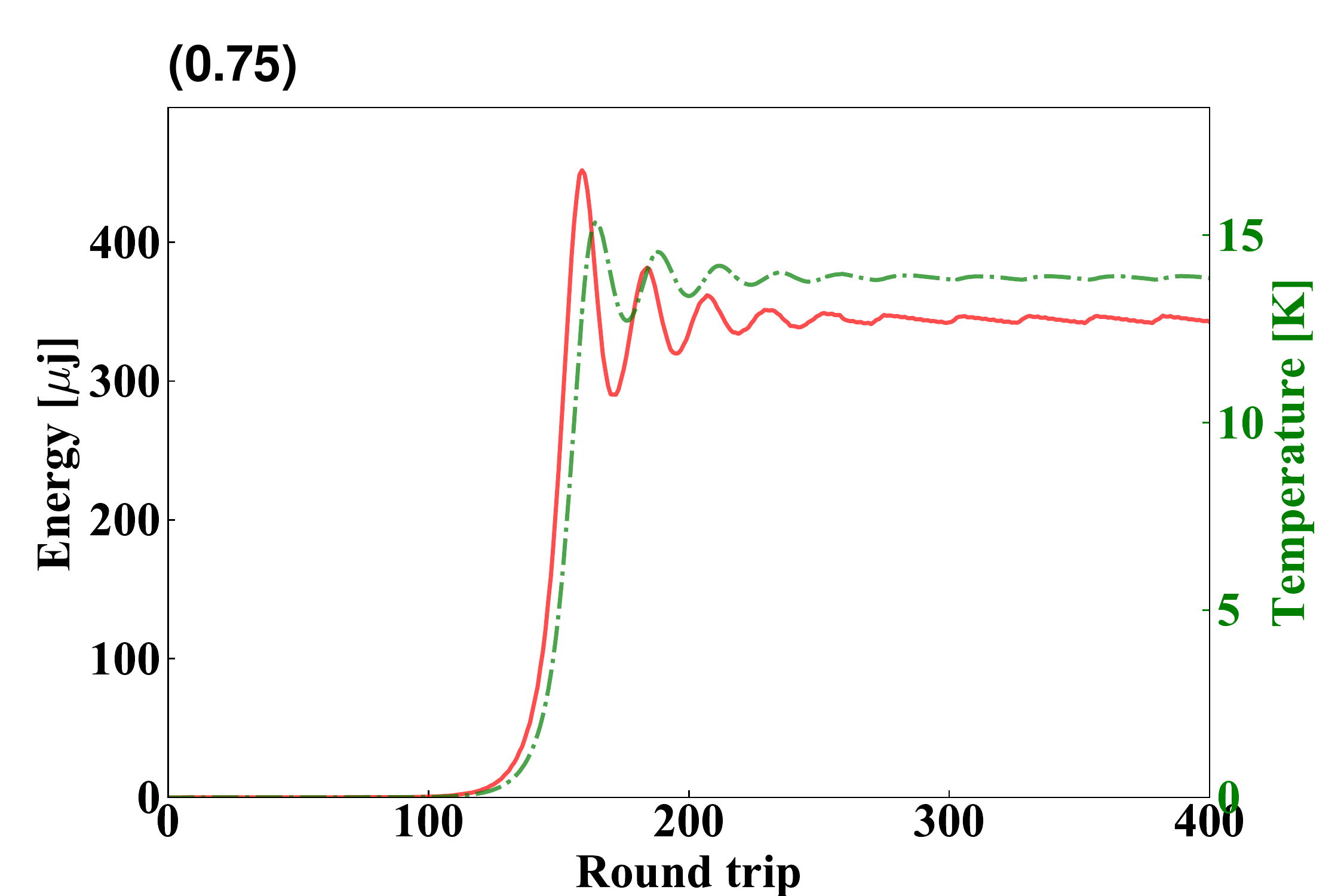}}
     \subfigure{\includegraphics*[width=220pt]{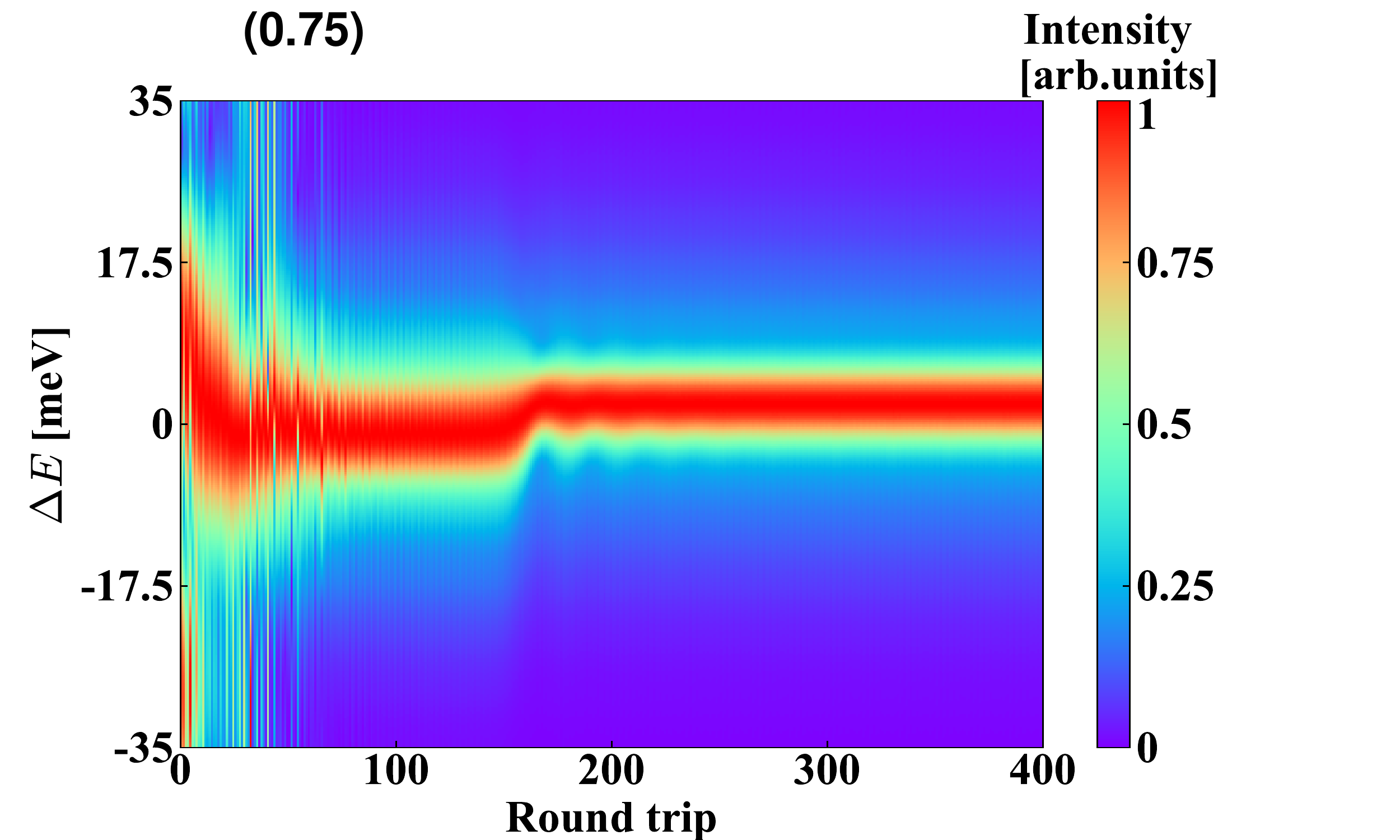}}
     \subfigure{\includegraphics*[width=200pt]{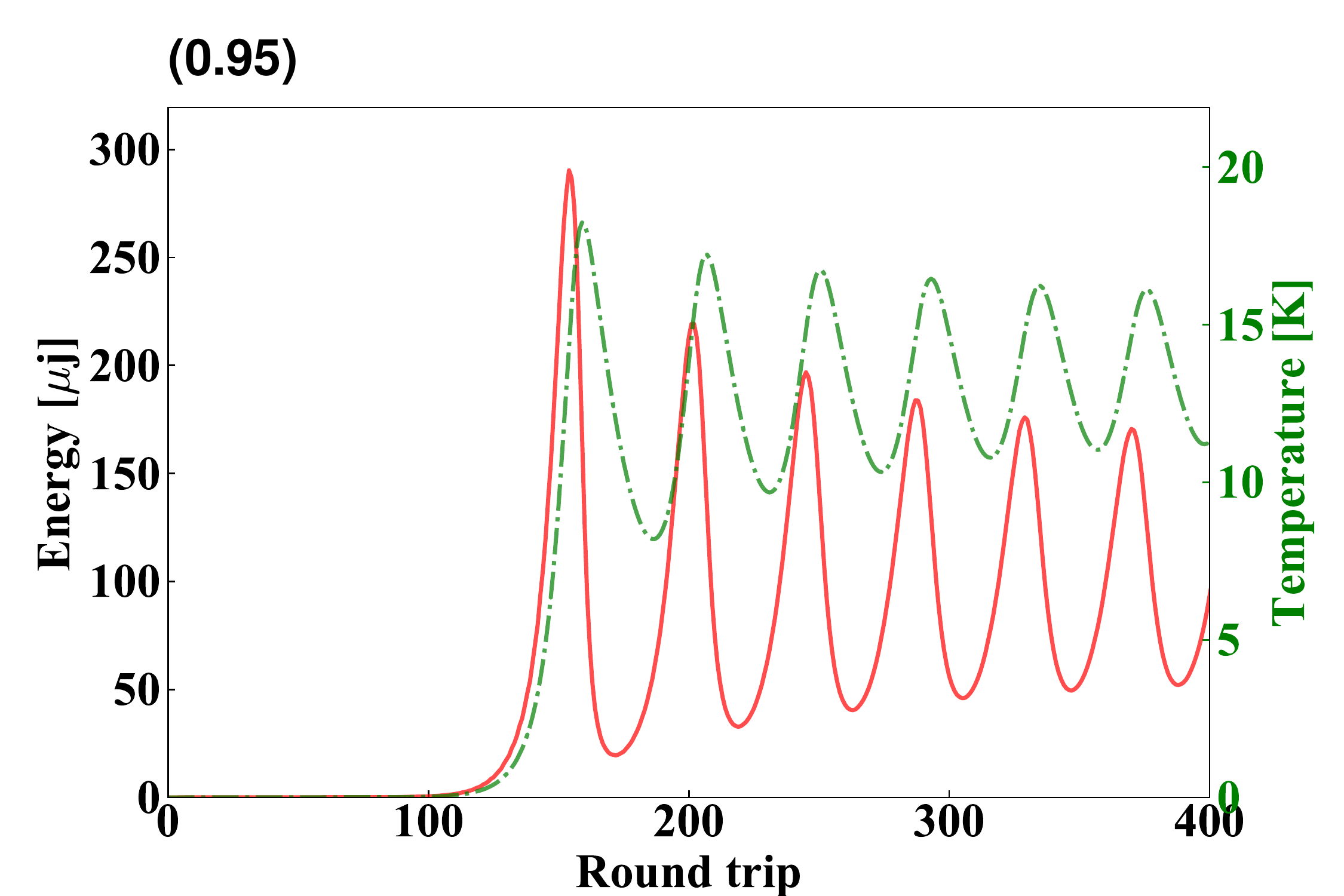}}
     \subfigure{\includegraphics*[width=220pt]{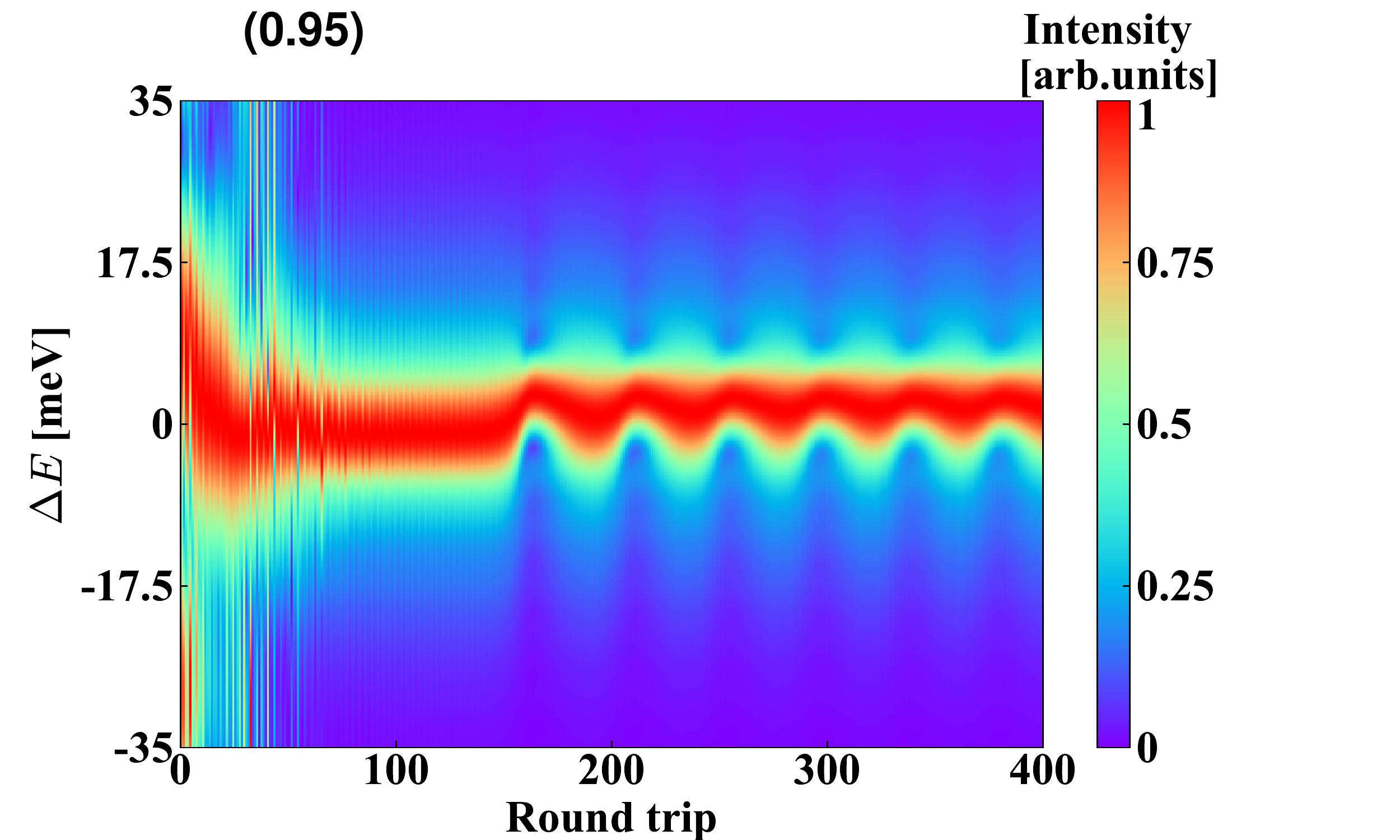}}
     \caption{The evolution of the pulse energy (red line) and the averaged temperature (minus 70~K, green line) at various $\eta$: 0.1, 0.5, 0.75, and 0.95. The evolution of the spectrum at various $\eta$: 0.1, 0.5, 0.75, and 0.95. In each round trip, the spectrum is normalized by its maximum value.}
     \label{fig:energy_evolution}
\end{figure*}

In order to investigate the influence of the thermal loading, the simulations were conducted while $\eta$ is 0.1, 0.5, 0.75, and 0.95. The results are presented in Fig.~\ref{fig:ref_400} and Fig.~\ref{fig:energy_evolution}. When $\eta = 0.95$, the reflectivity shift of the downstream mirror can be observed, shown by a green dash line in the top plot of Fig.~\ref{fig:ref_400}. In comparison, the Bragg reflectivity of the upstream mirror is shown by an orange dash line. A typical spectrum of the XFELO pulse at $\eta = 0.95$ is shown by a red line. As expected, the spectral intensities exist in the overlap of the two reflectivity curves from different mirrors. The total integrated reflectance changes dramatically while the shift of the Bragg reflectivity position exceeds half of the Darwin width. This effect would significantly influence the stability of optical cavity and gain process of the XFELO. Thus, the peak power decreases significantly as $\eta$ rises, as shown in the bottom plot of Fig.~\ref{fig:ref_400}. Another fact should be noted that the different reflectivity also results in a different time delay caused by Bragg reflection. The mismatch of time delay decreases the peak power further, for the X-ray pulses overlap on the electron beam imperfectly. It would be a crucial issue for electron beams with a short bunch length.

Fig.~\ref{fig:energy_evolution} shows the detailed simulation results, including the spectrum evolution and pulse energy growth with each $\eta$. The left panel of Fig.~\ref{fig:energy_evolution} presents the energy evolutions (red line) and its corresponding temperature change (green line). With a small $\eta$ (0.1), which gives an insignificant temperature change, the energy of the x-ray pulse grows as it is amplified by electron beams again and again. The pulse energy reaches 1~mJ at saturation, and the corresponding temperature change is only 0.8~K. At the steady-state, pulse energy has a very small oscillation, which is an XFELO nature. When $\eta$ is set to 0.5, the saturated pulse energy decrease to about 900~$\mu$J, and the temperature change increase to 10~K. The oscillation is also enlarged, as the reflectivity curve is more unstable. The decrement of the pulse energy results from the mismatch of time delay and the relatively large cavity loss due to the shift of the reflectivity curve.


While $\eta$ increases to 0.75, the large remaining heat would lead to a non-trivial temperature change that interrupts X-ray pulse growth through the reflectivity shift. When the total reflectivity can not maintain the stability of the system, the pulse energy decreases, and the temperature change follows it. However, when the temperature change is below the critical value, the pulse energy grows exponentially again as a comparatively intense seed is retained. The repetition of this process leads to that the temperature gradually approaches the critical value and that pulse energy has an attenuating oscillation. The steady-state pulse energy is about 350~$\mu$J, only a third of the value of $\eta=0.1$. 

A large oscillation occurs when much heat remains in the crystal with a large $\eta$. The oscillations are attenuated but stay at a relatively large value eventually. The root of the oscillation is the lingering negative feedback of the temperature change on the pulse energy. Besides, the temperature change is always behind the energy increase. The combined result is that temperature crest gradually approaches the troughs of the pulse energy and reach a balance where oscillation occurs.

The evolution of the spectrum is presented in the right panel of Fig.~\ref{fig:energy_evolution}. The simulations share the same initial random number. Thus, the result for each $\eta$ has nearly the same period of struggling to build up the longitudinal coherence of the X-ray pulse. As energy increases, the spectral width decreases further. The final FWHM spectral width is nearly 13~meV when $\eta$ is 0.1 or 0.5. This value decreases to about 9~mev at $\eta = 0.75$ because the overlap of the reflectivity between the upstream and downstream mirror is narrower. The maximum available shift is about 6~meV, which is nearly half of the Darwin width. 




\section{\label{sec:conclusion} conclusion}

In this paper, we have performed a numerical study to modelling heat load of diamond mirrors in the XFELO operation, including the light-material interaction, the thermal behavior analysis, and the corresponding heat-load coupled XFELO simulation. The light-material interactions were simulated by the particle tracking package GEANT4 with the dedicated Bragg reflection physical process. The advantage of this GEANT4 model is to give accurate and reliable predictions of X-ray absorptions compared to the theoretic attenuation model. The following transient thermal conduction was analyzed by the commercial finite-element analysis software ANSYS. With the 70~K initial temperature and crystal size of 800$\mu$m$\times$800$\mu$m$\times$70$\mu$m, cases for single and multiple pulse input are analyzed. The single-pulse input result indicates that the temperature gradient along the thickness would be negligible in 20~ns, and the temperature would be nearly homogeneous on the whole crystal in 150~ns. The multiple-pulse result indicates that the accumulated temperature increment would reach 3.5~K, which meets the requirement of mirrors for XFELO operation at 70~K in principle. To further investigate the impact of the heat load on XFELO operation, a simplified thermal loading model focusing on the temperature change averaged over the entire crystal is described. Within the simplified model, the parameter $\eta$ is defined to describe the ratio between remaining and initial temperature change. Then, heat-load coupled XFELO simulations are conducted with the parameters of SHINE, while $\eta$ is 0.1, 0.5 0.75, and 0.95. We study the generation of 14.33~keV X-ray pulse, whose energy was found to be 1~mJ in the cavity. With a large $\eta$, the pulse energy and peak power were found to decrease markedly due to the reflectivity shifting and time delay mismatching. The simulations also indicate that the heat load would induce the oscillation of the pulse energy and spectrum at a steady state. And the scope of the energy oscillation is determined by the heat load and the value of $\eta$. 

Beyond this work, the cooling system of mirrors for the XFELO operation of SHINE will be designed and optimized. With the optimized cooling system, the thermal loading analysis with non-Fourier heat conduction will be conducted, so as to provide a better understanding of various heat load effects.







\begin{acknowledgments}
The authors are grateful to J.~Yan and K.~Li for helpful discussions and useful comments. This work was partially supported by the National Key Research and Development Program of China (2018YFE0103100, 2016YFA0401900) and the National Natural Science Foundation of China (11935020, 11775293).
\end{acknowledgments}


\bibliographystyle{apsrev4-1}


\begin{thebibliography}{43}%
     \makeatletter
     \providecommand \@ifxundefined [1]{%
      \@ifx{#1\undefined}
     }%
     \providecommand \@ifnum [1]{%
      \ifnum #1\expandafter \@firstoftwo
      \else \expandafter \@secondoftwo
      \fi
     }%
     \providecommand \@ifx [1]{%
      \ifx #1\expandafter \@firstoftwo
      \else \expandafter \@secondoftwo
      \fi
     }%
     \providecommand \natexlab [1]{#1}%
     \providecommand \enquote  [1]{``#1''}%
     \providecommand \bibnamefont  [1]{#1}%
     \providecommand \bibfnamefont [1]{#1}%
     \providecommand \citenamefont [1]{#1}%
     \providecommand \href@noop [0]{\@secondoftwo}%
     \providecommand \href [0]{\begingroup \@sanitize@url \@href}%
     \providecommand \@href[1]{\@@startlink{#1}\@@href}%
     \providecommand \@@href[1]{\endgroup#1\@@endlink}%
     \providecommand \@sanitize@url [0]{\catcode `\\12\catcode `\$12\catcode
       `\&12\catcode `\#12\catcode `\^12\catcode `\_12\catcode `\%12\relax}%
     \providecommand \@@startlink[1]{}%
     \providecommand \@@endlink[0]{}%
     \providecommand \url  [0]{\begingroup\@sanitize@url \@url }%
     \providecommand \@url [1]{\endgroup\@href {#1}{\urlprefix }}%
     \providecommand \urlprefix  [0]{URL }%
     \providecommand \Eprint [0]{\href }%
     \providecommand \doibase [0]{http://dx.doi.org/}%
     \providecommand \selectlanguage [0]{\@gobble}%
     \providecommand \bibinfo  [0]{\@secondoftwo}%
     \providecommand \bibfield  [0]{\@secondoftwo}%
     \providecommand \translation [1]{[#1]}%
     \providecommand \BibitemOpen [0]{}%
     \providecommand \bibitemStop [0]{}%
     \providecommand \bibitemNoStop [0]{.\EOS\space}%
     \providecommand \EOS [0]{\spacefactor3000\relax}%
     \providecommand \BibitemShut  [1]{\csname bibitem#1\endcsname}%
     \let\auto@bib@innerbib\@empty
     \bibitem [{\citenamefont {Ackermann}\ \emph {et~al.}(2007)\citenamefont
       {Ackermann}, \citenamefont {Asova}, \citenamefont {Ayvazyan}, \citenamefont
       {Azima}, \citenamefont {Baboi}, \citenamefont {B{\"{a}}hr}, \citenamefont
       {Balandin}, \citenamefont {Beutner}, \citenamefont {Brandt}, \citenamefont
       {Bolzmann}, \citenamefont {Brinkmann}, \citenamefont {Brovko}, \citenamefont
       {Castellano}, \citenamefont {Castro}, \citenamefont {Catani}, \citenamefont
       {Chiadroni}, \citenamefont {Choroba}, \citenamefont {Cianchi}, \citenamefont
       {Costello}, \citenamefont {Cubaynes}, \citenamefont {Dardis}, \citenamefont
       {Decking}, \citenamefont {Delsim-Hashemi}, \citenamefont {Delserieys},
       \citenamefont {{Di Pirro}}, \citenamefont {Dohlus}, \citenamefont
       {D{\"{u}}sterer}, \citenamefont {Eckhardt}, \citenamefont {Edwards},
       \citenamefont {Faatz}, \citenamefont {Feldhaus}, \citenamefont
       {Fl{\"{o}}ttmann}, \citenamefont {Frisch}, \citenamefont {Fr{\"{o}}hlich},
       \citenamefont {Garvey}, \citenamefont {Gensch}, \citenamefont {Gerth},
       \citenamefont {G{\"{o}}rler}, \citenamefont {Golubeva}, \citenamefont
       {Grabosch}, \citenamefont {Grecki}, \citenamefont {Grimm}, \citenamefont
       {Hacker}, \citenamefont {Hahn}, \citenamefont {Han}, \citenamefont
       {Honkavaara}, \citenamefont {Hott}, \citenamefont {H{\"{u}}ning},
       \citenamefont {Ivanisenko}, \citenamefont {Jaeschke}, \citenamefont
       {Jalmuzna}, \citenamefont {Jezynski}, \citenamefont {Kammering},
       \citenamefont {Katalev}, \citenamefont {Kavanagh}, \citenamefont {Kennedy},
       \citenamefont {Khodyachykh}, \citenamefont {Klose}, \citenamefont
       {Kocharyan}, \citenamefont {K{\"{o}}rfer}, \citenamefont {Kollewe},
       \citenamefont {Koprek}, \citenamefont {Korepanov}, \citenamefont {Kostin},
       \citenamefont {Krassilnikov}, \citenamefont {Kube}, \citenamefont {Kuhlmann},
       \citenamefont {Lewis}, \citenamefont {Lilje}, \citenamefont {Limberg},
       \citenamefont {Lipka}, \citenamefont {L{\"{o}}hl}, \citenamefont {Luna},
       \citenamefont {Luong}, \citenamefont {Martins}, \citenamefont {Meyer},
       \citenamefont {Michelato}, \citenamefont {Miltchev}, \citenamefont
       {M{\"{o}}ller}, \citenamefont {Monaco}, \citenamefont {M{\"{u}}ller},
       \citenamefont {Napieralski}, \citenamefont {Napoly}, \citenamefont
       {Nicolosi}, \citenamefont {N{\"{o}}lle}, \citenamefont {Nũez}, \citenamefont
       {Oppelt}, \citenamefont {Pagani}, \citenamefont {Paparella}, \citenamefont
       {Pchalek}, \citenamefont {Pedregosa-Gutierrez}, \citenamefont {Petersen},
       \citenamefont {Petrosyan}, \citenamefont {Petrosyan}, \citenamefont
       {Petrosyan}, \citenamefont {Pfl{\"{u}}ger}, \citenamefont {Pl{\"{o}}njes},
       \citenamefont {Poletto}, \citenamefont {Pozniak}, \citenamefont {Prat},
       \citenamefont {Proch}, \citenamefont {Pucyk}, \citenamefont {Radcliffe},
       \citenamefont {Redlin}, \citenamefont {Rehlich}, \citenamefont {Richter},
       \citenamefont {Roehrs}, \citenamefont {Roensch}, \citenamefont {Romaniuk},
       \citenamefont {Ross}, \citenamefont {Rossbach}, \citenamefont {Rybnikov},
       \citenamefont {Sachwitz}, \citenamefont {Saldin}, \citenamefont {Sandner},
       \citenamefont {Schlarb}, \citenamefont {Schmidt}, \citenamefont {Schmitz},
       \citenamefont {Schm{\"{u}}ser}, \citenamefont {Schneider}, \citenamefont
       {Schneidmiller}, \citenamefont {Schnepp}, \citenamefont {Schreiber},
       \citenamefont {Seidel}, \citenamefont {Sertore}, \citenamefont {Shabunov},
       \citenamefont {Simon}, \citenamefont {Simrock}, \citenamefont {Sombrowski},
       \citenamefont {Sorokin}, \citenamefont {Spanknebel}, \citenamefont
       {Spesyvtsev}, \citenamefont {Staykov}, \citenamefont {Steffen}, \citenamefont
       {Stephan}, \citenamefont {Stulle}, \citenamefont {Thom}, \citenamefont
       {Tiedtke}, \citenamefont {Tischer}, \citenamefont {Toleikis}, \citenamefont
       {Treusch}, \citenamefont {Trines}, \citenamefont {Tsakov}, \citenamefont
       {Vogel}, \citenamefont {Weiland}, \citenamefont {Weise}, \citenamefont
       {Wellh{\"{o}}fer}, \citenamefont {Wendt}, \citenamefont {Will}, \citenamefont
       {Winter}, \citenamefont {Wittenburg}, \citenamefont {Wurth}, \citenamefont
       {Yeates}, \citenamefont {Yurkov}, \citenamefont {Zagorodnov},\ and\
       \citenamefont {Zapfe}}]{Ackermann.2007}%
       \BibitemOpen
       \bibfield  {author} {\bibinfo {author} {\bibfnamefont {W.}~\bibnamefont
       {Ackermann}}, \bibinfo {author} {\bibfnamefont {G.}~\bibnamefont {Asova}},
       \bibinfo {author} {\bibfnamefont {V.}~\bibnamefont {Ayvazyan}}, \bibinfo
       {author} {\bibfnamefont {A.}~\bibnamefont {Azima}}, \bibinfo {author}
       {\bibfnamefont {N.}~\bibnamefont {Baboi}}, \bibinfo {author} {\bibfnamefont
       {J.}~\bibnamefont {B{\"{a}}hr}}, \bibinfo {author} {\bibfnamefont
       {V.}~\bibnamefont {Balandin}}, \bibinfo {author} {\bibfnamefont
       {B.}~\bibnamefont {Beutner}}, \bibinfo {author} {\bibfnamefont
       {A.}~\bibnamefont {Brandt}}, \bibinfo {author} {\bibfnamefont
       {A.}~\bibnamefont {Bolzmann}}, \bibinfo {author} {\bibfnamefont
       {R.}~\bibnamefont {Brinkmann}}, \bibinfo {author} {\bibfnamefont {O.~I.}\
       \bibnamefont {Brovko}}, \bibinfo {author} {\bibfnamefont {M.}~\bibnamefont
       {Castellano}}, \bibinfo {author} {\bibfnamefont {P.}~\bibnamefont {Castro}},
       \bibinfo {author} {\bibfnamefont {L.}~\bibnamefont {Catani}}, \bibinfo
       {author} {\bibfnamefont {E.}~\bibnamefont {Chiadroni}}, \bibinfo {author}
       {\bibfnamefont {S.}~\bibnamefont {Choroba}}, \bibinfo {author} {\bibfnamefont
       {A.}~\bibnamefont {Cianchi}}, \bibinfo {author} {\bibfnamefont {J.~T.}\
       \bibnamefont {Costello}}, \bibinfo {author} {\bibfnamefont {D.}~\bibnamefont
       {Cubaynes}}, \bibinfo {author} {\bibfnamefont {J.}~\bibnamefont {Dardis}},
       \bibinfo {author} {\bibfnamefont {W.}~\bibnamefont {Decking}}, \bibinfo
       {author} {\bibfnamefont {H.}~\bibnamefont {Delsim-Hashemi}}, \bibinfo
       {author} {\bibfnamefont {A.}~\bibnamefont {Delserieys}}, \bibinfo {author}
       {\bibfnamefont {G.}~\bibnamefont {{Di Pirro}}}, \bibinfo {author}
       {\bibfnamefont {M.}~\bibnamefont {Dohlus}}, \bibinfo {author} {\bibfnamefont
       {S.}~\bibnamefont {D{\"{u}}sterer}}, \bibinfo {author} {\bibfnamefont
       {A.}~\bibnamefont {Eckhardt}}, \bibinfo {author} {\bibfnamefont {H.~T.}\
       \bibnamefont {Edwards}}, \bibinfo {author} {\bibfnamefont {B.}~\bibnamefont
       {Faatz}}, \bibinfo {author} {\bibfnamefont {J.}~\bibnamefont {Feldhaus}},
       \bibinfo {author} {\bibfnamefont {K.}~\bibnamefont {Fl{\"{o}}ttmann}},
       \bibinfo {author} {\bibfnamefont {J.}~\bibnamefont {Frisch}}, \bibinfo
       {author} {\bibfnamefont {L.}~\bibnamefont {Fr{\"{o}}hlich}}, \bibinfo
       {author} {\bibfnamefont {T.}~\bibnamefont {Garvey}}, \bibinfo {author}
       {\bibfnamefont {U.}~\bibnamefont {Gensch}}, \bibinfo {author} {\bibfnamefont
       {C.}~\bibnamefont {Gerth}}, \bibinfo {author} {\bibfnamefont
       {M.}~\bibnamefont {G{\"{o}}rler}}, \bibinfo {author} {\bibfnamefont
       {N.}~\bibnamefont {Golubeva}}, \bibinfo {author} {\bibfnamefont {H.~J.}\
       \bibnamefont {Grabosch}}, \bibinfo {author} {\bibfnamefont {M.}~\bibnamefont
       {Grecki}}, \bibinfo {author} {\bibfnamefont {O.}~\bibnamefont {Grimm}},
       \bibinfo {author} {\bibfnamefont {K.}~\bibnamefont {Hacker}}, \bibinfo
       {author} {\bibfnamefont {U.}~\bibnamefont {Hahn}}, \bibinfo {author}
       {\bibfnamefont {J.~H.}\ \bibnamefont {Han}}, \bibinfo {author} {\bibfnamefont
       {K.}~\bibnamefont {Honkavaara}}, \bibinfo {author} {\bibfnamefont
       {T.}~\bibnamefont {Hott}}, \bibinfo {author} {\bibfnamefont {M.}~\bibnamefont
       {H{\"{u}}ning}}, \bibinfo {author} {\bibfnamefont {Y.}~\bibnamefont
       {Ivanisenko}}, \bibinfo {author} {\bibfnamefont {E.}~\bibnamefont
       {Jaeschke}}, \bibinfo {author} {\bibfnamefont {W.}~\bibnamefont {Jalmuzna}},
       \bibinfo {author} {\bibfnamefont {T.}~\bibnamefont {Jezynski}}, \bibinfo
       {author} {\bibfnamefont {R.}~\bibnamefont {Kammering}}, \bibinfo {author}
       {\bibfnamefont {V.}~\bibnamefont {Katalev}}, \bibinfo {author} {\bibfnamefont
       {K.}~\bibnamefont {Kavanagh}}, \bibinfo {author} {\bibfnamefont {E.~T.}\
       \bibnamefont {Kennedy}}, \bibinfo {author} {\bibfnamefont {S.}~\bibnamefont
       {Khodyachykh}}, \bibinfo {author} {\bibfnamefont {K.}~\bibnamefont {Klose}},
       \bibinfo {author} {\bibfnamefont {V.}~\bibnamefont {Kocharyan}}, \bibinfo
       {author} {\bibfnamefont {M.}~\bibnamefont {K{\"{o}}rfer}}, \bibinfo {author}
       {\bibfnamefont {M.}~\bibnamefont {Kollewe}}, \bibinfo {author} {\bibfnamefont
       {W.}~\bibnamefont {Koprek}}, \bibinfo {author} {\bibfnamefont
       {S.}~\bibnamefont {Korepanov}}, \bibinfo {author} {\bibfnamefont
       {D.}~\bibnamefont {Kostin}}, \bibinfo {author} {\bibfnamefont
       {M.}~\bibnamefont {Krassilnikov}}, \bibinfo {author} {\bibfnamefont
       {G.}~\bibnamefont {Kube}}, \bibinfo {author} {\bibfnamefont {M.}~\bibnamefont
       {Kuhlmann}}, \bibinfo {author} {\bibfnamefont {C.~L.}\ \bibnamefont {Lewis}},
       \bibinfo {author} {\bibfnamefont {L.}~\bibnamefont {Lilje}}, \bibinfo
       {author} {\bibfnamefont {T.}~\bibnamefont {Limberg}}, \bibinfo {author}
       {\bibfnamefont {D.}~\bibnamefont {Lipka}}, \bibinfo {author} {\bibfnamefont
       {F.}~\bibnamefont {L{\"{o}}hl}}, \bibinfo {author} {\bibfnamefont
       {H.}~\bibnamefont {Luna}}, \bibinfo {author} {\bibfnamefont {M.}~\bibnamefont
       {Luong}}, \bibinfo {author} {\bibfnamefont {M.}~\bibnamefont {Martins}},
       \bibinfo {author} {\bibfnamefont {M.}~\bibnamefont {Meyer}}, \bibinfo
       {author} {\bibfnamefont {P.}~\bibnamefont {Michelato}}, \bibinfo {author}
       {\bibfnamefont {V.}~\bibnamefont {Miltchev}}, \bibinfo {author}
       {\bibfnamefont {W.~D.}\ \bibnamefont {M{\"{o}}ller}}, \bibinfo {author}
       {\bibfnamefont {L.}~\bibnamefont {Monaco}}, \bibinfo {author} {\bibfnamefont
       {W.~F.}\ \bibnamefont {M{\"{u}}ller}}, \bibinfo {author} {\bibfnamefont
       {O.}~\bibnamefont {Napieralski}}, \bibinfo {author} {\bibfnamefont
       {O.}~\bibnamefont {Napoly}}, \bibinfo {author} {\bibfnamefont
       {P.}~\bibnamefont {Nicolosi}}, \bibinfo {author} {\bibfnamefont
       {D.}~\bibnamefont {N{\"{o}}lle}}, \bibinfo {author} {\bibfnamefont
       {T.}~\bibnamefont {Nũez}}, \bibinfo {author} {\bibfnamefont
       {A.}~\bibnamefont {Oppelt}}, \bibinfo {author} {\bibfnamefont
       {C.}~\bibnamefont {Pagani}}, \bibinfo {author} {\bibfnamefont
       {R.}~\bibnamefont {Paparella}}, \bibinfo {author} {\bibfnamefont
       {N.}~\bibnamefont {Pchalek}}, \bibinfo {author} {\bibfnamefont
       {J.}~\bibnamefont {Pedregosa-Gutierrez}}, \bibinfo {author} {\bibfnamefont
       {B.}~\bibnamefont {Petersen}}, \bibinfo {author} {\bibfnamefont
       {B.}~\bibnamefont {Petrosyan}}, \bibinfo {author} {\bibfnamefont
       {G.}~\bibnamefont {Petrosyan}}, \bibinfo {author} {\bibfnamefont
       {L.}~\bibnamefont {Petrosyan}}, \bibinfo {author} {\bibfnamefont
       {J.}~\bibnamefont {Pfl{\"{u}}ger}}, \bibinfo {author} {\bibfnamefont
       {E.}~\bibnamefont {Pl{\"{o}}njes}}, \bibinfo {author} {\bibfnamefont
       {L.}~\bibnamefont {Poletto}}, \bibinfo {author} {\bibfnamefont
       {K.}~\bibnamefont {Pozniak}}, \bibinfo {author} {\bibfnamefont
       {E.}~\bibnamefont {Prat}}, \bibinfo {author} {\bibfnamefont {D.}~\bibnamefont
       {Proch}}, \bibinfo {author} {\bibfnamefont {P.}~\bibnamefont {Pucyk}},
       \bibinfo {author} {\bibfnamefont {P.}~\bibnamefont {Radcliffe}}, \bibinfo
       {author} {\bibfnamefont {H.}~\bibnamefont {Redlin}}, \bibinfo {author}
       {\bibfnamefont {K.}~\bibnamefont {Rehlich}}, \bibinfo {author} {\bibfnamefont
       {M.}~\bibnamefont {Richter}}, \bibinfo {author} {\bibfnamefont
       {M.}~\bibnamefont {Roehrs}}, \bibinfo {author} {\bibfnamefont
       {J.}~\bibnamefont {Roensch}}, \bibinfo {author} {\bibfnamefont
       {R.}~\bibnamefont {Romaniuk}}, \bibinfo {author} {\bibfnamefont
       {M.}~\bibnamefont {Ross}}, \bibinfo {author} {\bibfnamefont {J.}~\bibnamefont
       {Rossbach}}, \bibinfo {author} {\bibfnamefont {V.}~\bibnamefont {Rybnikov}},
       \bibinfo {author} {\bibfnamefont {M.}~\bibnamefont {Sachwitz}}, \bibinfo
       {author} {\bibfnamefont {E.~L.}\ \bibnamefont {Saldin}}, \bibinfo {author}
       {\bibfnamefont {W.}~\bibnamefont {Sandner}}, \bibinfo {author} {\bibfnamefont
       {H.}~\bibnamefont {Schlarb}}, \bibinfo {author} {\bibfnamefont
       {B.}~\bibnamefont {Schmidt}}, \bibinfo {author} {\bibfnamefont
       {M.}~\bibnamefont {Schmitz}}, \bibinfo {author} {\bibfnamefont
       {P.}~\bibnamefont {Schm{\"{u}}ser}}, \bibinfo {author} {\bibfnamefont
       {J.~R.}\ \bibnamefont {Schneider}}, \bibinfo {author} {\bibfnamefont {E.~A.}\
       \bibnamefont {Schneidmiller}}, \bibinfo {author} {\bibfnamefont
       {S.}~\bibnamefont {Schnepp}}, \bibinfo {author} {\bibfnamefont
       {S.}~\bibnamefont {Schreiber}}, \bibinfo {author} {\bibfnamefont
       {M.}~\bibnamefont {Seidel}}, \bibinfo {author} {\bibfnamefont
       {D.}~\bibnamefont {Sertore}}, \bibinfo {author} {\bibfnamefont {A.~V.}\
       \bibnamefont {Shabunov}}, \bibinfo {author} {\bibfnamefont {C.}~\bibnamefont
       {Simon}}, \bibinfo {author} {\bibfnamefont {S.}~\bibnamefont {Simrock}},
       \bibinfo {author} {\bibfnamefont {E.}~\bibnamefont {Sombrowski}}, \bibinfo
       {author} {\bibfnamefont {A.~A.}\ \bibnamefont {Sorokin}}, \bibinfo {author}
       {\bibfnamefont {P.}~\bibnamefont {Spanknebel}}, \bibinfo {author}
       {\bibfnamefont {R.}~\bibnamefont {Spesyvtsev}}, \bibinfo {author}
       {\bibfnamefont {L.}~\bibnamefont {Staykov}}, \bibinfo {author} {\bibfnamefont
       {B.}~\bibnamefont {Steffen}}, \bibinfo {author} {\bibfnamefont
       {F.}~\bibnamefont {Stephan}}, \bibinfo {author} {\bibfnamefont
       {F.}~\bibnamefont {Stulle}}, \bibinfo {author} {\bibfnamefont
       {H.}~\bibnamefont {Thom}}, \bibinfo {author} {\bibfnamefont {K.}~\bibnamefont
       {Tiedtke}}, \bibinfo {author} {\bibfnamefont {M.}~\bibnamefont {Tischer}},
       \bibinfo {author} {\bibfnamefont {S.}~\bibnamefont {Toleikis}}, \bibinfo
       {author} {\bibfnamefont {R.}~\bibnamefont {Treusch}}, \bibinfo {author}
       {\bibfnamefont {D.}~\bibnamefont {Trines}}, \bibinfo {author} {\bibfnamefont
       {I.}~\bibnamefont {Tsakov}}, \bibinfo {author} {\bibfnamefont
       {E.}~\bibnamefont {Vogel}}, \bibinfo {author} {\bibfnamefont
       {T.}~\bibnamefont {Weiland}}, \bibinfo {author} {\bibfnamefont
       {H.}~\bibnamefont {Weise}}, \bibinfo {author} {\bibfnamefont
       {M.}~\bibnamefont {Wellh{\"{o}}fer}}, \bibinfo {author} {\bibfnamefont
       {M.}~\bibnamefont {Wendt}}, \bibinfo {author} {\bibfnamefont
       {I.}~\bibnamefont {Will}}, \bibinfo {author} {\bibfnamefont {A.}~\bibnamefont
       {Winter}}, \bibinfo {author} {\bibfnamefont {K.}~\bibnamefont {Wittenburg}},
       \bibinfo {author} {\bibfnamefont {W.}~\bibnamefont {Wurth}}, \bibinfo
       {author} {\bibfnamefont {P.}~\bibnamefont {Yeates}}, \bibinfo {author}
       {\bibfnamefont {M.~V.}\ \bibnamefont {Yurkov}}, \bibinfo {author}
       {\bibfnamefont {I.}~\bibnamefont {Zagorodnov}}, \ and\ \bibinfo {author}
       {\bibfnamefont {K.}~\bibnamefont {Zapfe}},\ }\href {\doibase
       10.1038/nphoton.2007.76} {\bibfield  {journal} {\bibinfo  {journal} {Nat.
       Photonics}\ }\textbf {\bibinfo {volume} {1}},\ \bibinfo {pages} {336}
       (\bibinfo {year} {2007})}\BibitemShut {NoStop}%
     \bibitem [{\citenamefont {Emma}\ \emph {et~al.}(2010)\citenamefont {Emma},
       \citenamefont {Akre}, \citenamefont {Arthur}, \citenamefont {Bionta},
       \citenamefont {Bostedt}, \citenamefont {Bozek}, \citenamefont {Brachmann},
       \citenamefont {Bucksbaum}, \citenamefont {Coffee}, \citenamefont {Decker},
       \citenamefont {Ding}, \citenamefont {Dowell}, \citenamefont {Edstrom},
       \citenamefont {Fisher}, \citenamefont {Frisch}, \citenamefont {Gilevich},
       \citenamefont {Hastings}, \citenamefont {Hays}, \citenamefont {Hering},
       \citenamefont {Huang}, \citenamefont {Iverson}, \citenamefont {Loos},
       \citenamefont {Messerschmidt}, \citenamefont {Miahnahri}, \citenamefont
       {Moeller}, \citenamefont {Nuhn}, \citenamefont {Pile}, \citenamefont
       {Ratner}, \citenamefont {Rzepiela}, \citenamefont {Schultz}, \citenamefont
       {Smith}, \citenamefont {Stefan}, \citenamefont {Tompkins}, \citenamefont
       {Turner}, \citenamefont {Welch}, \citenamefont {White}, \citenamefont {Wu},
       \citenamefont {Yocky},\ and\ \citenamefont {Galayda}}]{Emma.2010}%
       \BibitemOpen
       \bibfield  {author} {\bibinfo {author} {\bibfnamefont {P.}~\bibnamefont
       {Emma}}, \bibinfo {author} {\bibfnamefont {R.}~\bibnamefont {Akre}}, \bibinfo
       {author} {\bibfnamefont {J.}~\bibnamefont {Arthur}}, \bibinfo {author}
       {\bibfnamefont {R.}~\bibnamefont {Bionta}}, \bibinfo {author} {\bibfnamefont
       {C.}~\bibnamefont {Bostedt}}, \bibinfo {author} {\bibfnamefont
       {J.}~\bibnamefont {Bozek}}, \bibinfo {author} {\bibfnamefont
       {A.}~\bibnamefont {Brachmann}}, \bibinfo {author} {\bibfnamefont
       {P.}~\bibnamefont {Bucksbaum}}, \bibinfo {author} {\bibfnamefont
       {R.}~\bibnamefont {Coffee}}, \bibinfo {author} {\bibfnamefont {F.~J.}\
       \bibnamefont {Decker}}, \bibinfo {author} {\bibfnamefont {Y.}~\bibnamefont
       {Ding}}, \bibinfo {author} {\bibfnamefont {D.}~\bibnamefont {Dowell}},
       \bibinfo {author} {\bibfnamefont {S.}~\bibnamefont {Edstrom}}, \bibinfo
       {author} {\bibfnamefont {A.}~\bibnamefont {Fisher}}, \bibinfo {author}
       {\bibfnamefont {J.}~\bibnamefont {Frisch}}, \bibinfo {author} {\bibfnamefont
       {S.}~\bibnamefont {Gilevich}}, \bibinfo {author} {\bibfnamefont
       {J.}~\bibnamefont {Hastings}}, \bibinfo {author} {\bibfnamefont
       {G.}~\bibnamefont {Hays}}, \bibinfo {author} {\bibfnamefont {P.}~\bibnamefont
       {Hering}}, \bibinfo {author} {\bibfnamefont {Z.}~\bibnamefont {Huang}},
       \bibinfo {author} {\bibfnamefont {R.}~\bibnamefont {Iverson}}, \bibinfo
       {author} {\bibfnamefont {H.}~\bibnamefont {Loos}}, \bibinfo {author}
       {\bibfnamefont {M.}~\bibnamefont {Messerschmidt}}, \bibinfo {author}
       {\bibfnamefont {A.}~\bibnamefont {Miahnahri}}, \bibinfo {author}
       {\bibfnamefont {S.}~\bibnamefont {Moeller}}, \bibinfo {author} {\bibfnamefont
       {H.~D.}\ \bibnamefont {Nuhn}}, \bibinfo {author} {\bibfnamefont
       {G.}~\bibnamefont {Pile}}, \bibinfo {author} {\bibfnamefont {D.}~\bibnamefont
       {Ratner}}, \bibinfo {author} {\bibfnamefont {J.}~\bibnamefont {Rzepiela}},
       \bibinfo {author} {\bibfnamefont {D.}~\bibnamefont {Schultz}}, \bibinfo
       {author} {\bibfnamefont {T.}~\bibnamefont {Smith}}, \bibinfo {author}
       {\bibfnamefont {P.}~\bibnamefont {Stefan}}, \bibinfo {author} {\bibfnamefont
       {H.}~\bibnamefont {Tompkins}}, \bibinfo {author} {\bibfnamefont
       {J.}~\bibnamefont {Turner}}, \bibinfo {author} {\bibfnamefont
       {J.}~\bibnamefont {Welch}}, \bibinfo {author} {\bibfnamefont
       {W.}~\bibnamefont {White}}, \bibinfo {author} {\bibfnamefont
       {J.}~\bibnamefont {Wu}}, \bibinfo {author} {\bibfnamefont {G.}~\bibnamefont
       {Yocky}}, \ and\ \bibinfo {author} {\bibfnamefont {J.}~\bibnamefont
       {Galayda}},\ }\href {\doibase 10.1038/nphoton.2010.176} {\bibfield  {journal}
       {\bibinfo  {journal} {Nat. Photonics}\ }\textbf {\bibinfo {volume} {4}},\
       \bibinfo {pages} {641} (\bibinfo {year} {2010})}\BibitemShut {NoStop}%
     \bibitem [{\citenamefont {Ishikawa}\ \emph {et~al.}(2012)\citenamefont
       {Ishikawa}, \citenamefont {Aoyagi}, \citenamefont {Asaka}, \citenamefont
       {Asano}, \citenamefont {Azumi}, \citenamefont {Bizen}, \citenamefont {Ego},
       \citenamefont {Fukami}, \citenamefont {Fukui}, \citenamefont {Furukawa},\
       and\ \citenamefont {Others}}]{Ishikawa.2012}%
       \BibitemOpen
       \bibfield  {author} {\bibinfo {author} {\bibfnamefont {T.}~\bibnamefont
       {Ishikawa}}, \bibinfo {author} {\bibfnamefont {H.}~\bibnamefont {Aoyagi}},
       \bibinfo {author} {\bibfnamefont {T.}~\bibnamefont {Asaka}}, \bibinfo
       {author} {\bibfnamefont {Y.}~\bibnamefont {Asano}}, \bibinfo {author}
       {\bibfnamefont {N.}~\bibnamefont {Azumi}}, \bibinfo {author} {\bibfnamefont
       {T.}~\bibnamefont {Bizen}}, \bibinfo {author} {\bibfnamefont
       {H.}~\bibnamefont {Ego}}, \bibinfo {author} {\bibfnamefont {K.}~\bibnamefont
       {Fukami}}, \bibinfo {author} {\bibfnamefont {T.}~\bibnamefont {Fukui}},
       \bibinfo {author} {\bibfnamefont {Y.}~\bibnamefont {Furukawa}}, \ and\
       \bibinfo {author} {\bibnamefont {Others}},\ }\href@noop {} {\bibfield
       {journal} {\bibinfo  {journal} {Nat. Photonics}\ }\textbf {\bibinfo {volume}
       {6}},\ \bibinfo {pages} {540} (\bibinfo {year} {2012})}\BibitemShut {NoStop}%
     \bibitem [{\citenamefont {Kang}\ \emph {et~al.}(2017)\citenamefont {Kang},
       \citenamefont {Min}, \citenamefont {Heo}, \citenamefont {Kim}, \citenamefont
       {Yang}, \citenamefont {Kim}, \citenamefont {Nam}, \citenamefont {Baek},
       \citenamefont {Choi}, \citenamefont {Mun}, \citenamefont {Park},
       \citenamefont {Suh}, \citenamefont {Shin}, \citenamefont {Hu}, \citenamefont
       {Hong}, \citenamefont {Jung}, \citenamefont {Kim}, \citenamefont {Kim},
       \citenamefont {Na}, \citenamefont {Park}, \citenamefont {Park}, \citenamefont
       {Han}, \citenamefont {Jung}, \citenamefont {Jeong}, \citenamefont {Lee},
       \citenamefont {Lee}, \citenamefont {Lee}, \citenamefont {Lee}, \citenamefont
       {Oh}, \citenamefont {Suh}, \citenamefont {Parc}, \citenamefont {Park},
       \citenamefont {Kim}, \citenamefont {Jung}, \citenamefont {Kim}, \citenamefont
       {Lee}, \citenamefont {Lee}, \citenamefont {Sung}, \citenamefont {Mok},
       \citenamefont {Yang}, \citenamefont {Lee}, \citenamefont {Shin},
       \citenamefont {Kim}, \citenamefont {Kim}, \citenamefont {Lee}, \citenamefont
       {Park}, \citenamefont {Kim}, \citenamefont {Park}, \citenamefont {Eom},
       \citenamefont {Rah}, \citenamefont {Kim}, \citenamefont {Nam}, \citenamefont
       {Park}, \citenamefont {Park}, \citenamefont {Kim}, \citenamefont {Kwon},
       \citenamefont {Park}, \citenamefont {Kim}, \citenamefont {Hyun},
       \citenamefont {Kim}, \citenamefont {Kim}, \citenamefont {Hwang},
       \citenamefont {Kim}, \citenamefont {Lim}, \citenamefont {Yu}, \citenamefont
       {Kim}, \citenamefont {Kang}, \citenamefont {Kim}, \citenamefont {Kim},
       \citenamefont {Lee}, \citenamefont {Lee}, \citenamefont {Park}, \citenamefont
       {Koo}, \citenamefont {Kim},\ and\ \citenamefont {Ko}}]{Kang.2017}%
       \BibitemOpen
       \bibfield  {author} {\bibinfo {author} {\bibfnamefont {H.~S.}\ \bibnamefont
       {Kang}}, \bibinfo {author} {\bibfnamefont {C.~K.}\ \bibnamefont {Min}},
       \bibinfo {author} {\bibfnamefont {H.}~\bibnamefont {Heo}}, \bibinfo {author}
       {\bibfnamefont {C.}~\bibnamefont {Kim}}, \bibinfo {author} {\bibfnamefont
       {H.}~\bibnamefont {Yang}}, \bibinfo {author} {\bibfnamefont {G.}~\bibnamefont
       {Kim}}, \bibinfo {author} {\bibfnamefont {I.}~\bibnamefont {Nam}}, \bibinfo
       {author} {\bibfnamefont {S.~Y.}\ \bibnamefont {Baek}}, \bibinfo {author}
       {\bibfnamefont {H.~J.}\ \bibnamefont {Choi}}, \bibinfo {author}
       {\bibfnamefont {G.}~\bibnamefont {Mun}}, \bibinfo {author} {\bibfnamefont
       {B.~R.}\ \bibnamefont {Park}}, \bibinfo {author} {\bibfnamefont {Y.~J.}\
       \bibnamefont {Suh}}, \bibinfo {author} {\bibfnamefont {D.~C.}\ \bibnamefont
       {Shin}}, \bibinfo {author} {\bibfnamefont {J.}~\bibnamefont {Hu}}, \bibinfo
       {author} {\bibfnamefont {J.}~\bibnamefont {Hong}}, \bibinfo {author}
       {\bibfnamefont {S.}~\bibnamefont {Jung}}, \bibinfo {author} {\bibfnamefont
       {S.~H.}\ \bibnamefont {Kim}}, \bibinfo {author} {\bibfnamefont {K.~H.}\
       \bibnamefont {Kim}}, \bibinfo {author} {\bibfnamefont {D.}~\bibnamefont
       {Na}}, \bibinfo {author} {\bibfnamefont {S.~S.}\ \bibnamefont {Park}},
       \bibinfo {author} {\bibfnamefont {Y.~J.}\ \bibnamefont {Park}}, \bibinfo
       {author} {\bibfnamefont {J.~H.}\ \bibnamefont {Han}}, \bibinfo {author}
       {\bibfnamefont {Y.~G.}\ \bibnamefont {Jung}}, \bibinfo {author}
       {\bibfnamefont {S.~H.}\ \bibnamefont {Jeong}}, \bibinfo {author}
       {\bibfnamefont {H.~G.}\ \bibnamefont {Lee}}, \bibinfo {author} {\bibfnamefont
       {S.}~\bibnamefont {Lee}}, \bibinfo {author} {\bibfnamefont {S.}~\bibnamefont
       {Lee}}, \bibinfo {author} {\bibfnamefont {W.~W.}\ \bibnamefont {Lee}},
       \bibinfo {author} {\bibfnamefont {B.}~\bibnamefont {Oh}}, \bibinfo {author}
       {\bibfnamefont {H.~S.}\ \bibnamefont {Suh}}, \bibinfo {author} {\bibfnamefont
       {Y.~W.}\ \bibnamefont {Parc}}, \bibinfo {author} {\bibfnamefont {S.~J.}\
       \bibnamefont {Park}}, \bibinfo {author} {\bibfnamefont {M.~H.}\ \bibnamefont
       {Kim}}, \bibinfo {author} {\bibfnamefont {N.~S.}\ \bibnamefont {Jung}},
       \bibinfo {author} {\bibfnamefont {Y.~C.}\ \bibnamefont {Kim}}, \bibinfo
       {author} {\bibfnamefont {M.~S.}\ \bibnamefont {Lee}}, \bibinfo {author}
       {\bibfnamefont {B.~H.}\ \bibnamefont {Lee}}, \bibinfo {author} {\bibfnamefont
       {C.~W.}\ \bibnamefont {Sung}}, \bibinfo {author} {\bibfnamefont {I.~S.}\
       \bibnamefont {Mok}}, \bibinfo {author} {\bibfnamefont {J.~M.}\ \bibnamefont
       {Yang}}, \bibinfo {author} {\bibfnamefont {C.~S.}\ \bibnamefont {Lee}},
       \bibinfo {author} {\bibfnamefont {H.}~\bibnamefont {Shin}}, \bibinfo {author}
       {\bibfnamefont {J.~H.}\ \bibnamefont {Kim}}, \bibinfo {author} {\bibfnamefont
       {Y.}~\bibnamefont {Kim}}, \bibinfo {author} {\bibfnamefont {J.~H.}\
       \bibnamefont {Lee}}, \bibinfo {author} {\bibfnamefont {S.~Y.}\ \bibnamefont
       {Park}}, \bibinfo {author} {\bibfnamefont {J.}~\bibnamefont {Kim}}, \bibinfo
       {author} {\bibfnamefont {J.}~\bibnamefont {Park}}, \bibinfo {author}
       {\bibfnamefont {I.}~\bibnamefont {Eom}}, \bibinfo {author} {\bibfnamefont
       {S.}~\bibnamefont {Rah}}, \bibinfo {author} {\bibfnamefont {S.}~\bibnamefont
       {Kim}}, \bibinfo {author} {\bibfnamefont {K.~H.}\ \bibnamefont {Nam}},
       \bibinfo {author} {\bibfnamefont {J.}~\bibnamefont {Park}}, \bibinfo {author}
       {\bibfnamefont {J.}~\bibnamefont {Park}}, \bibinfo {author} {\bibfnamefont
       {S.}~\bibnamefont {Kim}}, \bibinfo {author} {\bibfnamefont {S.}~\bibnamefont
       {Kwon}}, \bibinfo {author} {\bibfnamefont {S.~H.}\ \bibnamefont {Park}},
       \bibinfo {author} {\bibfnamefont {K.~S.}\ \bibnamefont {Kim}}, \bibinfo
       {author} {\bibfnamefont {H.}~\bibnamefont {Hyun}}, \bibinfo {author}
       {\bibfnamefont {S.~N.}\ \bibnamefont {Kim}}, \bibinfo {author} {\bibfnamefont
       {S.}~\bibnamefont {Kim}}, \bibinfo {author} {\bibfnamefont {S.~M.}\
       \bibnamefont {Hwang}}, \bibinfo {author} {\bibfnamefont {M.~J.}\ \bibnamefont
       {Kim}}, \bibinfo {author} {\bibfnamefont {C.~Y.}\ \bibnamefont {Lim}},
       \bibinfo {author} {\bibfnamefont {C.~J.}\ \bibnamefont {Yu}}, \bibinfo
       {author} {\bibfnamefont {B.~S.}\ \bibnamefont {Kim}}, \bibinfo {author}
       {\bibfnamefont {T.~H.}\ \bibnamefont {Kang}}, \bibinfo {author}
       {\bibfnamefont {K.~W.}\ \bibnamefont {Kim}}, \bibinfo {author} {\bibfnamefont
       {S.~H.}\ \bibnamefont {Kim}}, \bibinfo {author} {\bibfnamefont {H.~S.}\
       \bibnamefont {Lee}}, \bibinfo {author} {\bibfnamefont {H.~S.}\ \bibnamefont
       {Lee}}, \bibinfo {author} {\bibfnamefont {K.~H.}\ \bibnamefont {Park}},
       \bibinfo {author} {\bibfnamefont {T.~Y.}\ \bibnamefont {Koo}}, \bibinfo
       {author} {\bibfnamefont {D.~E.}\ \bibnamefont {Kim}}, \ and\ \bibinfo
       {author} {\bibfnamefont {I.~S.}\ \bibnamefont {Ko}},\ }\href {\doibase
       10.1038/s41566-017-0029-8} {\bibfield  {journal} {\bibinfo  {journal} {Nat.
       Photonics}\ }\textbf {\bibinfo {volume} {11}},\ \bibinfo {pages} {708}
       (\bibinfo {year} {2017})}\BibitemShut {NoStop}%
     \bibitem [{\citenamefont {Milne}\ \emph {et~al.}(2017)\citenamefont {Milne},
       \citenamefont {Schietinger}, \citenamefont {Aiba}, \citenamefont {Alarcon},
       \citenamefont {Alex}, \citenamefont {Anghel}, \citenamefont {Arsov},
       \citenamefont {Beard}, \citenamefont {Beaud}, \citenamefont {Bettoni},\ and\
       \citenamefont {Others}}]{milne2017swissfel}%
       \BibitemOpen
       \bibfield  {author} {\bibinfo {author} {\bibfnamefont {C.}~\bibnamefont
       {Milne}}, \bibinfo {author} {\bibfnamefont {T.}~\bibnamefont {Schietinger}},
       \bibinfo {author} {\bibfnamefont {M.}~\bibnamefont {Aiba}}, \bibinfo {author}
       {\bibfnamefont {A.}~\bibnamefont {Alarcon}}, \bibinfo {author} {\bibfnamefont
       {J.}~\bibnamefont {Alex}}, \bibinfo {author} {\bibfnamefont {A.}~\bibnamefont
       {Anghel}}, \bibinfo {author} {\bibfnamefont {V.}~\bibnamefont {Arsov}},
       \bibinfo {author} {\bibfnamefont {C.}~\bibnamefont {Beard}}, \bibinfo
       {author} {\bibfnamefont {P.}~\bibnamefont {Beaud}}, \bibinfo {author}
       {\bibfnamefont {S.}~\bibnamefont {Bettoni}}, \ and\ \bibinfo {author}
       {\bibnamefont {Others}},\ }\href@noop {} {\bibfield  {journal} {\bibinfo
       {journal} {Appl. Sci.}\ }\textbf {\bibinfo {volume} {7}},\ \bibinfo {pages}
       {720} (\bibinfo {year} {2017})}\BibitemShut {NoStop}%
     \bibitem [{\citenamefont {Kim}\ \emph {et~al.}(2008)\citenamefont {Kim},
       \citenamefont {Shvyd'ko},\ and\ \citenamefont {Reiche}}]{kim2008proposal}%
       \BibitemOpen
       \bibfield  {author} {\bibinfo {author} {\bibfnamefont {K.~J.}\ \bibnamefont
       {Kim}}, \bibinfo {author} {\bibfnamefont {Y.}~\bibnamefont {Shvyd'ko}}, \
       and\ \bibinfo {author} {\bibfnamefont {S.}~\bibnamefont {Reiche}},\ }\href
       {\doibase 10.1103/PhysRevLett.100.244802} {\bibfield  {journal} {\bibinfo
       {journal} {Phys. Rev. Lett.}\ }\textbf {\bibinfo {volume} {100}},\ \bibinfo
       {pages} {244802} (\bibinfo {year} {2008})}\BibitemShut {NoStop}%
     \bibitem [{\citenamefont {Lindberg}\ \emph {et~al.}(2011)\citenamefont
       {Lindberg}, \citenamefont {Kim}, \citenamefont {Shvyd'Ko},\ and\
       \citenamefont {Fawley}}]{Lindberg.2011}%
       \BibitemOpen
       \bibfield  {author} {\bibinfo {author} {\bibfnamefont {R.~R.}\ \bibnamefont
       {Lindberg}}, \bibinfo {author} {\bibfnamefont {K.~J.}\ \bibnamefont {Kim}},
       \bibinfo {author} {\bibfnamefont {Y.}~\bibnamefont {Shvyd'Ko}}, \ and\
       \bibinfo {author} {\bibfnamefont {W.~M.}\ \bibnamefont {Fawley}},\ }\href
       {\doibase 10.1103/PhysRevSTAB.14.010701} {\bibfield  {journal} {\bibinfo
       {journal} {Phys. Rev. Accel. Beams}\ }\textbf {\bibinfo {volume} {14}},\
       \bibinfo {pages} {010701} (\bibinfo {year} {2011})}\BibitemShut {NoStop}%
     \bibitem [{\citenamefont {Dai}\ \emph {et~al.}(2012)\citenamefont {Dai},
       \citenamefont {Deng},\ and\ \citenamefont {Dai}}]{Dai.2012}%
       \BibitemOpen
       \bibfield  {author} {\bibinfo {author} {\bibfnamefont {J.}~\bibnamefont
       {Dai}}, \bibinfo {author} {\bibfnamefont {H.}~\bibnamefont {Deng}}, \ and\
       \bibinfo {author} {\bibfnamefont {Z.}~\bibnamefont {Dai}},\ }\href {\doibase
       10.1103/PhysRevLett.108.034802} {\bibfield  {journal} {\bibinfo  {journal}
       {Phys. Rev. Lett.}\ }\textbf {\bibinfo {volume} {108}},\ \bibinfo {pages}
       {34802} (\bibinfo {year} {2012})}\BibitemShut {NoStop}%
     \bibitem [{\citenamefont {Authier}(2010)}]{AUTHIER.2003b}%
       \BibitemOpen
       \bibfield  {author} {\bibinfo {author} {\bibfnamefont {A.}~\bibnamefont
       {Authier}},\ }\href {\doibase 10.1093/acprof:oso/9780198528920.001.0001}
       {\emph {\bibinfo {title} {Dyn. Theory X-Ray Diffr.}}}\ (\bibinfo  {publisher}
       {Oxford University Press},\ \bibinfo {year} {2010})\BibitemShut {NoStop}%
     \bibitem [{\citenamefont {Shvyd'ko}\ \emph {et~al.}(2011)\citenamefont
       {Shvyd'ko}, \citenamefont {Stoupin}, \citenamefont {Blank},\ and\
       \citenamefont {Terentyev}}]{Shvydko.2011}%
       \BibitemOpen
       \bibfield  {author} {\bibinfo {author} {\bibfnamefont {Y.}~\bibnamefont
       {Shvyd'ko}}, \bibinfo {author} {\bibfnamefont {S.}~\bibnamefont {Stoupin}},
       \bibinfo {author} {\bibfnamefont {V.}~\bibnamefont {Blank}}, \ and\ \bibinfo
       {author} {\bibfnamefont {S.}~\bibnamefont {Terentyev}},\ }\href {\doibase
       10.1038/nphoton.2011.197} {\bibfield  {journal} {\bibinfo  {journal} {Nat.
       Photonics}\ }\textbf {\bibinfo {volume} {5}},\ \bibinfo {pages} {539}
       (\bibinfo {year} {2011})}\BibitemShut {NoStop}%
     \bibitem [{\citenamefont {Rousse}\ \emph {et~al.}(2001)\citenamefont {Rousse},
       \citenamefont {Rischel},\ and\ \citenamefont {Gauthier}}]{Rousse.2001}%
       \BibitemOpen
       \bibfield  {author} {\bibinfo {author} {\bibfnamefont {A.}~\bibnamefont
       {Rousse}}, \bibinfo {author} {\bibfnamefont {C.}~\bibnamefont {Rischel}}, \
       and\ \bibinfo {author} {\bibfnamefont {J.~C.}\ \bibnamefont {Gauthier}},\
       }\href {\doibase 10.1103/RevModPhys.73.17} {\bibfield  {journal} {\bibinfo
       {journal} {Rev. Mod. Phys.}\ }\textbf {\bibinfo {volume} {73}},\ \bibinfo
       {pages} {17} (\bibinfo {year} {2001})}\BibitemShut {NoStop}%
     \bibitem [{\citenamefont {Bargheer}\ \emph {et~al.}(2006)\citenamefont
       {Bargheer}, \citenamefont {Zhavoronkov}, \citenamefont {Woerner},\ and\
       \citenamefont {Elsaesser}}]{Bargheer.2006}%
       \BibitemOpen
       \bibfield  {author} {\bibinfo {author} {\bibfnamefont {M.}~\bibnamefont
       {Bargheer}}, \bibinfo {author} {\bibfnamefont {N.}~\bibnamefont
       {Zhavoronkov}}, \bibinfo {author} {\bibfnamefont {M.}~\bibnamefont
       {Woerner}}, \ and\ \bibinfo {author} {\bibfnamefont {T.}~\bibnamefont
       {Elsaesser}},\ }\href {\doibase 10.1002/cphc.200500591} {\bibfield  {journal}
       {\bibinfo  {journal} {ChemPhysChem}\ }\textbf {\bibinfo {volume} {7}},\
       \bibinfo {pages} {783} (\bibinfo {year} {2006})}\BibitemShut {NoStop}%
     \bibitem [{\citenamefont {Kojima}\ \emph {et~al.}(1994)\citenamefont {Kojima},
       \citenamefont {Llu}, \citenamefont {Kudo}, \citenamefont {Kawado},
       \citenamefont {Ishikawa},\ and\ \citenamefont {Matsushita}}]{Kojima.1994}%
       \BibitemOpen
       \bibfield  {author} {\bibinfo {author} {\bibfnamefont {S.}~\bibnamefont
       {Kojima}}, \bibinfo {author} {\bibfnamefont {K.~Y.}\ \bibnamefont {Llu}},
       \bibinfo {author} {\bibfnamefont {Y.}~\bibnamefont {Kudo}}, \bibinfo {author}
       {\bibfnamefont {S.}~\bibnamefont {Kawado}}, \bibinfo {author} {\bibfnamefont
       {T.}~\bibnamefont {Ishikawa}}, \ and\ \bibinfo {author} {\bibfnamefont
       {T.}~\bibnamefont {Matsushita}},\ }\href {\doibase 10.1143/JJAP.33.5612}
       {\bibfield  {journal} {\bibinfo  {journal} {Jpn. J. Appl. Phys.}\ }\textbf
       {\bibinfo {volume} {33}},\ \bibinfo {pages} {561} (\bibinfo {year}
       {1994})}\BibitemShut {NoStop}%
     \bibitem [{\citenamefont {Perrin}(2006)}]{Perrin.2006}%
       \BibitemOpen
       \bibfield  {author} {\bibinfo {author} {\bibfnamefont {B.}~\bibnamefont
       {Perrin}},\ }in\ \href {\doibase 10.1007/11767862_13} {\emph {\bibinfo
       {booktitle} {Top. Appl. Phys.}}},\ \bibinfo {series} {Topics in Applied
       Physics}, Vol.\ \bibinfo {volume} {107},\ \bibinfo {editor} {edited by\
       \bibinfo {editor} {\bibfnamefont {S.}~\bibnamefont {Volz}}}\ (\bibinfo
       {publisher} {Springer Berlin Heidelberg},\ \bibinfo {year} {2006})\ pp.\
       \bibinfo {pages} {333--359}\BibitemShut {NoStop}%
     \bibitem [{\citenamefont {{Van Aerenbergh}}\ \emph {et~al.}(2010)\citenamefont
       {{Van Aerenbergh}}, \citenamefont {Detlefs}, \citenamefont {H{\"{a}}rtwig},
       \citenamefont {Lafford}, \citenamefont {Masiello}, \citenamefont {Roth},
       \citenamefont {Schmid}, \citenamefont {Wattecamps},\ and\ \citenamefont
       {Zhang}}]{vanaerenbergh.2010}%
       \BibitemOpen
       \bibfield  {author} {\bibinfo {author} {\bibfnamefont {P.}~\bibnamefont {{Van
       Aerenbergh}}}, \bibinfo {author} {\bibfnamefont {C.}~\bibnamefont {Detlefs}},
       \bibinfo {author} {\bibfnamefont {J.}~\bibnamefont {H{\"{a}}rtwig}}, \bibinfo
       {author} {\bibfnamefont {T.~A.}\ \bibnamefont {Lafford}}, \bibinfo {author}
       {\bibfnamefont {F.}~\bibnamefont {Masiello}}, \bibinfo {author}
       {\bibfnamefont {T.}~\bibnamefont {Roth}}, \bibinfo {author} {\bibfnamefont
       {W.}~\bibnamefont {Schmid}}, \bibinfo {author} {\bibfnamefont
       {P.}~\bibnamefont {Wattecamps}}, \ and\ \bibinfo {author} {\bibfnamefont
       {L.}~\bibnamefont {Zhang}},\ }in\ \href {\doibase 10.1063/1.3463179} {\emph
       {\bibinfo {booktitle} {AIP Conf. Proc.}}},\ \bibinfo {series} {AIP Conference
       Proceedings}, Vol.\ \bibinfo {volume} {1234}\ (\bibinfo  {publisher} {AIP},\
       \bibinfo {year} {2010})\ pp.\ \bibinfo {pages} {229--232}\BibitemShut
       {NoStop}%
     \bibitem [{\citenamefont {Stoupin}\ \emph {et~al.}(2014)\citenamefont
       {Stoupin}, \citenamefont {Terentyev}, \citenamefont {Blank}, \citenamefont
       {Shvyd'Ko}, \citenamefont {Goetze}, \citenamefont {Assoufid}, \citenamefont
       {Polyakov}, \citenamefont {Kuznetsov}, \citenamefont {Kornilov},
       \citenamefont {Katsoudas}, \citenamefont {Alonso-Mori}, \citenamefont
       {Chollet}, \citenamefont {Feng}, \citenamefont {Glownia}, \citenamefont
       {Lemke}, \citenamefont {Robert}, \citenamefont {Sikorski}, \citenamefont
       {Song},\ and\ \citenamefont {Zhu}}]{Stoupin.2014}%
       \BibitemOpen
       \bibfield  {author} {\bibinfo {author} {\bibfnamefont {S.}~\bibnamefont
       {Stoupin}}, \bibinfo {author} {\bibfnamefont {S.~A.}\ \bibnamefont
       {Terentyev}}, \bibinfo {author} {\bibfnamefont {V.~D.}\ \bibnamefont
       {Blank}}, \bibinfo {author} {\bibfnamefont {Y.~V.}\ \bibnamefont {Shvyd'Ko}},
       \bibinfo {author} {\bibfnamefont {K.}~\bibnamefont {Goetze}}, \bibinfo
       {author} {\bibfnamefont {L.}~\bibnamefont {Assoufid}}, \bibinfo {author}
       {\bibfnamefont {S.~N.}\ \bibnamefont {Polyakov}}, \bibinfo {author}
       {\bibfnamefont {M.~S.}\ \bibnamefont {Kuznetsov}}, \bibinfo {author}
       {\bibfnamefont {N.~V.}\ \bibnamefont {Kornilov}}, \bibinfo {author}
       {\bibfnamefont {J.}~\bibnamefont {Katsoudas}}, \bibinfo {author}
       {\bibfnamefont {R.}~\bibnamefont {Alonso-Mori}}, \bibinfo {author}
       {\bibfnamefont {M.}~\bibnamefont {Chollet}}, \bibinfo {author} {\bibfnamefont
       {Y.}~\bibnamefont {Feng}}, \bibinfo {author} {\bibfnamefont {J.~M.}\
       \bibnamefont {Glownia}}, \bibinfo {author} {\bibfnamefont {H.}~\bibnamefont
       {Lemke}}, \bibinfo {author} {\bibfnamefont {A.}~\bibnamefont {Robert}},
       \bibinfo {author} {\bibfnamefont {M.}~\bibnamefont {Sikorski}}, \bibinfo
       {author} {\bibfnamefont {S.}~\bibnamefont {Song}}, \ and\ \bibinfo {author}
       {\bibfnamefont {D.}~\bibnamefont {Zhu}},\ }\href {\doibase
       10.1107/S1600576714013028} {\bibfield  {journal} {\bibinfo  {journal} {J.
       Appl. Crystallogr.}\ }\textbf {\bibinfo {volume} {47}},\ \bibinfo {pages}
       {1329} (\bibinfo {year} {2014})}\BibitemShut {NoStop}%
     \bibitem [{\citenamefont {Koz{\'{a}}k}\ \emph {et~al.}(2015)\citenamefont
       {Koz{\'{a}}k}, \citenamefont {Troj{\'{a}}nek},\ and\ \citenamefont
       {Mal{\'{y}}}}]{Kozak.2015}%
       \BibitemOpen
       \bibfield  {author} {\bibinfo {author} {\bibfnamefont {M.}~\bibnamefont
       {Koz{\'{a}}k}}, \bibinfo {author} {\bibfnamefont {F.}~\bibnamefont
       {Troj{\'{a}}nek}}, \ and\ \bibinfo {author} {\bibfnamefont {P.}~\bibnamefont
       {Mal{\'{y}}}},\ }\href {\doibase 10.1088/1367-2630/17/5/053027} {\bibfield
       {journal} {\bibinfo  {journal} {New J. Phys.}\ }\textbf {\bibinfo {volume}
       {17}},\ \bibinfo {pages} {53027} (\bibinfo {year} {2015})}\BibitemShut
       {NoStop}%
     \bibitem [{\citenamefont {Kolodziej}\ \emph {et~al.}(2018)\citenamefont
       {Kolodziej}, \citenamefont {Shvyd'ko}, \citenamefont {Shu}, \citenamefont
       {Kearney}, \citenamefont {Stoupin}, \citenamefont {Liu}, \citenamefont {Gog},
       \citenamefont {Walko}, \citenamefont {Wang}, \citenamefont {Said},
       \citenamefont {Roberts}, \citenamefont {Goetze}, \citenamefont {Baldini},
       \citenamefont {Yang}, \citenamefont {Fister}, \citenamefont {Blank},
       \citenamefont {Terentyev},\ and\ \citenamefont {Kim}}]{Kolodziej2018}%
       \BibitemOpen
       \bibfield  {author} {\bibinfo {author} {\bibfnamefont {T.}~\bibnamefont
       {Kolodziej}}, \bibinfo {author} {\bibfnamefont {Y.}~\bibnamefont {Shvyd'ko}},
       \bibinfo {author} {\bibfnamefont {D.}~\bibnamefont {Shu}}, \bibinfo {author}
       {\bibfnamefont {S.}~\bibnamefont {Kearney}}, \bibinfo {author} {\bibfnamefont
       {S.}~\bibnamefont {Stoupin}}, \bibinfo {author} {\bibfnamefont
       {W.}~\bibnamefont {Liu}}, \bibinfo {author} {\bibfnamefont {T.}~\bibnamefont
       {Gog}}, \bibinfo {author} {\bibfnamefont {D.~A.}\ \bibnamefont {Walko}},
       \bibinfo {author} {\bibfnamefont {J.}~\bibnamefont {Wang}}, \bibinfo {author}
       {\bibfnamefont {A.}~\bibnamefont {Said}}, \bibinfo {author} {\bibfnamefont
       {T.}~\bibnamefont {Roberts}}, \bibinfo {author} {\bibfnamefont
       {K.}~\bibnamefont {Goetze}}, \bibinfo {author} {\bibfnamefont
       {M.}~\bibnamefont {Baldini}}, \bibinfo {author} {\bibfnamefont
       {W.}~\bibnamefont {Yang}}, \bibinfo {author} {\bibfnamefont {T.}~\bibnamefont
       {Fister}}, \bibinfo {author} {\bibfnamefont {V.}~\bibnamefont {Blank}},
       \bibinfo {author} {\bibfnamefont {S.}~\bibnamefont {Terentyev}}, \ and\
       \bibinfo {author} {\bibfnamefont {K.~J.}\ \bibnamefont {Kim}},\ }\href
       {\doibase 10.1107/S1600577518007695} {\bibfield  {journal} {\bibinfo
       {journal} {J. Synchrotron Radiat.}\ }\textbf {\bibinfo {volume} {25}},\
       \bibinfo {pages} {1022} (\bibinfo {year} {2018})}\BibitemShut {NoStop}%
     \bibitem [{\citenamefont {Samoylova}\ \emph {et~al.}(2019)\citenamefont
       {Samoylova}, \citenamefont {Shu}, \citenamefont {Dong}, \citenamefont
       {Geloni}, \citenamefont {Karabekyan}, \citenamefont {Terentev}, \citenamefont
       {Blank}, \citenamefont {Liu}, \citenamefont {Wohlenberg}, \citenamefont
       {Decking},\ and\ \citenamefont {Sinn}}]{Samoylova.2019D}%
       \BibitemOpen
       \bibfield  {author} {\bibinfo {author} {\bibfnamefont {L.}~\bibnamefont
       {Samoylova}}, \bibinfo {author} {\bibfnamefont {D.}~\bibnamefont {Shu}},
       \bibinfo {author} {\bibfnamefont {X.}~\bibnamefont {Dong}}, \bibinfo {author}
       {\bibfnamefont {G.}~\bibnamefont {Geloni}}, \bibinfo {author} {\bibfnamefont
       {S.}~\bibnamefont {Karabekyan}}, \bibinfo {author} {\bibfnamefont
       {S.}~\bibnamefont {Terentev}}, \bibinfo {author} {\bibfnamefont
       {V.}~\bibnamefont {Blank}}, \bibinfo {author} {\bibfnamefont
       {S.}~\bibnamefont {Liu}}, \bibinfo {author} {\bibfnamefont {T.}~\bibnamefont
       {Wohlenberg}}, \bibinfo {author} {\bibfnamefont {W.}~\bibnamefont {Decking}},
       \ and\ \bibinfo {author} {\bibfnamefont {H.}~\bibnamefont {Sinn}},\ }\href
       {\doibase 10.1063/1.5084579} {\bibfield  {journal} {\bibinfo  {journal} {AIP
       Conference Proceedings}\ }\textbf {\bibinfo {volume} {2054}},\ \bibinfo
       {pages} {030016} (\bibinfo {year} {2019})},\ \Eprint
       {http://arxiv.org/abs/https://aip.scitation.org/doi/pdf/10.1063/1.5084579}
       {https://aip.scitation.org/doi/pdf/10.1063/1.5084579} \BibitemShut {NoStop}%
     \bibitem [{\citenamefont {Song}\ \emph {et~al.}(2016)\citenamefont {Song},
       \citenamefont {Zhang}, \citenamefont {Guo}, \citenamefont {Li},\ and\
       \citenamefont {Deng}}]{Song.2016}%
       \BibitemOpen
       \bibfield  {author} {\bibinfo {author} {\bibfnamefont {M.~Q.}\ \bibnamefont
       {Song}}, \bibinfo {author} {\bibfnamefont {Q.~M.}\ \bibnamefont {Zhang}},
       \bibinfo {author} {\bibfnamefont {Y.~H.}\ \bibnamefont {Guo}}, \bibinfo
       {author} {\bibfnamefont {K.}~\bibnamefont {Li}}, \ and\ \bibinfo {author}
       {\bibfnamefont {H.~X.}\ \bibnamefont {Deng}},\ }\href {\doibase
       10.1088/1674-1137/40/4/048101} {\bibfield  {journal} {\bibinfo  {journal}
       {Chinese Phys. C}\ }\textbf {\bibinfo {volume} {40}},\ \bibinfo {pages}
       {48101} (\bibinfo {year} {2016})}\BibitemShut {NoStop}%
     \bibitem [{\citenamefont {Yang}\ \emph {et~al.}(2018)\citenamefont {Yang},
       \citenamefont {Wang},\ and\ \citenamefont {Wu}}]{Yang.2018}%
       \BibitemOpen
       \bibfield  {author} {\bibinfo {author} {\bibfnamefont {B.}~\bibnamefont
       {Yang}}, \bibinfo {author} {\bibfnamefont {S.}~\bibnamefont {Wang}}, \ and\
       \bibinfo {author} {\bibfnamefont {J.}~\bibnamefont {Wu}},\ }\href {\doibase
       10.1107/S1600577517015466} {\bibfield  {journal} {\bibinfo  {journal} {J.
       Synchrotron Radiat.}\ }\textbf {\bibinfo {volume} {25}},\ \bibinfo {pages}
       {166} (\bibinfo {year} {2018})}\BibitemShut {NoStop}%
     \bibitem [{\citenamefont {Liu}\ \emph {et~al.}(2019)\citenamefont {Liu},
       \citenamefont {Decking}, \citenamefont {Kocharyan}, \citenamefont {Saldin},
       \citenamefont {Serkez}, \citenamefont {Shayduk}, \citenamefont {Sinn},\ and\
       \citenamefont {Geloni}}]{Liu.2019}%
       \BibitemOpen
       \bibfield  {author} {\bibinfo {author} {\bibfnamefont {S.}~\bibnamefont
       {Liu}}, \bibinfo {author} {\bibfnamefont {W.}~\bibnamefont {Decking}},
       \bibinfo {author} {\bibfnamefont {V.}~\bibnamefont {Kocharyan}}, \bibinfo
       {author} {\bibfnamefont {E.}~\bibnamefont {Saldin}}, \bibinfo {author}
       {\bibfnamefont {S.}~\bibnamefont {Serkez}}, \bibinfo {author} {\bibfnamefont
       {R.}~\bibnamefont {Shayduk}}, \bibinfo {author} {\bibfnamefont
       {H.}~\bibnamefont {Sinn}}, \ and\ \bibinfo {author} {\bibfnamefont
       {G.}~\bibnamefont {Geloni}},\ }\href {\doibase
       10.1103/PhysRevAccelBeams.22.060704} {\bibfield  {journal} {\bibinfo
       {journal} {Phys. Rev. Accel. Beams}\ }\textbf {\bibinfo {volume} {22}},\
       \bibinfo {pages} {60704} (\bibinfo {year} {2019})}\BibitemShut {NoStop}%
     \bibitem [{\citenamefont {Seltzer}(1996)}]{Seltzer.1995}%
       \BibitemOpen
       \bibfield  {author} {\bibinfo {author} {\bibfnamefont {S.}~\bibnamefont
       {Seltzer}},\ }\href {\doibase 10.18434/t4d01f} {\enquote {\bibinfo {title}
       {{Tables of X-Ray Mass Attenuation Coefficients and Mass Energy-Absorption
       Coefficients, NIST Standard Reference Database 126}},}\ } (\bibinfo {year}
       {1996})\BibitemShut {NoStop}%
     \bibitem [{\citenamefont {Henke}\ \emph {et~al.}(1993)\citenamefont {Henke},
       \citenamefont {Gullikson},\ and\ \citenamefont {Davis}}]{Henke.1993}%
       \BibitemOpen
       \bibfield  {author} {\bibinfo {author} {\bibfnamefont {B.~L.}\ \bibnamefont
       {Henke}}, \bibinfo {author} {\bibfnamefont {E.~M.}\ \bibnamefont
       {Gullikson}}, \ and\ \bibinfo {author} {\bibfnamefont {J.~C.}\ \bibnamefont
       {Davis}},\ }\href {\doibase 10.1006/adnd.1993.1013} {\bibfield  {journal}
       {\bibinfo  {journal} {At. Data Nucl. Data Tables}\ }\textbf {\bibinfo
       {volume} {54}},\ \bibinfo {pages} {181} (\bibinfo {year} {1993})}\BibitemShut
       {NoStop}%
     \bibitem [{\citenamefont {Lee}(2017)}]{ANSYS17}%
       \BibitemOpen
       \bibfield  {author} {\bibinfo {author} {\bibfnamefont {H.-H.}\ \bibnamefont
       {Lee}},\ }\href@noop {} {\emph {\bibinfo {title} {Finite Element Simulations
       with ANSYS Workbench 17}}}\ (\bibinfo  {publisher} {SDC Publications},\
       \bibinfo {year} {2017})\BibitemShut {NoStop}%
     \bibitem [{\citenamefont {Hubbell}\ and\ \citenamefont
       {Overbo}(1979)}]{Hubbell.1979}%
       \BibitemOpen
       \bibfield  {author} {\bibinfo {author} {\bibfnamefont {J.~H.}\ \bibnamefont
       {Hubbell}}\ and\ \bibinfo {author} {\bibfnamefont {I.}~\bibnamefont
       {Overbo}},\ }\href {\doibase 10.1063/1.555593} {\bibfield  {journal}
       {\bibinfo  {journal} {J. Phys. Chem. Ref. Data}\ }\textbf {\bibinfo {volume}
       {8}},\ \bibinfo {pages} {69} (\bibinfo {year} {1979})}\BibitemShut {NoStop}%
     \bibitem [{\citenamefont {Stoupin}\ and\ \citenamefont
       {Shvyd'Ko}(2011)}]{Stoupin.2011}%
       \BibitemOpen
       \bibfield  {author} {\bibinfo {author} {\bibfnamefont {S.}~\bibnamefont
       {Stoupin}}\ and\ \bibinfo {author} {\bibfnamefont {Y.~V.}\ \bibnamefont
       {Shvyd'Ko}},\ }\href {\doibase 10.1103/PhysRevB.83.104102} {\bibfield
       {journal} {\bibinfo  {journal} {Phys. Rev. B - Condens. Matter Mater. Phys.}\
       }\textbf {\bibinfo {volume} {83}},\ \bibinfo {pages} {1} (\bibinfo {year}
       {2011})}\BibitemShut {NoStop}%
     \bibitem [{\citenamefont {Honkanen}\ \emph {et~al.}(2018)\citenamefont
       {Honkanen}, \citenamefont {Ferrero}, \citenamefont {Guigay},\ and\
       \citenamefont {Mocella}}]{Honkanen.2018}%
       \BibitemOpen
       \bibfield  {author} {\bibinfo {author} {\bibfnamefont {A.~P.}\ \bibnamefont
       {Honkanen}}, \bibinfo {author} {\bibfnamefont {C.}~\bibnamefont {Ferrero}},
       \bibinfo {author} {\bibfnamefont {J.~P.}\ \bibnamefont {Guigay}}, \ and\
       \bibinfo {author} {\bibfnamefont {V.}~\bibnamefont {Mocella}},\ }\href
       {\doibase 10.1107/S1600576718001930} {\bibfield  {journal} {\bibinfo
       {journal} {J. Appl. Crystallogr.}\ }\textbf {\bibinfo {volume} {51}},\
       \bibinfo {pages} {514} (\bibinfo {year} {2018})}\BibitemShut {NoStop}%
     \bibitem [{\citenamefont {Inyushkin}\ \emph {et~al.}(2018)\citenamefont
       {Inyushkin}, \citenamefont {Taldenkov}, \citenamefont {Ralchenko},
       \citenamefont {Bolshakov}, \citenamefont {Koliadin},\ and\ \citenamefont
       {Katrusha}}]{Inyushkin.2018}%
       \BibitemOpen
       \bibfield  {author} {\bibinfo {author} {\bibfnamefont {A.~V.}\ \bibnamefont
       {Inyushkin}}, \bibinfo {author} {\bibfnamefont {A.~N.}\ \bibnamefont
       {Taldenkov}}, \bibinfo {author} {\bibfnamefont {V.~G.}\ \bibnamefont
       {Ralchenko}}, \bibinfo {author} {\bibfnamefont {A.~P.}\ \bibnamefont
       {Bolshakov}}, \bibinfo {author} {\bibfnamefont {A.~V.}\ \bibnamefont
       {Koliadin}}, \ and\ \bibinfo {author} {\bibfnamefont {A.~N.}\ \bibnamefont
       {Katrusha}},\ }\href {\doibase 10.1103/PhysRevB.97.144305} {\bibfield
       {journal} {\bibinfo  {journal} {Phys. Rev. B}\ }\textbf {\bibinfo {volume}
       {97}},\ \bibinfo {pages} {144305} (\bibinfo {year} {2018})}\BibitemShut
       {NoStop}%
     \bibitem [{\citenamefont {Allison}\ \emph {et~al.}(2016)\citenamefont
       {Allison}, \citenamefont {Amako}, \citenamefont {Apostolakis}, \citenamefont
       {Arce}, \citenamefont {Asai}, \citenamefont {Aso}, \citenamefont {Bagli},
       \citenamefont {Bagulya}, \citenamefont {Banerjee}, \citenamefont {Barrand},
       \citenamefont {Beck}, \citenamefont {Bogdanov}, \citenamefont {Brandt},
       \citenamefont {Brown}, \citenamefont {Burkhardt}, \citenamefont {Canal},
       \citenamefont {Cano-Ott}, \citenamefont {Chauvie}, \citenamefont {Cho},
       \citenamefont {Cirrone}, \citenamefont {Cooperman}, \citenamefont
       {Cort{\'{e}}s-Giraldo}, \citenamefont {Cosmo}, \citenamefont {Cuttone},
       \citenamefont {Depaola}, \citenamefont {Desorgher}, \citenamefont {Dong},
       \citenamefont {Dotti}, \citenamefont {Elvira}, \citenamefont {Folger},
       \citenamefont {Francis}, \citenamefont {Galoyan}, \citenamefont {Garnier},
       \citenamefont {Gayer}, \citenamefont {Genser}, \citenamefont {Grichine},
       \citenamefont {Guatelli}, \citenamefont {Gu{\`{e}}ye}, \citenamefont
       {Gumplinger}, \citenamefont {Howard}, \citenamefont
       {Hřivn{\'{a}}{\v{c}}ov{\'{a}}}, \citenamefont {Hwang}, \citenamefont
       {Incerti}, \citenamefont {Ivanchenko}, \citenamefont {Ivanchenko},
       \citenamefont {Jones}, \citenamefont {Jun}, \citenamefont {Kaitaniemi},
       \citenamefont {Karakatsanis}, \citenamefont {Karamitrosi}, \citenamefont
       {Kelsey}, \citenamefont {Kimura}, \citenamefont {Koi}, \citenamefont
       {Kurashige}, \citenamefont {Lechner}, \citenamefont {Lee}, \citenamefont
       {Longo}, \citenamefont {Maire}, \citenamefont {Mancusi}, \citenamefont
       {Mantero}, \citenamefont {Mendoza}, \citenamefont {Morgan}, \citenamefont
       {Murakami}, \citenamefont {Nikitina}, \citenamefont {Pandola}, \citenamefont
       {Paprocki}, \citenamefont {Perl}, \citenamefont {Petrovi{\'{c}}},
       \citenamefont {Pia}, \citenamefont {Pokorski}, \citenamefont {Quesada},
       \citenamefont {Raine}, \citenamefont {Reis}, \citenamefont {Ribon},
       \citenamefont {{Risti{\'{c}} Fira}}, \citenamefont {Romano}, \citenamefont
       {Russo}, \citenamefont {Santin}, \citenamefont {Sasaki}, \citenamefont
       {Sawkey}, \citenamefont {Shin}, \citenamefont {Strakovsky}, \citenamefont
       {Taborda}, \citenamefont {Tanaka}, \citenamefont {Tom{\'{e}}}, \citenamefont
       {Toshito}, \citenamefont {Tran}, \citenamefont {Truscott}, \citenamefont
       {Urban}, \citenamefont {Uzhinsky}, \citenamefont {Verbeke}, \citenamefont
       {Verderi}, \citenamefont {Wendt}, \citenamefont {Wenzel}, \citenamefont
       {Wright}, \citenamefont {Wright}, \citenamefont {Yamashita}, \citenamefont
       {Yarba},\ and\ \citenamefont {Yoshida}}]{Allison.2016}%
       \BibitemOpen
       \bibfield  {author} {\bibinfo {author} {\bibfnamefont {J.}~\bibnamefont
       {Allison}}, \bibinfo {author} {\bibfnamefont {K.}~\bibnamefont {Amako}},
       \bibinfo {author} {\bibfnamefont {J.}~\bibnamefont {Apostolakis}}, \bibinfo
       {author} {\bibfnamefont {P.}~\bibnamefont {Arce}}, \bibinfo {author}
       {\bibfnamefont {M.}~\bibnamefont {Asai}}, \bibinfo {author} {\bibfnamefont
       {T.}~\bibnamefont {Aso}}, \bibinfo {author} {\bibfnamefont {E.}~\bibnamefont
       {Bagli}}, \bibinfo {author} {\bibfnamefont {A.}~\bibnamefont {Bagulya}},
       \bibinfo {author} {\bibfnamefont {S.}~\bibnamefont {Banerjee}}, \bibinfo
       {author} {\bibfnamefont {G.}~\bibnamefont {Barrand}}, \bibinfo {author}
       {\bibfnamefont {B.~R.}\ \bibnamefont {Beck}}, \bibinfo {author}
       {\bibfnamefont {A.~G.}\ \bibnamefont {Bogdanov}}, \bibinfo {author}
       {\bibfnamefont {D.}~\bibnamefont {Brandt}}, \bibinfo {author} {\bibfnamefont
       {J.~M.}\ \bibnamefont {Brown}}, \bibinfo {author} {\bibfnamefont
       {H.}~\bibnamefont {Burkhardt}}, \bibinfo {author} {\bibfnamefont
       {P.}~\bibnamefont {Canal}}, \bibinfo {author} {\bibfnamefont
       {D.}~\bibnamefont {Cano-Ott}}, \bibinfo {author} {\bibfnamefont
       {S.}~\bibnamefont {Chauvie}}, \bibinfo {author} {\bibfnamefont
       {K.}~\bibnamefont {Cho}}, \bibinfo {author} {\bibfnamefont {G.~A.}\
       \bibnamefont {Cirrone}}, \bibinfo {author} {\bibfnamefont {G.}~\bibnamefont
       {Cooperman}}, \bibinfo {author} {\bibfnamefont {M.~A.}\ \bibnamefont
       {Cort{\'{e}}s-Giraldo}}, \bibinfo {author} {\bibfnamefont {G.}~\bibnamefont
       {Cosmo}}, \bibinfo {author} {\bibfnamefont {G.}~\bibnamefont {Cuttone}},
       \bibinfo {author} {\bibfnamefont {G.}~\bibnamefont {Depaola}}, \bibinfo
       {author} {\bibfnamefont {L.}~\bibnamefont {Desorgher}}, \bibinfo {author}
       {\bibfnamefont {X.}~\bibnamefont {Dong}}, \bibinfo {author} {\bibfnamefont
       {A.}~\bibnamefont {Dotti}}, \bibinfo {author} {\bibfnamefont {V.~D.}\
       \bibnamefont {Elvira}}, \bibinfo {author} {\bibfnamefont {G.}~\bibnamefont
       {Folger}}, \bibinfo {author} {\bibfnamefont {Z.}~\bibnamefont {Francis}},
       \bibinfo {author} {\bibfnamefont {A.}~\bibnamefont {Galoyan}}, \bibinfo
       {author} {\bibfnamefont {L.}~\bibnamefont {Garnier}}, \bibinfo {author}
       {\bibfnamefont {M.}~\bibnamefont {Gayer}}, \bibinfo {author} {\bibfnamefont
       {K.~L.}\ \bibnamefont {Genser}}, \bibinfo {author} {\bibfnamefont {V.~M.}\
       \bibnamefont {Grichine}}, \bibinfo {author} {\bibfnamefont {S.}~\bibnamefont
       {Guatelli}}, \bibinfo {author} {\bibfnamefont {P.}~\bibnamefont
       {Gu{\`{e}}ye}}, \bibinfo {author} {\bibfnamefont {P.}~\bibnamefont
       {Gumplinger}}, \bibinfo {author} {\bibfnamefont {A.~S.}\ \bibnamefont
       {Howard}}, \bibinfo {author} {\bibfnamefont {I.}~\bibnamefont
       {Hřivn{\'{a}}{\v{c}}ov{\'{a}}}}, \bibinfo {author} {\bibfnamefont
       {S.}~\bibnamefont {Hwang}}, \bibinfo {author} {\bibfnamefont
       {S.}~\bibnamefont {Incerti}}, \bibinfo {author} {\bibfnamefont
       {A.}~\bibnamefont {Ivanchenko}}, \bibinfo {author} {\bibfnamefont {V.~N.}\
       \bibnamefont {Ivanchenko}}, \bibinfo {author} {\bibfnamefont {F.~W.}\
       \bibnamefont {Jones}}, \bibinfo {author} {\bibfnamefont {S.~Y.}\ \bibnamefont
       {Jun}}, \bibinfo {author} {\bibfnamefont {P.}~\bibnamefont {Kaitaniemi}},
       \bibinfo {author} {\bibfnamefont {N.}~\bibnamefont {Karakatsanis}}, \bibinfo
       {author} {\bibfnamefont {M.}~\bibnamefont {Karamitrosi}}, \bibinfo {author}
       {\bibfnamefont {M.}~\bibnamefont {Kelsey}}, \bibinfo {author} {\bibfnamefont
       {A.}~\bibnamefont {Kimura}}, \bibinfo {author} {\bibfnamefont
       {T.}~\bibnamefont {Koi}}, \bibinfo {author} {\bibfnamefont {H.}~\bibnamefont
       {Kurashige}}, \bibinfo {author} {\bibfnamefont {A.}~\bibnamefont {Lechner}},
       \bibinfo {author} {\bibfnamefont {S.~B.}\ \bibnamefont {Lee}}, \bibinfo
       {author} {\bibfnamefont {F.}~\bibnamefont {Longo}}, \bibinfo {author}
       {\bibfnamefont {M.}~\bibnamefont {Maire}}, \bibinfo {author} {\bibfnamefont
       {D.}~\bibnamefont {Mancusi}}, \bibinfo {author} {\bibfnamefont
       {A.}~\bibnamefont {Mantero}}, \bibinfo {author} {\bibfnamefont
       {E.}~\bibnamefont {Mendoza}}, \bibinfo {author} {\bibfnamefont
       {B.}~\bibnamefont {Morgan}}, \bibinfo {author} {\bibfnamefont
       {K.}~\bibnamefont {Murakami}}, \bibinfo {author} {\bibfnamefont
       {T.}~\bibnamefont {Nikitina}}, \bibinfo {author} {\bibfnamefont
       {L.}~\bibnamefont {Pandola}}, \bibinfo {author} {\bibfnamefont
       {P.}~\bibnamefont {Paprocki}}, \bibinfo {author} {\bibfnamefont
       {J.}~\bibnamefont {Perl}}, \bibinfo {author} {\bibfnamefont {I.}~\bibnamefont
       {Petrovi{\'{c}}}}, \bibinfo {author} {\bibfnamefont {M.~G.}\ \bibnamefont
       {Pia}}, \bibinfo {author} {\bibfnamefont {W.}~\bibnamefont {Pokorski}},
       \bibinfo {author} {\bibfnamefont {J.~M.}\ \bibnamefont {Quesada}}, \bibinfo
       {author} {\bibfnamefont {M.}~\bibnamefont {Raine}}, \bibinfo {author}
       {\bibfnamefont {M.~A.}\ \bibnamefont {Reis}}, \bibinfo {author}
       {\bibfnamefont {A.}~\bibnamefont {Ribon}}, \bibinfo {author} {\bibfnamefont
       {A.}~\bibnamefont {{Risti{\'{c}} Fira}}}, \bibinfo {author} {\bibfnamefont
       {F.}~\bibnamefont {Romano}}, \bibinfo {author} {\bibfnamefont
       {G.}~\bibnamefont {Russo}}, \bibinfo {author} {\bibfnamefont
       {G.}~\bibnamefont {Santin}}, \bibinfo {author} {\bibfnamefont
       {T.}~\bibnamefont {Sasaki}}, \bibinfo {author} {\bibfnamefont
       {D.}~\bibnamefont {Sawkey}}, \bibinfo {author} {\bibfnamefont {J.~I.}\
       \bibnamefont {Shin}}, \bibinfo {author} {\bibfnamefont {I.~I.}\ \bibnamefont
       {Strakovsky}}, \bibinfo {author} {\bibfnamefont {A.}~\bibnamefont {Taborda}},
       \bibinfo {author} {\bibfnamefont {S.}~\bibnamefont {Tanaka}}, \bibinfo
       {author} {\bibfnamefont {B.}~\bibnamefont {Tom{\'{e}}}}, \bibinfo {author}
       {\bibfnamefont {T.}~\bibnamefont {Toshito}}, \bibinfo {author} {\bibfnamefont
       {H.~N.}\ \bibnamefont {Tran}}, \bibinfo {author} {\bibfnamefont {P.~R.}\
       \bibnamefont {Truscott}}, \bibinfo {author} {\bibfnamefont {L.}~\bibnamefont
       {Urban}}, \bibinfo {author} {\bibfnamefont {V.}~\bibnamefont {Uzhinsky}},
       \bibinfo {author} {\bibfnamefont {J.~M.}\ \bibnamefont {Verbeke}}, \bibinfo
       {author} {\bibfnamefont {M.}~\bibnamefont {Verderi}}, \bibinfo {author}
       {\bibfnamefont {B.~L.}\ \bibnamefont {Wendt}}, \bibinfo {author}
       {\bibfnamefont {H.}~\bibnamefont {Wenzel}}, \bibinfo {author} {\bibfnamefont
       {D.~H.}\ \bibnamefont {Wright}}, \bibinfo {author} {\bibfnamefont {D.~M.}\
       \bibnamefont {Wright}}, \bibinfo {author} {\bibfnamefont {T.}~\bibnamefont
       {Yamashita}}, \bibinfo {author} {\bibfnamefont {J.}~\bibnamefont {Yarba}}, \
       and\ \bibinfo {author} {\bibfnamefont {H.}~\bibnamefont {Yoshida}},\ }\href
       {\doibase 10.1016/j.nima.2016.06.125} {\bibfield  {journal} {\bibinfo
       {journal} {Nucl. Instruments Methods Phys. Res. Sect. A Accel. Spectrometers,
       Detect. Assoc. Equip.}\ }\textbf {\bibinfo {volume} {835}},\ \bibinfo {pages}
       {186} (\bibinfo {year} {2016})}\BibitemShut {NoStop}%
     \bibitem [{\citenamefont {Cirrone}\ \emph {et~al.}(2010)\citenamefont
       {Cirrone}, \citenamefont {Cuttone}, \citenamefont {{Di Rosa}}, \citenamefont
       {Pandola}, \citenamefont {Romano},\ and\ \citenamefont
       {Zhang}}]{Cirrone.2010}%
       \BibitemOpen
       \bibfield  {author} {\bibinfo {author} {\bibfnamefont {G.~A.}\ \bibnamefont
       {Cirrone}}, \bibinfo {author} {\bibfnamefont {G.}~\bibnamefont {Cuttone}},
       \bibinfo {author} {\bibfnamefont {F.}~\bibnamefont {{Di Rosa}}}, \bibinfo
       {author} {\bibfnamefont {L.}~\bibnamefont {Pandola}}, \bibinfo {author}
       {\bibfnamefont {F.}~\bibnamefont {Romano}}, \ and\ \bibinfo {author}
       {\bibfnamefont {Q.}~\bibnamefont {Zhang}},\ }\href {\doibase
       10.1016/j.nima.2010.02.112} {\bibfield  {journal} {\bibinfo  {journal} {Nucl.
       Instruments Methods Phys. Res. Sect. A Accel. Spectrometers, Detect. Assoc.
       Equip.}\ }\textbf {\bibinfo {volume} {618}},\ \bibinfo {pages} {315}
       (\bibinfo {year} {2010})}\BibitemShut {NoStop}%
     \bibitem [{\citenamefont {Shvyd'Ko}\ and\ \citenamefont
       {Lindberg}(2012)}]{Shvydko.2012}%
       \BibitemOpen
       \bibfield  {author} {\bibinfo {author} {\bibfnamefont {Y.}~\bibnamefont
       {Shvyd'Ko}}\ and\ \bibinfo {author} {\bibfnamefont {R.}~\bibnamefont
       {Lindberg}},\ }\href {\doibase 10.1103/PhysRevSTAB.15.100702} {\bibfield
       {journal} {\bibinfo  {journal} {Phys. Rev. Accel. Beams}\ }\textbf {\bibinfo
       {volume} {15}},\ \bibinfo {pages} {100702} (\bibinfo {year}
       {2012})}\BibitemShut {NoStop}%
     \bibitem [{\citenamefont {Zhao}\ \emph {et~al.}(2018)\citenamefont {Zhao},
       \citenamefont {Wang}, \citenamefont {Yang},\ and\ \citenamefont
       {Yin}}]{Zhao:2018lcl}%
       \BibitemOpen
       \bibfield  {author} {\bibinfo {author} {\bibfnamefont {Z.}~\bibnamefont
       {Zhao}}, \bibinfo {author} {\bibfnamefont {D.}~\bibnamefont {Wang}}, \bibinfo
       {author} {\bibfnamefont {Z.-H.}\ \bibnamefont {Yang}}, \ and\ \bibinfo
       {author} {\bibfnamefont {L.}~\bibnamefont {Yin}},\ }in\ \href {\doibase
       10.18429/JACoW-FEL2017-MOP055} {\emph {\bibinfo {booktitle} {{Proceedings,
       38th International Free Electron Laser Conference, FEL2017}}}}\ (\bibinfo
       {year} {2018})\ p.\ \bibinfo {pages} {MOP055}\BibitemShut {NoStop}%
     \bibitem [{\citenamefont {Yan}\ and\ \citenamefont {Deng}(2019)}]{Yan.2019M}%
       \BibitemOpen
       \bibfield  {author} {\bibinfo {author} {\bibfnamefont {J.}~\bibnamefont
       {Yan}}\ and\ \bibinfo {author} {\bibfnamefont {H.}~\bibnamefont {Deng}},\
       }\href {\doibase 10.1103/PhysRevAccelBeams.22.090701} {\bibfield  {journal}
       {\bibinfo  {journal} {Phys. Rev. Accel. Beams}\ }\textbf {\bibinfo {volume}
       {22}},\ \bibinfo {pages} {090701} (\bibinfo {year} {2019})}\BibitemShut
       {NoStop}%
     \bibitem [{\citenamefont {Li}\ and\ \citenamefont {Deng}(2018)}]{Li.2018}%
       \BibitemOpen
       \bibfield  {author} {\bibinfo {author} {\bibfnamefont {K.}~\bibnamefont
       {Li}}\ and\ \bibinfo {author} {\bibfnamefont {H.}~\bibnamefont {Deng}},\
       }\href {\doibase 10.1063/1.5037180} {\bibfield  {journal} {\bibinfo
       {journal} {Appl. Phys. Lett.}\ }\textbf {\bibinfo {volume} {113}},\ \bibinfo
       {pages} {61106} (\bibinfo {year} {2018})}\BibitemShut {NoStop}%
     \bibitem [{\citenamefont {Medvedev}\ \emph {et~al.}(2018)\citenamefont
       {Medvedev}, \citenamefont {Tkachenko}, \citenamefont {Lipp}, \citenamefont
       {Li},\ and\ \citenamefont {Ziaja}}]{Medvedev.2018}%
       \BibitemOpen
       \bibfield  {author} {\bibinfo {author} {\bibfnamefont {N.}~\bibnamefont
       {Medvedev}}, \bibinfo {author} {\bibfnamefont {V.}~\bibnamefont {Tkachenko}},
       \bibinfo {author} {\bibfnamefont {V.}~\bibnamefont {Lipp}}, \bibinfo {author}
       {\bibfnamefont {Z.}~\bibnamefont {Li}}, \ and\ \bibinfo {author}
       {\bibfnamefont {B.}~\bibnamefont {Ziaja}},\ }\href {\doibase
       10.1051/fopen/2018003} {\bibfield  {journal} {\bibinfo  {journal} {4Open}\
       }\textbf {\bibinfo {volume} {1}},\ \bibinfo {pages} {3} (\bibinfo {year}
       {2018})},\ \Eprint {http://arxiv.org/abs/1805.07524} {arXiv:1805.07524}
       \BibitemShut {NoStop}%
     \bibitem [{\citenamefont {Sokolowski-Tinten}\ and\ \citenamefont {{Von Der
       Linde}}(2004)}]{SokolowskiTinten.2004}%
       \BibitemOpen
       \bibfield  {author} {\bibinfo {author} {\bibfnamefont {K.}~\bibnamefont
       {Sokolowski-Tinten}}\ and\ \bibinfo {author} {\bibfnamefont {D.}~\bibnamefont
       {{Von Der Linde}}},\ }\href {\doibase 10.1088/0953-8984/16/49/R04} {\bibfield
        {journal} {\bibinfo  {journal} {J. Phys. Condens. Matter}\ }\textbf
       {\bibinfo {volume} {16}},\ \bibinfo {pages} {R1517} (\bibinfo {year}
       {2004})}\BibitemShut {NoStop}%
     \bibitem [{\citenamefont {Wei}\ \emph {et~al.}(1993)\citenamefont {Wei},
       \citenamefont {Kuo}, \citenamefont {Thomas}, \citenamefont {Anthony},\ and\
       \citenamefont {Banholzer}}]{Wei.1993}%
       \BibitemOpen
       \bibfield  {author} {\bibinfo {author} {\bibfnamefont {L.}~\bibnamefont
       {Wei}}, \bibinfo {author} {\bibfnamefont {P.~K.}\ \bibnamefont {Kuo}},
       \bibinfo {author} {\bibfnamefont {R.~L.}\ \bibnamefont {Thomas}}, \bibinfo
       {author} {\bibfnamefont {T.~R.}\ \bibnamefont {Anthony}}, \ and\ \bibinfo
       {author} {\bibfnamefont {W.~F.}\ \bibnamefont {Banholzer}},\ }\href {\doibase
       10.1103/PhysRevLett.70.3764} {\bibfield  {journal} {\bibinfo  {journal}
       {Phys. Rev. Lett.}\ }\textbf {\bibinfo {volume} {70}},\ \bibinfo {pages}
       {3764} (\bibinfo {year} {1993})}\BibitemShut {NoStop}%
     \bibitem [{\citenamefont {Reeber}\ and\ \citenamefont
       {Wang}(1996)}]{Reeber.1996}%
       \BibitemOpen
       \bibfield  {author} {\bibinfo {author} {\bibfnamefont {R.~R.}\ \bibnamefont
       {Reeber}}\ and\ \bibinfo {author} {\bibfnamefont {K.}~\bibnamefont {Wang}},\
       }\href {\doibase 10.1007/BF02666175} {\bibfield  {journal} {\bibinfo
       {journal} {J. Electron. Mater.}\ }\textbf {\bibinfo {volume} {25}},\ \bibinfo
       {pages} {63} (\bibinfo {year} {1996})}\BibitemShut {NoStop}%
     \bibitem [{\citenamefont {Shorr}(2015)}]{Shorr.2015}%
       \BibitemOpen
       \bibfield  {author} {\bibinfo {author} {\bibfnamefont {B.~F.}\ \bibnamefont
       {Shorr}},\ }\href {\doibase 10.1007/978-3-662-46968-2} {\emph {\bibinfo
       {title} {{Foundations of Engineering Mechanics Thermal Integrity in Mechanics
       and Engineering}}}}\ (\bibinfo  {publisher} {Springer},\ \bibinfo {year}
       {2015})\BibitemShut {NoStop}%
     \bibitem [{\citenamefont {Reiche}(1999)}]{Reiche.1999}%
       \BibitemOpen
       \bibfield  {author} {\bibinfo {author} {\bibfnamefont {S.}~\bibnamefont
       {Reiche}},\ }\href {\doibase 10.1016/S0168-9002(99)00114-X} {\bibfield
       {journal} {\bibinfo  {journal} {Nucl. Instruments Methods Phys. Res. Sect. A
       Accel. Spectrometers, Detect. Assoc. Equip.}\ }\textbf {\bibinfo {volume}
       {429}},\ \bibinfo {pages} {243} (\bibinfo {year} {1999})}\BibitemShut
       {NoStop}%
     \bibitem [{\citenamefont {Karssenberg}\ \emph {et~al.}(2006)\citenamefont
       {Karssenberg}, \citenamefont {{Van Der Slot}}, \citenamefont {Volokhine},
       \citenamefont {Verschuur},\ and\ \citenamefont {Boller}}]{Karssenberg.2006}%
       \BibitemOpen
       \bibfield  {author} {\bibinfo {author} {\bibfnamefont {J.~G.}\ \bibnamefont
       {Karssenberg}}, \bibinfo {author} {\bibfnamefont {P.~J.}\ \bibnamefont {{Van
       Der Slot}}}, \bibinfo {author} {\bibfnamefont {I.~V.}\ \bibnamefont
       {Volokhine}}, \bibinfo {author} {\bibfnamefont {J.~W.}\ \bibnamefont
       {Verschuur}}, \ and\ \bibinfo {author} {\bibfnamefont {K.~J.}\ \bibnamefont
       {Boller}},\ }\href {\doibase 10.1063/1.2363253} {\bibfield  {journal}
       {\bibinfo  {journal} {J. Appl. Phys.}\ }\textbf {\bibinfo {volume} {100}},\
       \bibinfo {pages} {93106} (\bibinfo {year} {2006})}\BibitemShut {NoStop}%
     \bibitem [{\citenamefont {Huang}\ \emph {et~al.}(2019)\citenamefont {Huang},
       \citenamefont {Li},\ and\ \citenamefont {Deng}}]{Huang.2019}%
       \BibitemOpen
       \bibfield  {author} {\bibinfo {author} {\bibfnamefont {N.~S.}\ \bibnamefont
       {Huang}}, \bibinfo {author} {\bibfnamefont {K.}~\bibnamefont {Li}}, \ and\
       \bibinfo {author} {\bibfnamefont {H.~X.}\ \bibnamefont {Deng}},\ }\href
       {\doibase 10.1007/s41365-019-0559-5} {\bibfield  {journal} {\bibinfo
       {journal} {Nucl. Sci. Tech.}\ }\textbf {\bibinfo {volume} {30}},\ \bibinfo
       {pages} {39} (\bibinfo {year} {2019})}\BibitemShut {NoStop}%
     \end{thebibliography}

%

\end{document}